\documentclass[journal=cgdefu,manuscript=article]{achemso}
\usepackage[utf8]{inputenc}

\usepackage{amsmath,amssymb}
\usepackage{booktabs}
\usepackage{array}
\usepackage{multirow}
\usepackage{rotating}
\usepackage{xcolor}
\usepackage{lineno}
\usepackage{hyperref}
\hypersetup{
    colorlinks=true,
    urlcolor=blue,
    citecolor=blue,
    linkcolor=blue
}
\providecommand{\doi}[1]{\href{https://doi.org/#1}{\textcolor{blue}{#1}}}
\setkeys{acs}{articletitle=true,doi=true}

\usepackage{xr}
\newif\ifdraftnotes
\draftnotesfalse         

\definecolor{ownercol}{RGB}{110,45,140}
\definecolor{todocol}{RGB}{190,30,30}
\definecolor{linkcol}{RGB}{0,90,180}
\definecolor{citecol}{RGB}{0,115,65}

\ifdraftnotes
  \hypersetup{colorlinks=true,
              linkcolor=linkcol,   
              citecolor=citecol,   
              urlcolor=citecol,
              linktoc=all,
              bookmarksnumbered=true}
\else
  \hypersetup{colorlinks=true,linkcolor=blue,citecolor=blue,urlcolor=blue}
\fi

\newcommand{\kjmol}{kJ\,mol$^{-1}$}

\newcommand{\Zp}{\ensuremath{Z'}}

\newcommand{\rmsdtw}{RMSD$_{20}$}

\SectionNumbersOn

\title{The Acridine Polymorph Landscape: Characterization and Selective Control by Additives}

\author{Adam Lahouari}
\affiliation[]{Department of Chemistry, New York University, New York, NY 10003, USA}

\author{Hengyu Zhou}
\affiliation[]{Department of Chemistry, New York University, New York, NY 10003, USA}

\author{Nikolaos Galanakis}
\affiliation[]{Department of Chemistry, New York University, New York, NY 10003, USA}

\author{Michael Kilgour}
\affiliation[]{Department of Chemistry, New York University, New York, NY 10003, USA}

\author{Daisuke Kuroshima}
\affiliation[]{Department of Chemistry, New York University, New York, NY 10003, USA}

\author{Jihye Han}
\affiliation[]{Department of Chemistry, New York University, New York, NY 10003, USA}

\author{Bart Kahr}
\affiliation[]{Department of Chemistry, New York University, New York, NY 10003, USA}

\author{Mark E. Tuckerman}
\affiliation{Department of Chemistry, New York University, New York, NY 10003, USA}
\alsoaffiliation{Simons Center for Theoretical Computational Chemistry, New York University, New York, New York 10003, USA}
\alsoaffiliation{NYU-ECNU Center for Computational Chemistry at NYU Shanghai, Shanghai 200062, China}

\author{Alexander G. Shtukenberg}
\affiliation[]{Department of Chemistry, New York University, New York, NY 10003, USA}

\author{Jutta Rogal}
\email{jrogal@flatironinstitute.org}
\affiliation[]{Initiative for Computational Catalysis, Flatiron Institute, New York, NY 10010, USA}


\begin{document}

\ifdraftnotes\linenumbers\fi

\begin{tocentry}
\includegraphics[width=10cm]{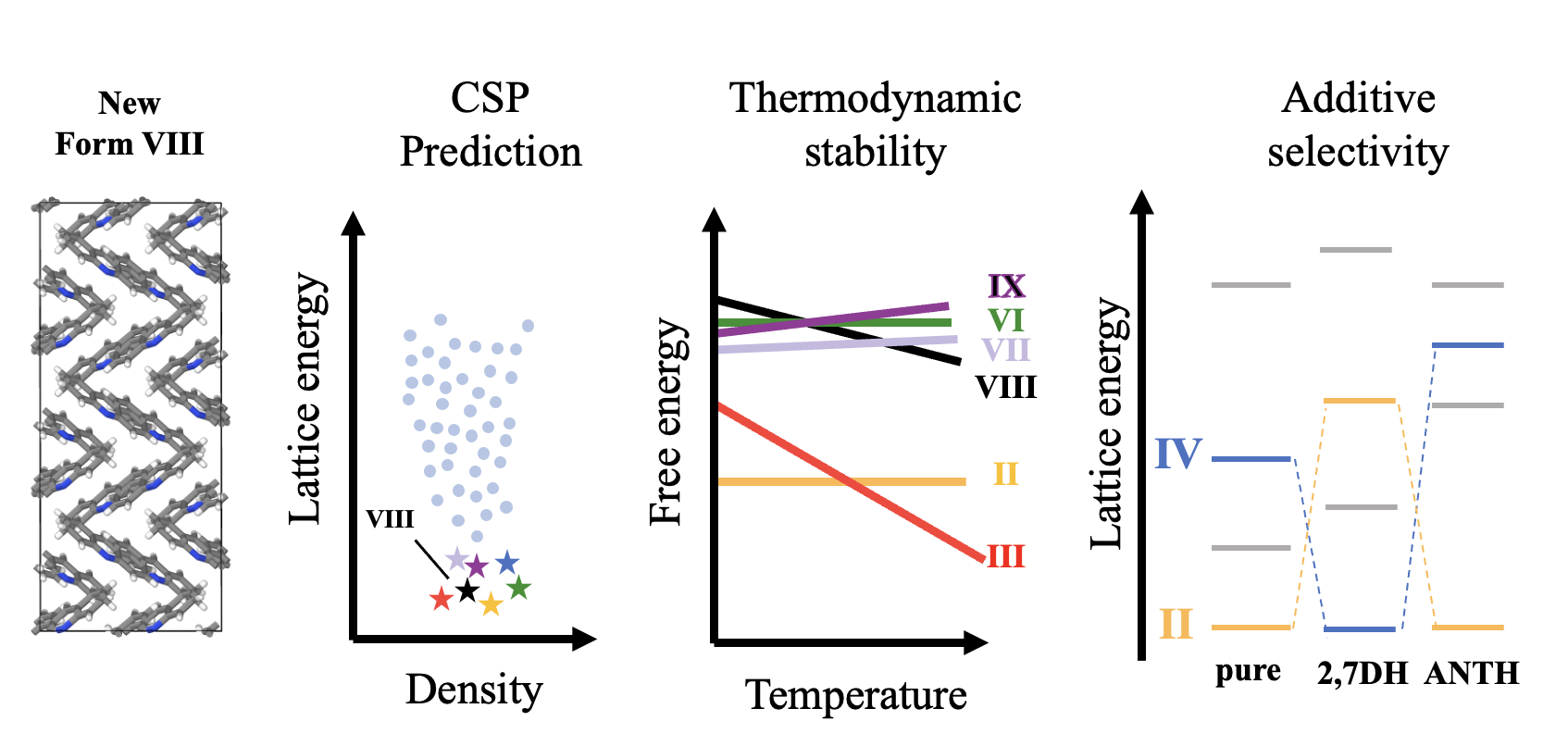}
\end{tocentry}

\begin{abstract}
Polymorphism in molecular crystals presents a major challenge for controlling crystallization. Here, we combine experiment and theory to characterize the polymorph landscape of acridine and investigate its selective control by additives. Melt crystallization experiments resolve the previously unknown structure of form VIII, while topological crystal structure prediction recovers all seven known unsolvated polymorphs with determined crystal structures. Combining a fine-tuned machine-learned interatomic potential with anharmonic free energy calculations reproduces the experimentally observed stability relationships and reveals a shallow free energy landscape. Finally, experimentally observed additive-induced polymorph preferences correlate with mixed-crystal substitutional defect energies and topological similarities between additive and acridine crystal structures. Together, these results provide a comprehensive picture of acridine polymorphism, linking crystal structure, thermodynamic stability, and additive-controlled crystallization.
\end{abstract}

\section{Introduction}
\label{sec:intro}

Polymorphism, the ability of a molecule to crystallize in different solid-state structures, is ubiquitous in molecular crystals. Different polymorphs can exhibit markedly different physical and chemical properties despite their identical molecular composition, making polymorphism highly relevant for applications ranging from pharmaceuticals~\cite{Hilfiker06,Brittain18} to organic electronic materials~\cite{Chung16,Gentili19} and agrochemicals~\cite{Yang2020,Zhu2021}.  A central challenge is therefore not only to identify and characterize possible crystal forms, but also to control which polymorph forms under a given set of conditions.
Polymorph formation can be influenced by a variety of  factors, including the choice of solvent,~\cite{Belenguer16,Threlfall00} temperature and pressure,~\cite{Chinnasamy22,Nyman16,Reyes22} confinement,~\cite{Fellah24,Hamilton12} and the presence of additives.~\cite{Black18,Zhang22} Additives are particularly intriguing in this context, as even small concentrations can substantially alter nucleation and growth and thereby selectively promote or suppress individual crystal forms.

Acridine, C$_{13}$H$_{9}$N, provides an attractive model system for investigating such effects because of its rich polymorphism.  
One hydrated and eight unsolvated crystal forms have been reported experimentally.
The crystallization conditions associated with the different acridine polymorphs have been summarized previously by Schur \emph{et al}.~\cite{Schur19} Most studies have focused on solution crystallization, where the common forms II, III, and IV are obtained from a wide range of solvents, frequently as mixtures. The less common forms VI and VII were initially discovered using additives,~\cite{Mei04} although form VII was subsequently crystallized without additives from several solvents.~\cite{Phillips56,Schur13,Stephens19} Form IX, in contrast, has only been obtained from toluene,~\cite{Stephens19} while the previously reported form V~\cite{Clarke75} has not been reproduced in more recent studies.~\cite{Schur13} 
Attempts to direct polymorph formation through solvent choice,~\cite{Phillips56} additives,~\cite{Mei04} or deuteration~\cite{Kupka12} have achieved selectivity in individual cases, but control generally remains limited, with multiple forms accessible under similar crystallization conditions. This rich polymorphism extends beyond solution crystallization: several forms have been obtained by vacuum sublimation~\cite{Clarke75} and crystallization from the melt.~\cite{Kofler43,Braga10,Fellah22} In particular, forms II, IV, VII, and VIII have been observed to grow directly from the melt in our previous study.~\cite{Fellah22} 
Acridine is also known to form solid solutions with the structurally related molecules anthracene and phenazine.~\cite{Lusi15,Schur15,Miasnikova58} 
Finally, several solid-state transformations between acridine polymorphs have also been reported,\cite{Lowde53,Phillips56,Braga10} further highlighting the complex relationships among its crystal forms.
This combination of pronounced polymorphism and its sensitivity to crystallization conditions and additives makes acridine a useful model system for investigating the molecular and thermodynamic factors governing polymorph formation.

Despite these extensive experimental studies, the molecular and thermodynamic origins of acridine polymorph selectivity, and particularly the mechanisms by which additives favor specific forms, remain poorly understood. Machine-learned interatomic potentials (MLIPs) are increasingly used to bring Density Functional Theory (DFT) accuracy to the large number of structures needed to rank organic crystal polymorphs,~\cite{lu2026omc,hanseroth2026fine} but general foundation models such as MACE-MP-0 and MACE-OFF23 still misrank the polymorphs of the 20 molecules of the CPOSS209 data set, which includes acridine.~\cite{Price25}
Machine-learning corrections to force-field energies have also recently been used to predict mixed crystals of acridine with anthracene and phenazine.~\cite{payne2026structure}
Here, we combine experiment with a range of computational approaches to characterize the polymorph landscape of acridine and investigate its selective crystallization by additives.  
Our topological crystal structure prediction approach~\cite{CrystalMath} is used to explore the accessible crystal structure landscape, identifying all experimentally known polymorphs among the predicted structures, while fine-tuned MLIPs~\cite{MolCryst,AMLP} and regularized thermodynamic integration (Reg-TI)~\cite{regti-kapil} allow the evaluation of their relative thermodynamic stabilities with high accuracy. 
The experimentally observed effects of additives are then examined from two complementary theoretical perspectives: substitutional defect energies quantify how additives alter the relative stability of the acridine polymorphs, while topological comparisons between the bulk crystal structures of the acridine polymorphs and additives provide a measure of their structural compatibility. 
For the additives exhibiting the strongest polymorph selectivity, clear correlations emerge between the experimentally preferred polymorphs and both their energetic and topological signatures, providing a theoretical rationale for the observed additive-controlled polymorph formation.

\section{Results and Discussion}
\label{sec:results}

\subsection{Experimental observation of acridine polymorphs}
\label{sec:pure-exp}
The main focus of the current experimental study was on acridine crystallization from the melt. Yet, to obtain some amounts of specific polymorphs, limited solution screening of acridine polymorphs was performed. Recrystallization from water-ethanol mixtures, toluene, and acetone typically provided forms III, IV, and VII, and in some cases also forms II and VIII. This is the first time form VIII was obtained from solution, it had previously only been crystallized from the melt. 

During melt crystallization, four different polymorphs -- forms III, IV, VI, and VIII -- were observed to directly nucleate from the melt above 273~K. In our earlier work,~\cite{Fellah22} four slightly different polymorphs were obtained from the melt, likely a result of different impurities in acridine chemicals. Our current findings together with our previous study suggest that most acridine polymorphs can directly crystallize from the melt. An overview of experimentally observed polymorphs together with results from the literature is compiled in Table~\ref{tab:cryst-methods}.

\begin{table}[H]
\centering
\caption{Crystallization of acridine polymorphs reported in the literature
and in this work (W).} 
\label{tab:cryst-methods}
\small
\setlength{\tabcolsep}{3pt}
\begin{tabular}{l cccccccc}
\toprule
Polymorph & II & III & IV & V & VI & VII & VIII & IX \\
\midrule
\multicolumn{9}{l}{\emph{Recrystallization from a solvent}}\\
Water                      & \citenum{Clarke75} &   & \citenum{Clarke75} &   &   &   &   &   \\
Ethanol                    & \citenum{Mei04,Kupka12,Schur15,Braga10,Clarke75} & \citenum{Clarke75,Kupka12,Schur13} & \citenum{Clarke75,Kupka12} & \citenum{Clarke75} &   &   &   &   \\
Ethanol--water mixture     & \citenum{Lowde53} & \citenum{Lowde53},W & W &   &   & W & W &   \\
Methanol                   & \citenum{Lusi15}$^{\&}$ & \citenum{Lusi15}$^{\&,\#}$,\citenum{Phillips56} &   &   &   &   &   &   \\
Methanol--water mixture    &   & \citenum{Schur13} &   &   &   &   &   &   \\
Ethanol--methanol mixture  & \citenum{Mei04,Schur15,Schur13} &   & \citenum{Schur13} &   & \citenum{Mei04}$^{*}$ & \citenum{Schur13}$^{*}$ &   &   \\
Acetone                    & W & \citenum{Schur13},W & \citenum{Schur13},W &   &   & \citenum{Phillips56,Schur13},W &   &   \\
Acetonitrile               & \citenum{Braga10} &   &   &   &   &   &   &   \\
Dimethylformamide          & \citenum{Mei04,Schur13} &   &   &   &   & \citenum{Mei04}$^{**}$ &   &   \\
Dimethylsulfoxide          &   & \citenum{Schur13} & \citenum{Braga10,Schur13} &   &   & \citenum{Schur13} &   &   \\
Tetrahydrofuran            & \citenum{Braga10} &   &   &   &   &   &   &   \\
Dioxane                    & \citenum{Braga10} &   &   &   &   &   &   &   \\
Ethylene glycol dimethyl ether &   & \citenum{Braga10} &   &   &   &   &   &   \\
Toluene                    & \citenum{Braga10,Schur13} & \citenum{Stephens19},W &   &   &   & \citenum{Stephens19},W &   & \citenum{Stephens19} \\
$p$-Xylene                 & \citenum{Braga10} &   & \citenum{Braga10} &   &   &   &   &   \\
Dichloromethane            &   & \citenum{Schur13} &   &   &   &   &   &   \\
Chloroform                 &   & \citenum{Phillips56} &   &   &   &   &   &   \\
Carbon tetrachloride       &   & \citenum{Phillips56} &   &   &   &   &   &   \\
\midrule
\emph{Sublimation crystallization} & \citenum{Clarke75} & \citenum{Clarke75} & \citenum{Clarke75} & \citenum{Clarke75} &   &   &   &   \\
\midrule
\emph{Melt crystallization}       & \citenum{Braga10,Fellah22} & W & W,\citenum{Fellah22} &  & W & \citenum{Fellah22} & W,\citenum{Fellah22} &   \\
\bottomrule
\end{tabular}

\vspace{4pt}
\begin{minipage}{0.95\textwidth}
\footnotesize
$^{\&}$Crystallization in the presence of anthracene. $^{\#}$Crystallization in the presence of anthracene and phenazine.
$^{*}$Crystallization in the presence of terephthalic acid. $^{**}$Crystallization in the presence of \emph{cis},\emph{trans}-muconic acid.
\end{minipage}
\end{table}

\begin{figure}[H]
    \centering
    \includegraphics[width=0.8\linewidth]{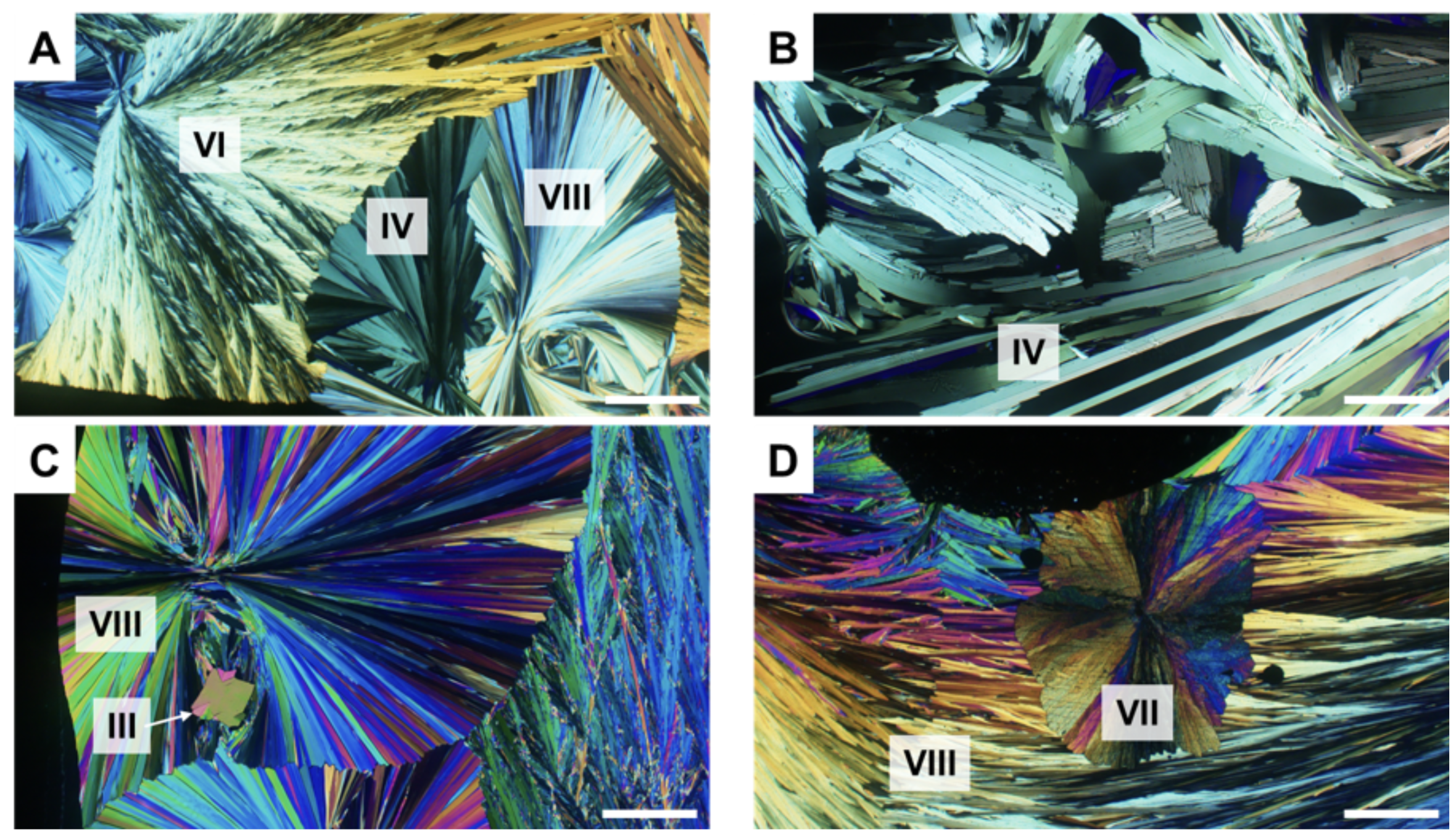}
    \caption{Polarized light micrographs of acridine polymorphs grown from the melt. 
    (A) Spherulites of forms IV and VIII grown from the melt at room temperature, with form VI cross-nucleated from form VIII. 
    (B) Form IV grown at 382~K. 
    (C) Growth of form III inside the ``eye'' of form VIII at room temperature. 
    (D) Form VII formed by phase transformation from Form VIII at room temperature. Scale bars are 500~$\mu$m.
    }
    \label{fig:micrographs}
\end{figure}

The most commonly observed polymorph is form VIII which nucleates from the melt at any temperature below the melting point. It usually forms straight, coarse-rayed spherulites with distinctive double-leaf structures and circular ``eyes''~\cite{Shtukenberg12}(Figure~\ref{fig:micrographs}A,C). 
Form IV nucleates either directly from the melt or cross-nucleates from forms VI and VIII below 373~K but is less common at temperatures close to the melting point. It forms either spherulites or single thin, straight fibers with grey, green, or pink anomalous interference colors (Figure~\ref{fig:micrographs}B). 
Form III rarely nucleates below 343~K and extends over only small areas of the film as it grows more slowly than the other polymorphs. It is usually seen near the nucleation centers of form VIII, in the regions of the spherulite ``eyes'' (Figure~\ref{fig:micrographs}C). Nonetheless, form III is the most common polymorph observed after phase transformation below room temperature. 
Polymorph VI grows only by cross-nucleation from forms VIII and IV, usually forming fibers thinner than those of forms VIII and IV grown under the same conditions.

Since the crystal structure of form VIII had not been previously reported, a single crystal of this polymorph was grown from the melt near its melting temperature and the structure was determined by single-crystal X-ray diffraction. Acridine VIII crystallizes in the space group $Cc$ with two molecules in the asymmetric unit (Table~\ref{tab:xrd-viii}).

\begin{table}[htbp]
\centering
\caption{Crystallographic data for acridine form VIII.} 
\label{tab:xrd-viii}
\small
\begin{tabular}{ll}
\toprule
empirical formula & C$_{13}$H$_{9}$N \\
formula weight, g/mol & 179.21 \\
crystal system & monoclinic \\
space group & $Cc$ \\
\Zp, $Z$ & 2, 8 \\
$a$, \AA & 12.582(11) \\
$b$, \AA & 13.644(13) \\
$c$, \AA & 11.950(10) \\
$\beta$, deg & 116.120(16) \\
volume, \AA$^3$ & 1842(3) \\
calculated density, g/cm$^{3}$ & 1.292 \\
temperature, °C & 100(2) \\
$R_1$ ($I > 2\sigma{_I}$)  & 0.0534  \\
$wR_2$ (all refl.) & 0.1547 \\
GoF & 1.029 \\
CCDC deposition number & 2588913 \\
\bottomrule
\end{tabular}
\end{table}

\subsection{Topological crystal structure prediction}
\label{sec:csp-topo}

Our theoretical exploration of acridine polymorphism begins with the identification of suitable crystal structures.
To this end, crystal structure prediction (CSP)~\cite{Price2014,lommerse2000,motherwell2002,day2005,day2009,bardwell2011,reilly2016,Hunnisett2024gen,Hunnisett2024rank} for acridine was carried out with CrystalMath,~\cite{CrystalMath} a topological approach to CSP that predicts stable molecular crystal structures and polymorphs from geometric and physical descriptors, without requiring energy evaluations.
CrystalMath is built on two central ideas: that in stable crystal structures the principal axes of the molecular inertia tensors and the normal vectors of rigid ring planes are aligned with specific crystallographic directions; and that heavy atoms occupy positions corresponding to the zeroes of a set of three-dimensional
geometric order parameters. Details are provided in Section~\ref{sec:methodTCSP}.

Acridine was treated as a single rigid fragment, consistent with its planar, fully fused tricyclic geometry. The search covered $Z' = 1$, 2, and 3 structures across the 19 most common space groups in the triclinic, monoclinic, and orthorhombic crystal systems in the Cambridge Structural Database (CSD)~\cite{CSD}. 
Across the three \Zp\ runs, 341{,}683 candidate structures were generated, of which 603 passed the initial geometric filters and were carried forward. 

\begin{figure}[H]
    \centering
    \includegraphics[width=0.7\linewidth]{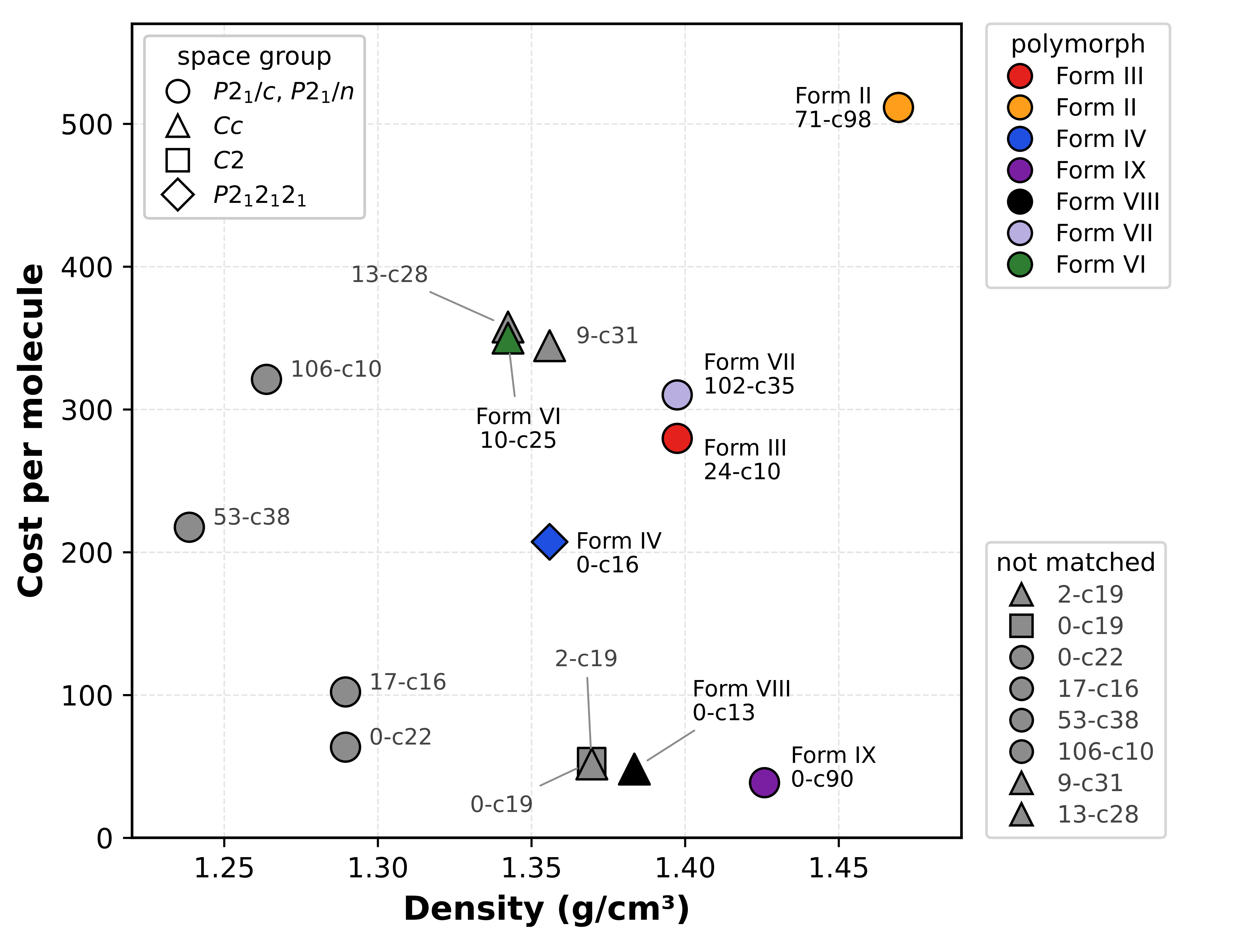}
    \caption{CrystalMath crystal structure prediction landscape for acridine. Each point is a structure in the final accepted pool, plotted as cost per molecule against density. Marker shape indicates the space group. Structures that match a known experimental polymorph at the 20/20 molecule level in a COMPACK packing-similarity comparison are colored by form. The eight unassigned structures are shown in grey.} 
    \label{fig:cmlandscape}
\end{figure}

After a multistage refinement and clustering to identify unique structures (see Section~\ref{sec:methodTCSP}), the final accepted pool contained 15 distinct structures across five space groups: eight in $P2_1/c$, five in $Cc$, one in $C2$, and one in $P2_12_12_1$. These 15 structures represent the complete set of CrystalMath predictions for acridine (the corresponding CIF files are provided in the Supporting Information). The corresponding landscape is shown in Figure~\ref{fig:cmlandscape} where the total cost is given by the topological cost function applied in CrystalMath.

The predicted structures were compared against seven acridine polymorphs, forms II, III, IV, VI, VII, VIII, and IX~\cite{Schur19}, using the COMPACK algorithm,~\cite{COMPACK} evaluating packing similarity over clusters of 20 molecules (\rmsdtw). 
All seven polymorphs are recovered within a final pool of only 15 structures, selected from 341{,}683 generated candidates without any energy evaluation. Recovering every known form of such a highly polymorphic system within so small a set of candidates is a demanding test of the method.

All seven polymorphs are matched at the 20/20 molecule level by at least one predicted structure, demonstrating complete coverage of the known acridine polymorph landscape. The best-match \rmsdtw\ values range from $0.207-0.381$~\AA\, as summarized in Table~\ref{tab:rmsd20-best}. The full $15 \times 7$ comparison matrix is given in supplemental Table~ref{tab:si-compack}. 
Finding form VIII is notable and suggests that the topological rules encoded in CrystalMath capture packing preferences that generalize beyond the training data. The eight remaining predicted structures in the final pool do not match any currently known acridine polymorph and may be forms that have not yet been found experimentally, which is consistent with the unusually large number of polymorphs already known for acridine.~\cite{Schur19}.

\begin{table}[H]
\centering
\caption{Best packing-similarity match for each experimental polymorph among the 15 CrystalMath predictions$^{a}$} 
\label{tab:rmsd20-best}
\small
\begin{tabular}{lcc}
\toprule
experimental form & best-matching prediction & \rmsdtw\ (\AA) \\
\midrule
Form II   & $P2_1/n$ 071-c98   & 0.364 \\
Form III  & $P2_1/c$ -024-c10  & 0.273 \\
Form IV   & $P2_12_12_1$ 0-c16 & 0.267 \\
Form VI   & $Cc$ 10-c25        & 0.207 \\
Form VII  & $P2_1/n$ 102-c35   & 0.252 \\
Form VIII & $Cc$ 0-c13         & 0.234 \\
Form IX   & $P2_1/n$ 0-c90     & 0.381 \\
\bottomrule
\end{tabular}

\vspace{4pt}
\footnotesize $^{a}$All matches are at the full 20/20 molecule level. The complete comparison matrix is given in supplemental Table~\ref{S-tab:si-compack}.
\end{table}

\begin{figure}[H]
    \centering
    \includegraphics[width=1.0\linewidth]{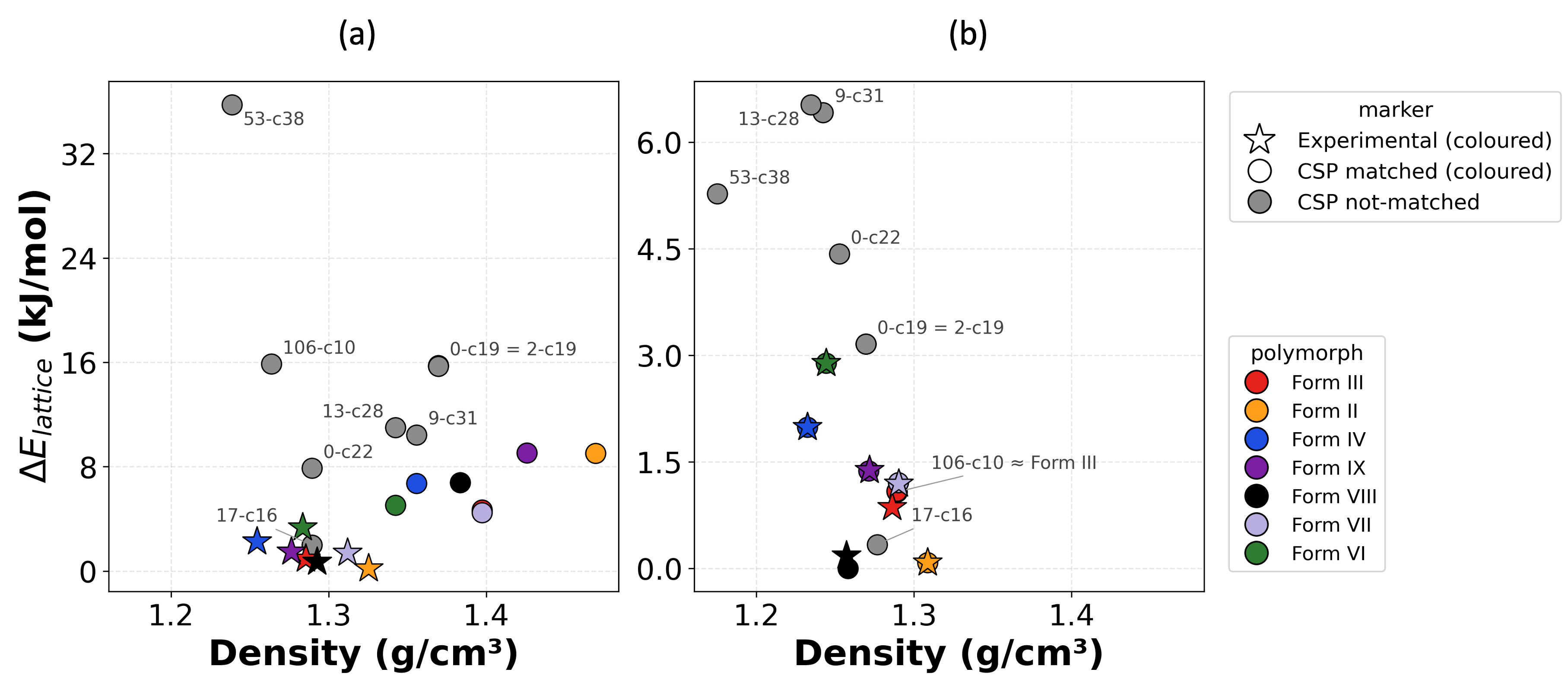}
    \caption{Relative lattice energy \(\Delta E_{lattice} \)versus density for the experimental acridine polymorphs (stars) and the CSP structures (circles), after (a) fixed-cell geometry optimization and (b) full cell optimization with the fine-tuned MLIP. CSP structures assigned to a known polymorph by \rmsdtw\ are shown in the same color as the corresponding experimental polymorph; unassigned CSP structures are shown in grey and labeled as index-ccell (e.g. 17-c16 for structure 000017 cell 16). Energies are per molecule and referenced to the lowest-energy structure.}
    \label{fig:csp_relative_Elatt}
\end{figure}

Using an MLIP fine-tuned for acridine (Section~\ref{sec:pure-mlip}), the lattice energies of the experimental and predicted structures can be compared.
For all structures a fixed-cell geometry optimization was performed, followed by a full relaxation of the atomic positions and lattice parameters (details in Section~\ref{sec:methodMLIP}). 
The corresponding energy vs. density plots are shown in Figure~\ref{fig:csp_relative_Elatt}. 
After the fixed-cell relaxation, the CSP structures are 0.06--0.15 g/cm$^3$ denser than their experimental counterparts and 1.7--8.8~kJ/mol higher in energy, indicating that CrystalMath predicted slightly compressed cell geometries.
After full relaxation, all predicted structures coincide with their corresponding experimental polymorph.  
The already low \rmsdtw\ values of the CSP candidate pool therefore translate into both structural and energetic agreement once the structures are fully relaxed. 

Two of the CSP structures that were not assigned to a known polymorph by \rmsdtw\ (grey markers in Figure~\ref{fig:csp_relative_Elatt}) stand out. Structure 106-c10 starts about 16~kJ/mol above the lowest energy polymorph and at a lower density than form III, yet it converges to form III upon relaxation, a match the \rmsdtw\ assignment of the unrelaxed structures did not pick up. The fully relaxed 17-c16 structure is only 0.33~kJ/mol above the lowest-energy structure, at 1.277 g/cm$^3$. This places it lower in energy than every known polymorph except forms II and VIII, suggesting that it may represent a polymorph that has yet to be observed. The six other unassigned structures lie 3.2--6.5~kJ/mol higher in energy, and two of them, 0-c19 and 2-c19, converge to the same $C2/c$ structure.

\subsection{Free energies of acridine polymorphs}
\label{sec:pure-mlip}

To go beyond $T=0$~K lattice energies and include finite temperature effects, we evaluate the full anharmonic free energies of acridine polymorphs.
The high computational cost of these calculations requires fast energy and force evaluations while maintaining near first-principles accuracy.
Here, we utilize an MLIP from the MolCryst database,~\cite{MolCryst} developed by fine-tuning the MACE-Multi Head 1 (MH1) foundation model with an omol head~\cite{omol, foundation-mh1} on a custom acridine dataset, using the Automated Machine Learning Pipeline (AMLP) code.~\cite{AMLP} 
Details regarding the MLIP fit are given in Section~\ref{sec:methodMLIP}.

\begin{figure}[htbp]
    \centering
    \includegraphics[width=0.8\linewidth]{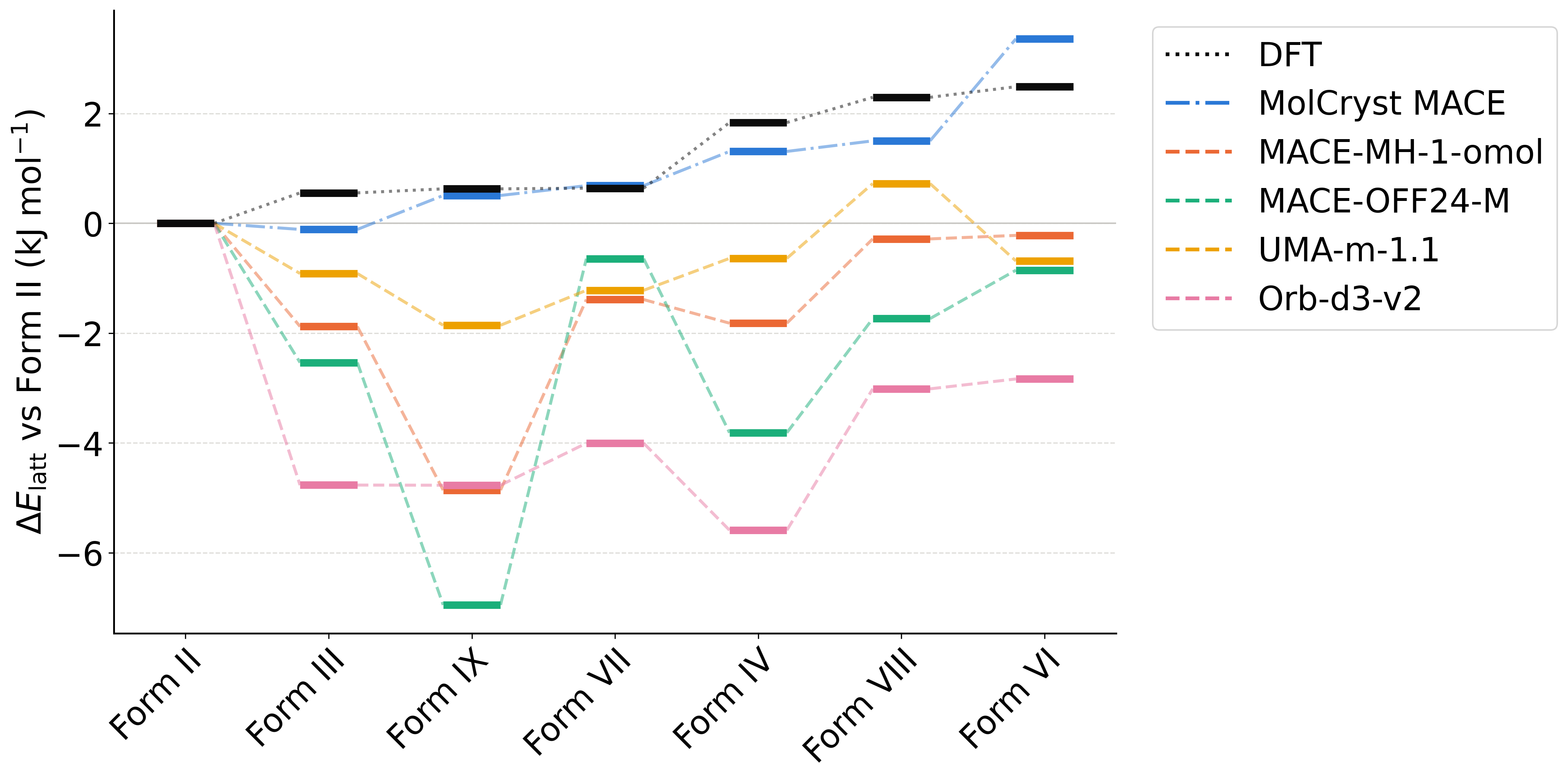}
    \caption{Relative lattice energies $\Delta E_{\text{latt}}$ of seven acridine polymorphs, referenced to form II, from single-point calculations at the PBE-D4 optimized geometries. Horizontal bars mark each form's energy, and dashed lines connect the forms for each model. Forms are ordered by increasing DFT energy.}
    \label{fig:ranking_ladder}
\end{figure}

The subtle energy differences between acridine polymorphs cannot be captured with any of the standard MLIP foundation models, making fine-tuning a necessary step to obtain reliable energetics.
In Figure~\ref{fig:ranking_ladder},  the lattice energies obtained with various foundation models are compared to DFT and our fine-tuned MolCryst MACE MLIP. 

Experimentally, forms~II and~III are the most stable acridine polymorphs which is correctly captured by the DFT lattice energies as well as the fine-tuned MolCryst MACE.  Except for a small shift between forms~II and~III (form III is more stable than form II by $0.1$~kJ/mol), the fine-tuned MLIP  gives the same energetic ordering of the polymorphs. In contrast, all four foundation models show large deviations in the energetic ordering as well as the magnitude in lattice energies.
The overall quality of the fine-tuned and foundation models as compared to the DFT reference is summarized in Table~\ref{tab:agree_mlip}. Clearly, fine-tuning significantly improved the accuracy of the applied MLIP which can then be employed to calculate  free energies.

\begin{table}[htbp]
    \centering
    \caption{Agreement of MLIP single-point lattice energies with DFT PBE-D4 for seven acridine polymorphs, evaluated at the DFT geometries. All metrics are computed over the seven forms, with form~II as the zero of energy.}
    \label{tab:agree_mlip}
    \begin{tabular}{lcccc}
        \toprule
        Model & Pearson $r$ & Spearman $\rho$ & Kendall $\tau$ & MAE (kJ\,mol$^{-1}$) \\
        \midrule
        MolCryst        & $0.88$  & $0.96$  & $0.90$  & $0.43$ \\
        MACE-MH-1-omol  & $0.31$  & $0.18$  & $0.24$  & $2.70$ \\
        MACE-OFF24-M    & $0.05$  & $-0.11$ & $-0.05$ & $3.57$ \\
        UMA-m-1.1       & $0.38$  & $0.21$  & $0.14$  & $1.86$ \\
        Orb-d3-v2       & $-0.21$ & $0.00$  & $0.05$  & $4.77$ \\
        \bottomrule
    \end{tabular}
\end{table}

\begin{figure}[H]
    \centering
    \includegraphics[width=0.8\linewidth]{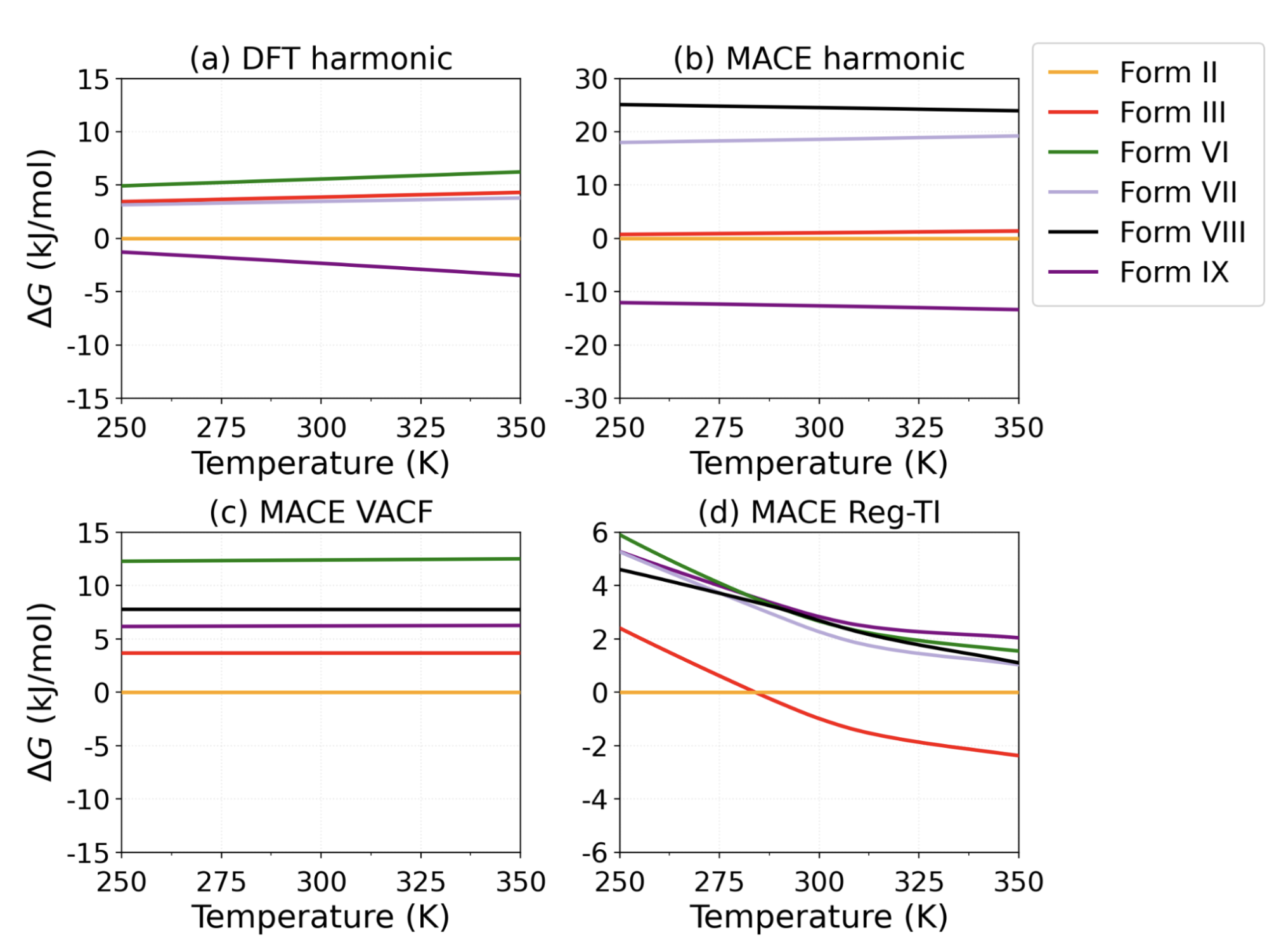}
    \caption{Free energy differences $\Delta G(T)$ between acridine polymorphs relative to form II. 
    (a) Harmonic approximation using DFT forces with the small-displacement method. 
    (b) Harmonic approximation using forces from the MolCryst fine-tuned MACE potential, evaluated on the same unit cells with the small-displacement method.
    (c) Quasi-harmonic approximation using the velocity autocorrelation function computed from NVT molecular dynamics trajectories with the MolCryst MACE model. 
    (d) Anharmonic Gibbs free energy from Reg-TI with MBAR reweighting of NPT trajectories, using the MolCryst fine-tuned MACE potential. 
    } 
    \label{fig:freeenergy}
\end{figure}

Three different approaches are used to evaluate free energies: the HA based on finite displacements, the Quasi Harmonic Approximation (QHA) derived from the Velocity Autocorrelation Function (VCAF) in the NVT ensemble, and a full anharmonic treatment based on Reg-TI. Details can be found in Section~\ref{sec:methodFE}.
The results are summarized in Figure~\ref{fig:freeenergy}, with the Harmonic Approximation (HA) based on DFT Perdew–Burke–Ernzerhof (PBE)-D4 and the fine-tuned MLIP shown in (a) and (b), respectively, the QHA in (c) and the full anharmonic results in (d). The HA and QHA yield free energy differences with respect to form~II that hardly change with temperature, indicating that they cannot resolve the small entropy differences between closely related crystal forms. They also do not reproduce any enantiotropic crossovers observed experimentally.
The anharmonic Gibbs free energies in Figure~\ref{fig:freeenergy}(d), on the other hand, show a rich free energy landscape with several enantiotropic relationships.

At 300~K, form III is the most stable polymorph, 0.99~kJ/mol below form II, becoming increasingly more stable at higher temperatures. This is consistent with the enantiotropic relationship between these two forms reported experimentally. The predicted form II-III crossover is close to 290~K, slightly below the
experimental estimate of $\sim$318~K.~\cite{Schur19} 

The remaining polymorphs, VI, VII, VIII, and IX, comprise a metastable group about 2.3--2.8~kJ/mol higher in energy than form II at 300~K (Table~\ref{tab:fe300}). Similar to form III, these polymorphs are stabilized with respect to form II with increasing temperature.
Several crossovers are observed within this group, with form IX becoming the least stable form above 300~K, which may explain why it is rarely obtained experimentally and hard to crystallize from the melt.
At higher temperature, form VIII becomes the most stable metastable form, crossing form VII and coming within about 1~kJ/mol of form II near 350~K because of its favorable entropy. All four metastable forms stay monotropic with respect to form II over the investigated temperature range.

\begin{table}[htbp]
\centering
\caption{Gibbs free energies at 300~K relative to form II, obtained from Reg-TI with MBAR reweighting. }
\label{tab:fe300}
\small
\begin{tabular}{lc}
\toprule
polymorph & $\Delta G_{300\,\mathrm{K}}$ (\kjmol) \\
\midrule
Form III  & $-0.99$ \\
Form II   & 0 (reference) \\
Form VII  & $+2.27$ \\
Form VI   & $+2.66$ \\
Form VIII & $+2.68$ \\
Form IX   & $+2.83$ \\
\bottomrule
\end{tabular}
\end{table}

In our experiments, phase transformations of acridine were followed with a hot-stage mounted on a polarized light microscope and confirmed by powder X-ray diffraction.
At room temperature, form VIII converts to forms III, VI, IV, and VII. Form VII looks mosaic-like and is darker between crossed polarizers than the other phases, and it is the only phase that can be identified from its optical appearance alone because it is opaque. Among these transformations, form VIII is monotropic with respect to form IV and enantiotropic with respect to forms III, VI, and VII, with enantiotropic points of about 371, 378, and 368~K. 
This is partially consistent with the free energy curves in Figure~\ref{fig:freeenergy}(d), predicting crossovers between form VIII and VI and VII. Form III, on the other hand, is more stable over the entire temperature range.
Furthermore, form VI converts monotropically to form IV above 358~K, and no transformation between these two forms is seen in our experiments below that temperature. There are also no transformations observed between forms IV and III, IV and VII, or VII and III. While form IV was not included in our free energy calculations, the latter observation is consistent with the theoretical predictions.

Overall, the polymorph landscape in acridine is complex and challenging to assess both experimentally and theoretically. Employing a fine-tuned MLIP together with full anharmonic free energy calculations, we were able to capture key features of relative polymorph stability and enantiotropic transformations.

\subsection{Effect of additives on polymorph formation}
To assess the effect of additives on polymorphic outcomes, we combine experimental observations of polymorph nucleation from the melt with the calculation of defect formation energies and a topological analysis of the crystalline phases of acridine and the additives.

\subsubsection{Polymorph nucleation statistics}
\label{sec:add-exp}

\begin{figure}[H]
    \centering
    \includegraphics[width=0.95\linewidth]{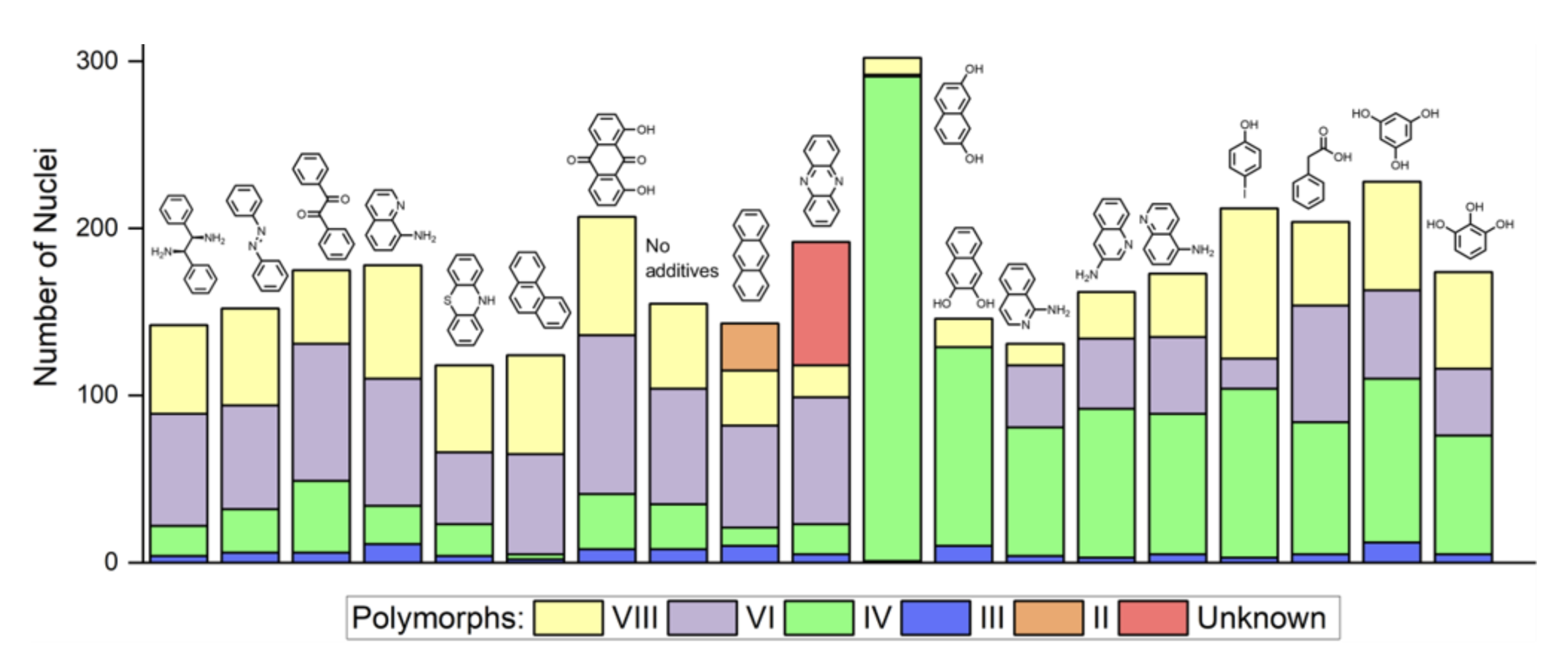}
    \caption{Nucleation statistics of acridine grown from the melt at 273~K in the presence of 18 additives at 5~wt~\%. Bars are stacked by polymorph. } 
    \label{fig:nucleation}
\end{figure}

Experimentally, we studied acridine crystallization from the melt in the presence of 18 compounds at 5~wt~\%.  The additives were chosen based on the similarity of their molecular structures and melting points to acridine.
The number of nucleation centers of each polymorph were counted for films crystallized at 273~K, room temperature, and 333~K. Similar trends were observed at  all three temperatures. 
Figure~\ref{fig:nucleation} presents the results obtained at 273~K, where slower crystal growth allows more nuclei to form from the melt, before the space is fully occupied.

Seven additives had no effect on the relative appearance of acridine polymorphs (left side in Figure~\ref{fig:nucleation}). Seven others increased the fraction of polymorph IV, though not dramatically (right side in Figure~\ref{fig:nucleation}). Of the four remaining additives, 2,3-dihydroxynaphthalene and 2,7-dihydroxynaphthalene strongly favored the nucleation of form IV, while anthracene promoted nucleation of form II that is not seen 
without additives. When adding phenazine, a new polymorph or mixed crystal was observed, labeled unknown, that transforms into a structure resembling form VIII above 363~K.
Its crystal structure is likely close to that of polymorph VIII, since their powder diffraction patterns are very similar (supplemental Figure~\ref{S-fig:si-pxrd}).

The notable shift towards form IV when adding 2,3-dihydroxynaphthalene and 2,7-dihydroxynaphthalene, as well as the nucleation of form II in the presence of anthracene mark the most pronounced effects that were further explored from a theoretical point of view.

\subsubsection{Mixed-crystal substitutional defect energies}
\label{sec:add-mlip}

\begin{figure}[htbp]
    \centering
    \includegraphics[width=0.8\linewidth]{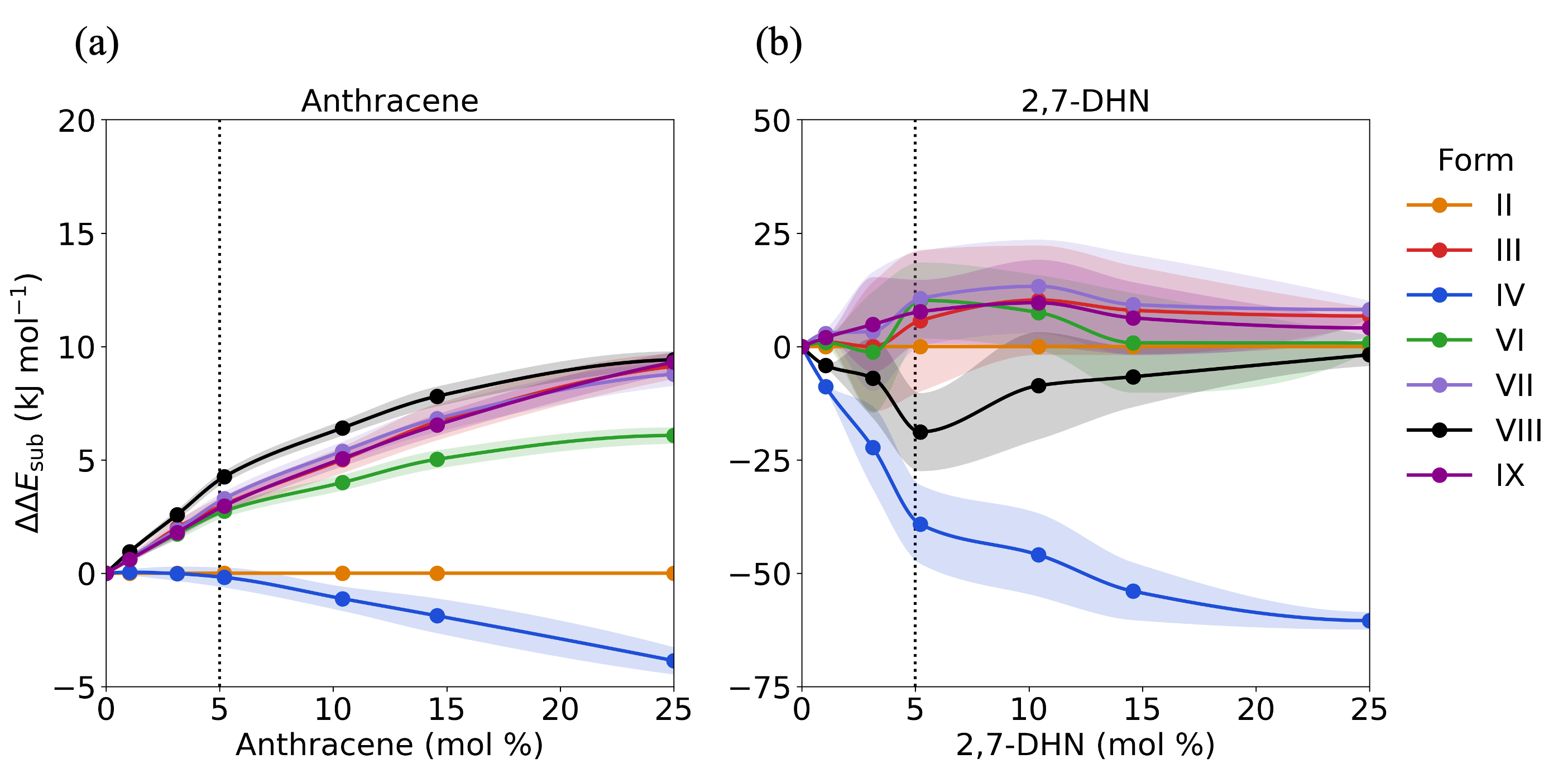}
    \caption{Substitution energy of the acridine polymorphs relative to form II, $\Delta\Delta E_{\mathrm{sub}}^{\text{poly}}(c)$ (Eq.~\ref{eq:ddesub}), as a function of the additive concentration $c$ (mol\,\%) for anthracene (a) and 2,7-dihydroxynaphthalene (b). Values are MACE single-point energies of 96-molecule supercells, averaged over random substitution patterns (10 for anthracene, up to 20 for 2,7-dihydroxynaphthalene); shaded areas show standard deviation. The dotted vertical line marks the additive concentration used in the crystallization experiments.
    } 
    \label{fig:substitution}
\end{figure}

The small additive concentrations require rather large simulation cells that cannot be evaluated directly using electronic structure calculations. We therefore extended the fine-tuned MACE MLIP to two of the additives with the largest effect on polymorph selectivity, anthracene and 2,7-dihydroxynaphthalene. Details regarding the MLIP training are given in Section~\ref{sec:methodMLIP}.
Mixed crystal supercells were set up for all seven known acridine polymorphs, forms II, III, IV, VI, VII, VIII, and IX, at six additive concentrations (1, 3, 5, 10, 15, and 25~ mol \%), with anthracene or 2,7-dihydroxynaphthalene molecules substituted at random acridine sites, respectively. The effect of the additives is  measured using a substitutional defect energy defined as
\begin{equation}
\Delta E_{\mathrm{sub}}(c) =
 \frac{\left(E_{\mathrm{mix}}(n) - E_{\mathrm{pure}} \right) - n\,\Delta\mu}{N_\text{mol}} \quad ,
\label{eq:esub}
\end{equation}
where $E_{\mathrm{mix}}(n)$ is the total energy of the supercell doped with $n$ additive molecules, $E_{\mathrm{pure}}$ is the total energy of the undoped acridine supercell, $N_{\mathrm{mol}}$ is the total number of molecules in the supercell, $c = n/N_\text{mol}$ is the additive concentration, and $\Delta\mu = E^{\mathrm{mol}}_{\mathrm{add}} -  E^{\mathrm{mol}}_{\mathrm{acr}}$ is the energy difference between the isolated additive and acridine molecules in the gas phase. 
This is equivalent to the difference in lattice energy between acridine with and without additives.
To assess the relative effect of additive concentration on the different polymorphs, the substitutional defect energies are referenced to form II
\begin{equation}
\Delta\Delta E_{\mathrm{sub}}^{\text{poly}}(c) =
\Delta E_{\mathrm{sub}}^{\text{poly}}(c) 
-
\Delta E_{\mathrm{sub}}^{\text{II}}(c)  \quad.
\label{eq:ddesub}
\end{equation}
$\Delta\Delta E_{\mathrm{sub}}^{\text{poly}}(c) < 0$ indicates that additives in a given polymorph reduce the difference in lattice energy to form II, making that polymorph more favorable, and $\Delta\Delta E_{\mathrm{sub}}^{\text{poly}}(c) > 0$ means a polymorph becomes less favorable.
Figure~\ref{fig:substitution} shows the relative substitutional defect formation energies as a function of concentration for anthracene (a) and 2,7-dihydroxynaphthalene (b). 
Clear trends emerge for both additives. 

Substituting anthracene increases $\Delta\Delta E_{\mathrm{sub}}^{\text{poly}}(c)$ for forms III, VI, VII, VIII, and IX, in particular for the low concentrations probed in experiment, with 0.6--1.0~kJ/mol at 1~mol\,\% and 2.7--4.3~kJ/mol at 5~mol\,\% (SI, Table~\ref{S-tab:si-subst-anth}). 
Form IV remains comparable to form II up to 5~mol\,\% and becomes slightly more favorable at higher concentrations, though these concentrations should rather be considered as co-crystals than substitutional defect.
The increase in relative lattice energies up to $3-5$~kJ/mol around 5~\% additive concentration is consistent with the observation that anthracene promotes the nucleation of form II.

The addition of 2,7-dihydroxynaphthalene, on the other hand, strongly stabilizes form IV relative to form II at all concentrations: by $8.9 \pm 0.6$~kJ/mol already at 1~mol\,\% and $39 \pm 9$~kJ/mol at 5~mol\,\%. 
(SI, Table~\ref{S-tab:si-subst-dhn}).
Form VIII is slightly stabilized with respect to from II, 
while forms III, VI, VII and IX show an increase in  $\Delta\Delta E_{\mathrm{sub}}^{\text{poly}}(c)$.
The larger variance in the defect formation energies between 3 and 15~mol\,\% results from the random substitution patterns, in which additive molecules can sit next to each other. 
Experimentally, the nucleation of form IV from the melt is promoted in the presence of 2,7-dihydroxynaphthalene, which is again consistent with the trend in stability of the mixed crystals.

The mixed-crystal energetics thus provide a first theoretical rationale for the experimentally observed promotion of form II by anthracene and form IV by 2,7-dihydroxynaphthalene.

\subsubsection{Topological similarity between bulk crystals of acridine polymorphs and additives}
\label{sec:add-topo}

\begin{figure}[htbp]
  \centering
  \includegraphics[width=\linewidth]{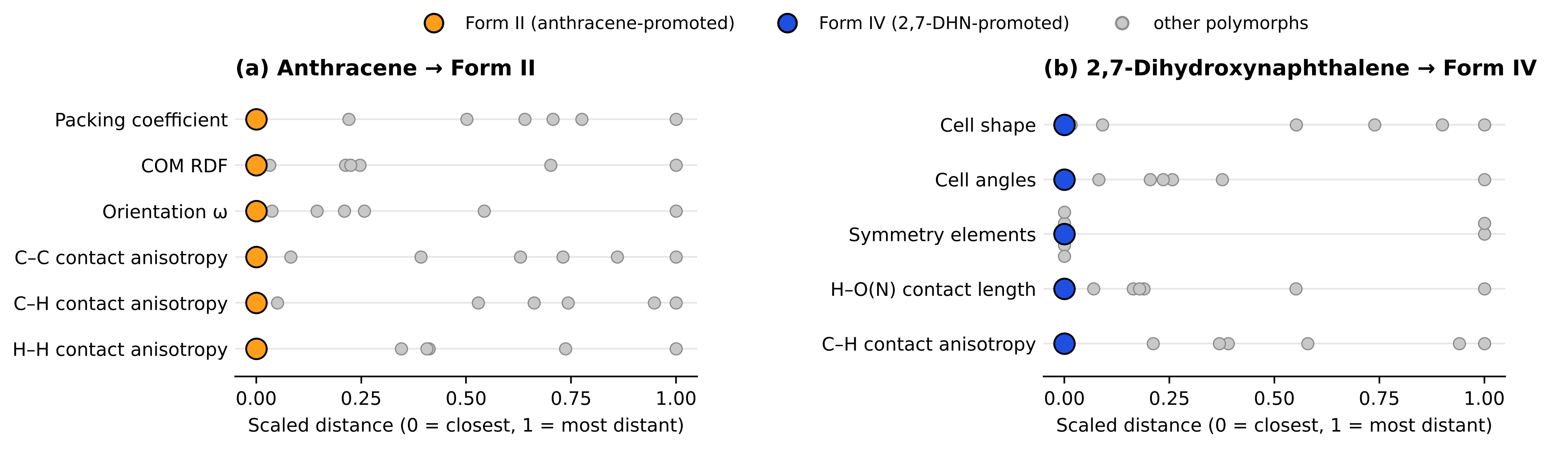}
  \caption{Topological similarity between the bulk crystals of the two most selective additives and the seven acridine polymorphs, for the descriptors discussed in the text. (a) Anthracene and (b) 2,7-dihydroxynaphthalene. Each descriptor is min--max scaled across the seven polymorphs, with 0  the closest and 1 the most distant match to the additive crystal. The experimentally promoted form is highlighted; the remaining polymorphs are shown in grey. The experimentally promoted forms, II for anthracene and IV for 2,7-dihydroxynaphthalene, are the closest match for every descriptor shown. The full set of descriptors is given in SI Figures~\ref{S-fig:si-radar-anth} and~\ref{S-fig:si-radar-27dh} and Tables~\ref{S-tab:si-anth-t1}--\ref{S-tab:si-cono}.}
  \label{fig:add-topo}
\end{figure}

As a complementary theoretical indicator of additive-induced polymorph selectivity, topological similarities between the bulk crystal structures of the acridine polymorphs and additives were assessed in terms of crystal packing, molecular orientation, symmetry, and intermolecular interactions (details regarding the similarity measures are given in the SI section~\ref{S-si:topo-sim}). 
Rather than identifying a single universal descriptor of polymorph selectivity, the analysis reveals that the relevant structural similarity measures depend strongly on the molecular topology of the additive. The additives were therefore classified into four categories according to features such as ring number and fusion, molecular shape, and heteroatom placement: (1) three-fused-ring aromatic molecules, (2) amino-substituted double-fused-ring positional isomers, (3) dihydroxynaphthalene positional isomers, and (4) molecules comprising two non-fused aromatic rings connected by rotatable bonds. Different combinations of topological descriptors emerge as most informative for these different molecule categories. The complete topological analysis of all additives and descriptors is provided in the Supplemental Information. 
As in the energetic analysis, we focus here on anthracene and 2,7-dihydroxynaphthalene, which exhibit the strongest polymorph selectivity.
The descriptors most relevant to these two additives are summarized in Figure~\ref{fig:add-topo}.
 
For anthracene, which selectively promotes the nucleation of form II, the topological analysis reveals a particularly clear correlation. As a rigid three-ring aromatic molecule closely resembling acridine, its crystalline structure displays several independent similarities specifically with form II, including an almost identical packing density, compatible molecular shape and symmetry, and similarities in local packing and molecular orientation. The consistent preference for form II across  these descriptors suggests that the observed selectivity may originate from a high degree of structural compatibility between bulk anthracene and form II, consistent with an isomorphic-substitution mechanism.  Conversely, the structural similarities observed for anthracene are not reinforced across multiple descriptors for the other polymorphs, rationalizing its selectivity toward form II. Similarly, the other additives in this category lack such consistent similarity with any single polymorph, in line with their absence of polymorph selectivity.

A different set of structural correlations emerges for 2,7-dihydroxynaphthalene, which experimentally promotes form IV  (Figure~\ref{fig:add-topo}(b)). Here, the strongest correspondence is found in the cell geometry: form IV is the only orthorhombic acridine polymorph and provides the closest match in cell shape and cell angles to the additive crystal. Form IV also provides the closest match in the directionality of its C--H contacts. Moreover, form IV is unique among the acridine polymorphs in having more hydrogen- than carbon-mediated contacts to its ring nitrogen, and the characteristic length scale of this network closely resembles the hydroxyl–oxygen interaction network in crystalline 2,7-dihydroxynaphthalene. This provides an additional possible connection between the topology of the additive crystal and its experimentally observed preference for form IV. 

Overall, these examples suggest that the bulk crystal structures of the additives can provide an intuitive understanding of which acridine polymorph may be preferentially promoted, based on distinct combinations of topological similarities for different additive classes.

\section{Conclusions}
\label{sec:conclusions}

We have mapped the acridine polymorph landscape by combining melt crystallization experiments with topological crystal structure prediction and  anharmonic free energy calculations based on a fine-tuned MLIP, and extended this framework to investigate the selective control of polymorph formation by additives.
Four results stand out. 
First, we report the crystal structure of form VIII for the first time and show that four polymorphs, forms III, IV, VI, and VIII, grow directly from the melt above 273~K. 

Second, topological crystal structure prediction identifies all seven known forms at the 20/20 packing-similarity level without evaluating an interatomic potential, including the newly solved form VIII, and proposes eight additional structures that may be experimentally realizable but have not yet been characterized. Notably, form VIII was unknown when the statistical distributions underlying the topological CSP were constructed, demonstrating that its identification does not result from prior knowledge of its crystal structure.

Third, the experimentally observed stability ordering is recovered only by combining a system-specific fine-tuned MLIP with fully anharmonic free energy calculations. An off-the-shelf foundation model gives an incorrect 0~K ranking, while free energies based on harmonic and quasi-harmonic approximations fail to reproduce the observed enantiotropic crossovers. The full anharmonic free energy landscape is consistent with experiment, reproducing the enantiotropic relationship between forms II and III and placing the four metastable forms within only 0.6~kJ/mol at 300~K, consistent with the concomitant crystallization, cross-nucleation, and variability observed experimentally.

Fourth, the experimentally observed impact of additives on polymorph nucleation is consistent with both the mixed-crystal substitutional defect energies and the topological similarities between the bulk crystal structures of the additives and acridine polymorphs.

Taken together, this work provides a comprehensive experimental and theoretical picture of acridine polymorphism, linking its crystal structure landscape and thermodynamic stability to experimentally observed crystallization behavior and additive-induced polymorph selectivity. These results provide a basis for understanding and controlling polymorph formation in acridine and, more broadly, in highly polymorphic molecular crystals.

\section{Methods}
\label{{sec:methods}}

\subsection{Experimental methods}
Acridine (Thermo Fisher, 97 \%), 1-aminoisoquinoline (Sigma-Aldrich, 99 \%), 3-aminoquinoline (Sigma-Aldrich, 98 \%), 5-aminoquinoline (Sigma-Aldrich, 97 \%), 8-aminoquinoline (Sigma-Aldrich, 98 \%), anthracene (Sigma-Aldrich, 97 \%), azobenzene (Sigma-Aldrich, 98 \%), benzil (Sigma-Aldrich, 98 \%), 1,8-dihydroxyanthraquinone (Sigma-Aldrich, 96 \%), 2,3-di\-hydroxy\-naphthalene (Eastman Chemical, no purity indicated), 2,7-dihydroxynaphthalene (Sigma-Aldrich, 97 \%), (1R,2R)-(+)-1,2-diphenyl-ethylenediamine (Sigma-Aldrich, 97 \%), 4-iodophenol (Sigma-Aldrich, 99 \%), orcinol (Sigma-Aldrich, 97 \%), phenanthrene (Acros Organics, 97 \%), phenazine (Acros Organics, 98 \%), phenothiazine (Sigma-Aldrich, 98 \%), phenylacetic acid (MCB, 99 \% ), pyrogallol (Thermo Fisher, 99 \%) were used without additional purification. Ethanol, toluene, and acetone used for solution growth are all reagent-grade. Acridine-additive mixtures were prepared by grinding acridine together with 5 wt\% of an additive.

Several milligrams of acridine or acridine-additive mixtures were sandwiched between two 0.1 mm glass cover slips, then melted to form a thin film. The thickness of these films was several microns. The melts were crystallized using a hot stage (Model FP90, Mettler-Toledo), a Kofler bench, or by rapid cooling between two metal blocks in the temperature range between melting point and 273~K. The sample was held isothermally to observe crystallization and possible phase transformations. Crystallization and phase transformations were monitored using polarized light optical microscopes Olympus BX50 and BX53 equipped with digital cameras. 

The approximate ratios between the nucleation rates of different polymorphs were obtained for crystallization at 273~K, room temperature (approximately 296~K), and 333~K counting the nucleation centers and cross nucleation for each polymorph in 1 to 5 slides. The total number of nuclei exceeded 100 to obtain representative statistics. 

Polymorphs were identified using a Bruker D8 Discover General Area Detector Diffraction System (GADDS) equipped with a V\AA NTEC-2000 2D detector and Cu-K$\alpha$ source ($\lambda$ = 1.54178 \AA). The X-ray beam was monochromated with a graphite crystal and collimated with a 0.5 mm capillary collimator (MONOCAP). X-ray powder diffraction patterns were collected from powder loaded into 0.8 mm Kapton capillaries in transmission mode or from as-grown crystalline films on glass slides in reflection mode.

Higher resolution powder X-ray diffraction patterns were collected using a Rigaku SmartLab diffractometer equipped with a rotating anode Cu-K$\alpha$ source ($\lambda$ = 1.54178 \AA) and HyPix-3000 detector using a convergent beam (CBO-E) optics. The material was placed into 0.8 mm Kapton capillaries. The patterns were recorded in 1D scan mode with $2\theta$ scans, $2\theta = 3-40 ^{\circ}$, scanning rate 0.3 $^{\circ}$/min. 

Single crystal X-ray diffraction data sets were acquired on a Bruker D8 SMART APEX II 'three-circle diffractometer' system equipped with a Incoatec 'microfocus sealed X-ray tube' (Mo-K$\alpha$ radiation, $\lambda$ = 0.71073 \AA), a multilayer optics monochromator and a PHOTON-II-C14 detector. The crystal temperature (100 K) was controlled by an Oxford Cryosystems 700+ Cooler. Crystals were mounted on a 0.2 mm MicroMount (MiTeGen) with Type B immersion oil (Cargille Labs). Data were collected and processed using the APEX4 software for data reduction, data correction, and cell refinement.~\cite{APEX4} The structures were solved by intrinsic phasing methods SHELXT~\cite{SHELXT} and refined with full-matrix least square methods of $F^2$ by SHELXL~\cite{SHELXL}. All non-hydrogen atoms were refined with anisotropic displacement parameters, and hydrogen atoms on carbons were placed in idealized positions (C-H = 0.95-1.00 \AA) and included using a riding model with $U_\mathrm{iso}$(H) = 1.2 or 1.5 $U_{eq}$(non-H). Crystallographic data have been deposited with the Cambridge Crystallographic Data Centre with the deposition number 2588913. Copies of these data can be requested from, free of charge, the CCDC website at \url{https://www.ccdc.cam.ac.uk/structures/}.

\subsection{Topological crystal structure prediction}
\label{sec:methodTCSP}
The version of CrystalMath used here includes several methodological improvements over the originally published protocol.~\cite{CrystalMath} Candidate crystals are now constructed from rigid molecular fragments using geometric statistics mined from the CSD for chemically comparable fragments, rather than by random sampling, and the resulting cells are expanded across 19 common space-group settings. Candidate structures are then relaxed with a flexible unit-cell optimization and ranked by a cost function that measures how closely their packing coefficient and intermolecular contacts resemble those observed for related molecules in the CSD.

The $Z' = 3$ search produced the largest pool, with 187{,}696 candidate structures, compared with 93{,}448 for $Z' = 1$ and 60{,}539 for $Z' = 2$; of these, 271, 290, and 42 structures, respectively, passed the initial geometric checks. The distribution of these structures across space groups is given in SI Table~\ref{S-tab:sg_distribution}. The total wall-clock time for this stage was about 6.0~h on a GPU-accelerated mobile workstation (NVIDIA RTX A-1000 6GB).

This initial pool was then put through the CrystalMath filtering and optimization pipeline, described in SI Section~\ref{S-si:crystalmath}. Structures are first screened against geometric order parameter constraints that require heavy atoms to sit at or near preferred crystallographic positions for the target space group. The surviving structures are then relaxed with a flexible unit-cell optimization that drives their packing coefficient and intermolecular close-contact distances toward values observed in the CSD, a step that takes the place of energy minimization. The relaxed structures are screened again for intermolecular contacts, and the structures that pass all filters are clustered by packing similarity and deduplicated.

\subsection{Machine-learned interatomic potential}
\label{sec:methodMLIP}

\subsubsection{Pure acridine}
All reference calculations were performed with DFT in VASP,\cite{vasp1,vasp2,vasp3,vasp4} using the PBE functional\cite{PBE} with the D4 dispersion correction\cite{d4-periodic} and a plane-wave cutoff of 850~eV; $k$-point meshes, convergence settings and the relaxation protocol are given in the Supporting Information (see \ref{S-si:dft}). The potential for pure acridine is a MACE model\cite{MACE} fine-tuned from the MACE-Multi Head 1 (MH1) foundation model with the \texttt{omol} head.\cite{foundation-mh1,omol} The training set comprises 21\,685 DFT configurations from the seven polymorphs: NVT ab initio MD at 50--600~K, phonon-displacement configurations (used for the harmonic approximation), and configurations from active-learning NVT simulations recomputed with single-point DFT. The model reaches a mean absolute error of 3.7~meV\,atom$^{-1}$ for energies and 8.3~meV\,\AA$^{-1}$ for forces. The reference calculations, the dataset and the training settings are described in the Supporting Information (see \ref{S-si:mlip}). 

\subsubsection{Acridine and additives}
To put the additive effect on an energetic footing, we developed separate fine-tuned MACE MLIPs for the two additive systems with the clearest experimental selectivity, acridine/anthracene and acridine/2,7-dihydroxynaphthalene, the two additive systems with the clearest experimental selectivity. Both start from the MACE-MH-1 foundation model with the \textsc{omol} head and use the same architecture and training settings as the pure-acridine model. 
The reference data were built from the unit cells of the seven acridine polymorphs, including all combinations of substituted sites. Each additive molecule was placed on the site of the acridine molecule it replaces, with the same position and orientation. All mixed cells were relaxed with DFT (using the same parameters than the relaxed pure acridine), and the structures along each relaxation were added to the training set, together with gas-phase configurations of acridine and of the additive. Both datasets were then extended by AL and split 85/15 into training and validation sets, giving 26\,968 training configurations for acridine/anthracene and 18\,790 for acridine/2,7-DHN.
On the validation sets, the energy MAEs are 0.13 and 0.17~\kjmol\,atom$^{-1}$ and the force MAEs 1.07 and 0.76~\kjmol\,\AA$^{-1}$ for the anthracene and 2,7-DHN models (SI Table~\ref{S-tab:si-mlip-acc}, Figures~\ref{S-fig:si-parity-anth} and~\ref{S-fig:si-parity-dhn}).

\subsubsection{Structure relaxation}
The seven polymorphs were first relaxed with DFT (PBE-D4), starting from their CSD structures and optimizing atomic positions and cell together. These relaxed structures are the starting point for the lattice energies and for all free energy
calculations below. The mixed crystals were relaxed in the same way by starting from the DFT optimized structure, with the corresponding additive model.

\subsection{Free energy calculations}
\label{sec:methodFE}

\subsubsection{Finite displacements}
HA free energies were computed with the finite-displacement method as implemented in
\texttt{phonopy},\cite{phonopy} using forces from both DFT and the MACE potential. Force constants were built from displaced
configurations of the relaxed cells, and the vibrational free energy was obtained from the resulting phonon frequencies on a mesh of wavevectors. Cell sizes, displacement amplitudes and meshes are given in the Supporting Information (see \ref{S-si:ha}).

\subsubsection{Velocity autocorrelation function}
To include finite-temperature effects on the vibrational spectrum, the vibrational density of states (VDOS) was also obtained from the Fourier transform of the mass-weighted VCAF of 100~ps MACE-driven NVT trajectories at 300~K. Vibrational free energies between 250 and 350~K were computed from this DOS with the quantum harmonic oscillator expression, which includes the zero-point energy (ZPE). Converged spectra were obtained for forms II,
III, VI, VIII and IX; details are given in the Supporting Information (see \ref{S-si:ha}).

\subsubsection{Regularized thermodynamic integration}
We computed finite-temperature Gibbs free energies for six acridine polymorphs, Forms II, III, VI, VII, VIII, and IX, using Reg-TI along a three-step Frenkel--Ladd path,\cite{FrenkelLadd} together with multistate Bennett acceptance ratio estimator (MBAR) reweighting\cite{MBAR} of NPT trajectories across 280--340~K; the method is set out in SI Section~\ref{S-si:regti}. The statistical uncertainty of the MBAR free energy differences is small, below 0.005~\kjmol\ per polymorph, which reflects the large number of NPT configurations sampled, so the remaining uncertainty is systematic rather than statistical.

\section*{Acknowledgments}
\subsection*{Author Contributions}
{\bf Adam Lahouari}:
Conceptualization (equal); Data Curation (lead); Formal Analysis (equal); Investigation (equal); Methodology (equal); Software (equal); Validation (lead); Visualization (equal); Writing -- original draft (equal); Writing -- review \& editing (equal).
{\bf Hengyu Zhou}:
Investigation (equal); Methodology (equal); Formal Analysis (equal); Visualization (equal); Writing -- review \& edition (supporting).
{\bf Nikolaos Galanakis}:
Conceptualization (supporting); Data Curation (supporting); Formal Analysis (equal); Investigation (equal); Methodology (equal); Software (equal); Validation (supporting); Visualization (equal); Writing -- original draft (equal); Writing -- review \& editing (equal).
{\bf Michael Kilgour}:
Investigation (supporting); Writing -- review \& editing (supporting).
{\bf Daisuke Kuroshima}:
Investigation (supporting); Writing -- review \& editing (supporting).
{\bf Jihye Han}:
Investigation (supporting); Writing -- review \& editing (supporting).
{\bf Bart Kahr}:
Funding Acquisition (equal); Supervision (equal); Writing – review \& editing (supporting).
{\bf Mark E. Tuckerman}:
Conceptualization (equal); Project administration (equal); Funding Acquisition (lead); Investigation (equal); Supervision (equal); Writing -- review \& edition (supporting).
{\bf Alexander G. Shtukenberg}:
Conceptualization (equal); Project administration (equal); Funding Acquisition (equal); Investigation (equal); Supervision (equal); Writing -- original draft (equal); Writing -- review \& editing (equal).
{\bf Jutta Rogal}:
Conceptualization (equal); Project administration (equal); Funding Acquisition (lead); Investigation (equal); Supervision (equal); Writing -- original draft (lead); Writing -- review \& editing (equal).

\subsection*{Funding}
This work was supported by the National Science Foundation under grant DMR-2118890. This work was supported by a grant from the Simons Foundation [MPS-T-MPS-00839534, MET]. Computational resources were provided in part by the NYU IT High Performance Computing facility. The Flatiron Institute is a division of the Simons Foundation.

\subsection*{Conflicts of Interest}
The authors declare no conflict of interest.

\subsection*{Data Availability}
The DFT training data sets (pure acridine and its mixed crystals with anthracene and 2,7-dihydroxynaphthalene) and the fine-tuned MACE models used in this work are available in the MolCryst-MLIPs database~\cite{MolCryst} at \url{https://github.com/adamlaho/MolCryst}.

\section*{Supporting Information}
The Supporting Information is available free of charge at \url{https://pubs.acs.org/doi/...}.
DFT calculation details and construction of the training data sets for acridine and its mixed crystals with anthracene and 2,7-dihydroxynaphthalene. Training and fine-tuning of the MACE models, with parity plots of energies and forces against DFT. Reg-TI free-energy protocol and the resulting lattice and free energies of the acridine polymorphs. Construction of the mixed-crystal supercells and substitution energies of anthracene and 2,7-dihydroxynaphthalene in all polymorphs.

\bibliography{references}

\end{document}


\begin{center}
  {\Large\bfseries Supporting Information}\\[6pt]
  {\large The Acridine Polymorph Landscape: Characterization and Selective Control by Additives}
\end{center}

\tableofcontents

\section{Reference electronic structure calculations}
\label{si:dft}

All reference calculations were performed with VASP-6.5.1 using the PBE functional with the D4 dispersion correction, PAW potentials, and a plane-wave cutoff of 850~eV. The Brillouin zone was sampled with $\Gamma$-centred meshes chosen from
the length $L$ of each lattice vector: 7 points for $L \le 5$~\AA, 6 for $5 < L \le 10$~\AA, 5 for $10 < L \le 20$~\AA, and 4 for $L > 20$~\AA. The resulting meshes are listed in Table~\ref{tab:si-kpoints}.

\begin{table}[htbp]
  \centering
  \caption{$k$-point meshes used for each polymorph.}
  \label{tab:si-kpoints}
  \begin{tabular}{llcc}
    \toprule
    Form & CSD refcode & $Z$ & $k$-mesh \\
    \midrule
    II   & ACRDIN04   & 4  & $5\times6\times5$ \\
    III  & ACRDIN07   & 8  & $6\times5\times5$ \\
    IV   & ACRDIN08   & 12 & $6\times5\times4$ \\
    VI   & ACRDIN05   & 8  & $6\times4\times5$ \\
    VII  & ACRDIN06   & 8  & $6\times4\times5$ \\
    VIII & ACRDINVIII & 8  & $5\times5\times5$ \\
    IX   & ACRDIN12   & 4  & $5\times5\times6$ \\
    \bottomrule
  \end{tabular}
\end{table}

The seven polymorphs were relaxed starting from their CSD structures. Atomic positions, cell shape and cell volume were optimized together with a conjugate-gradient algorithm, using an electronic convergence of $10^{-6}$~eV, Fermi smearing of 0.05~eV and real-space projection. The isolated-molecule reference was obtained by relaxing the atomic positions of a single acridine molecule in a fixed cubic cell of 30~\AA{} at the same level of theory.

\textit{Ab initio} molecular dynamics used for the training set was run in the NVT ensemble with a Langevin thermostat (friction 10~ps$^{-1}$) and a 1~fs timestep, with an electronic convergence of $10^{-6}$~eV, Fermi smearing of 0.05~eV and real-space projection. Configurations selected by active learning were recomputed with single-point calculations using an electronic convergence of $10^{-6}$~eV, Gaussian smearing of 0.05~eV and reciprocal-space projection. 

\begin{table}[htbp]
  \centering
  \caption{Number of DFT configurations per polymorph in the
  pure-acridine dataset before cleaning.}
  \label{tab:si-dataset}
  \begin{tabular}{lrr}
    \toprule
    Form & Configurations & Share (\%) \\
    \midrule
    II    & 8\,855  & 39.3 \\
    III   & 4\,432  & 19.7 \\
    IV    & 1\,229  & 5.5  \\
    VI    & 1\,334  & 5.9  \\
    VII   & 1\,154  & 5.1  \\
    VIII  & 705     & 3.1  \\
    IX    & 4\,795  & 21.3 \\
    \midrule
    Total & 22\,504 & 100  \\
    \bottomrule
  \end{tabular}
\end{table}

\section{Machine-learned interatomic potential}
\label{si:mlip}

\subsection{Architecture and fine-tuning}
The potential is a MACE equivariant message-passing model~\cite{MACE}, obtained by fine-tuning the MACE-MH-1 foundation model with an \textsc{omol} head using the AMLP code~\cite{AMLP}, and distributed as part of the MolCryst database~\cite{MolCryst}. The architecture is inherited from the foundation model: two
interaction layers with correlation order 3 (body order 4), spherical harmonics up to $l=3$, hidden irreps $512\times0e + 512\times1o$ (512 channels), and a radial cutoff of 6.0~\AA{} (receptive field of 12.0~\AA), for a total of $6.18\times10^{6}$ parameters. The dataset was split 85/15 into training (18\,430 configurations) and validation sets, and the atomic reference energies $E_0$ used in the fit are listed in Table~\ref{tab:e0s}. The model was trained for up to 300 epochs with the AMSGrad variant of Adam, a batch size of 12, a weight decay of $5\times10^{-7}$, gradient clipping at 10, and an exponential moving average of the weights with decay 0.99. The initial learning rate of $2\times10^{-3}$ was reduced by a factor of 0.8 after 15 epochs without improvement of the validation loss. A weighted energy--force loss was used with weights of 100 (energy) and 10 (forces) for the first 200 epochs; for the remaining epochs the force
weight was raised to 100 and the learning rate lowered to $5\times10^{-5}$. Double precision (float64) was used for both training and inference.

\begin{table}[h]
  \centering
  \caption{Atomic reference energies $E_0$ used for fine-tuning.}
  \label{tab:e0s}
  \begin{tabular}{lr}
    \hline
    Element & $E_0$ (eV) \\
    \hline
    H & $-0.00292689$ \\
    C & $-1.24732389$ \\
    N & $-3.12548797$ \\
    \hline
  \end{tabular}
\end{table}

\subsection{Training dataset}
The dataset contains 22\,504 configurations before cleaning (Table~\ref{tab:si-dataset}). Outliers were removed with the DBSCAN clustering algorithm applied. The configurations not assigned to any cluster were discarded (noise). This removed 819 configurations, leaving 21\,685, which were split into training and validation sets

\begin{figure}[H]
  \centering
  \includegraphics[width=0.6\linewidth]{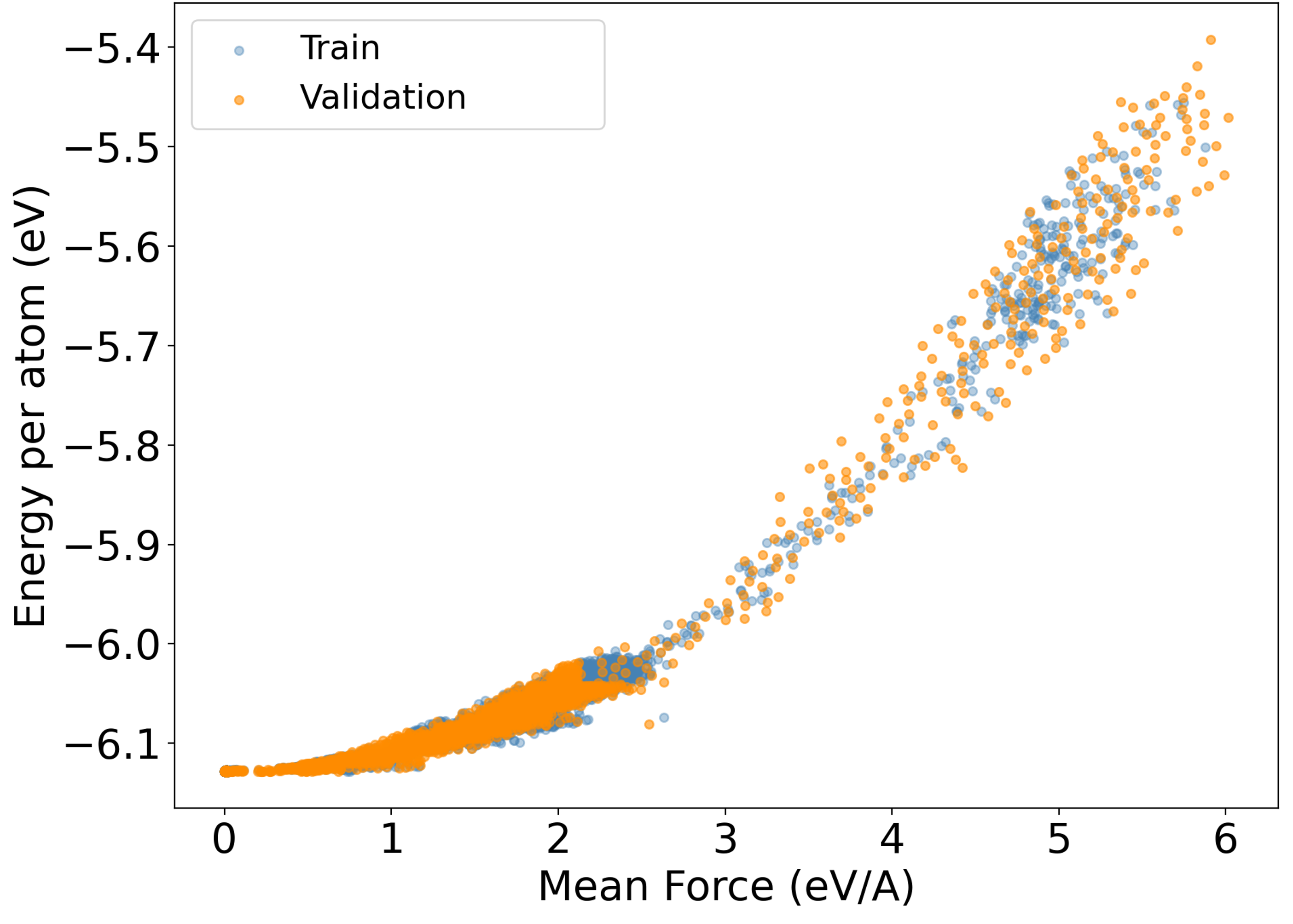}
  \caption{Energy per atom against the mean atomic force magnitude
  (in eV\,\AA$^{-1}$) for the 18\,435 training and 3\,250 validation
  configurations of the pure-acridine dataset after cleaning.}
  \label{fig:si-dataset-split}
\end{figure}


\subsection{Additive models}
The construction of the mixed-composition data is summarized in the main text. The substitution gives additive concentrations of 12.5--50\% for Forms III, VI, VII and VIII, 25 and 50\% for Forms II and IX, and 8--50\% for Form IV. For 2,7-DHN, two orientations of the molecule on each site were considered, and the hydroxyl hydrogens were placed by a cone search at an O--H distance of 0.98~\AA. Architecture and training settings are those of the pure-acridine model, except for the batch size (8 for anthracene, 13 for 2,7-DHN) and, for 2,7-DHN, a second training stage starting at epoch 250 instead of 200. The 2,7-DHN model uses an additional atomic reference energy for oxygen, $E_0 = -1.5314622$~eV.

\subsection{Validation}
\begin{table}[htbp]
\centering
\caption{Accuracy of the fine-tuned potentials against DFT on the
held-out configurations.}
\label{tab:si-mlip-acc}
\small
\begin{tabular}{lcc}
\toprule
Model & Energy MAE (\kjmol\,atom$^{-1}$) & Force MAE (\kjmol\,\AA$^{-1}$) \\
\midrule
Pure acridine (MolCryst)     & 0.15 & 1.06 \\
Acridine/anthracene          & 0.13 & 1.07 \\
Acridine/2,7-DHN             & 0.17 & 0.76 \\
\bottomrule
\end{tabular}
\end{table}

The accuracy of each model was measured on the 15\% of its dataset held out from training (Table~\ref{tab:si-mlip-acc}). Parity plots of energies and forces for the two additive models are shown in Figures~\ref{fig:si-parity-anth} and~\ref{fig:si-parity-dhn}. Float32
and float64 inference agree to within 0.015~\kjmol\ per molecule. 

\begin{figure}[H]
  \centering
  \includegraphics[width=0.8\linewidth]{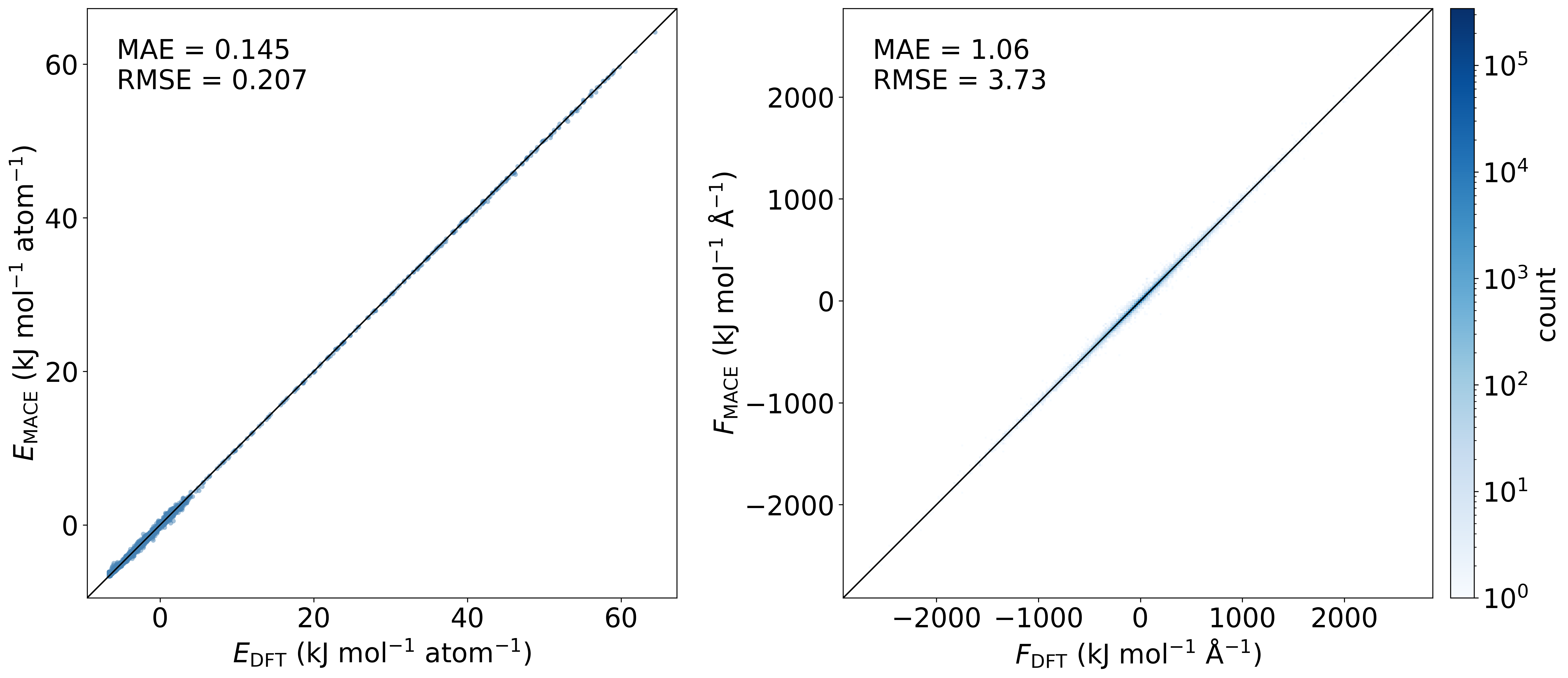}
  \caption{Parity between DFT and the pure-acridine MACE model on the
  held-out configurations: energies per atom, shifted by their
  mean (left), and force components (right, logarithmic colour scale).}
  \label{fig:si-parity-acr}
\end{figure}

\begin{figure}[H]
  \centering
  \includegraphics[width=0.8\linewidth]{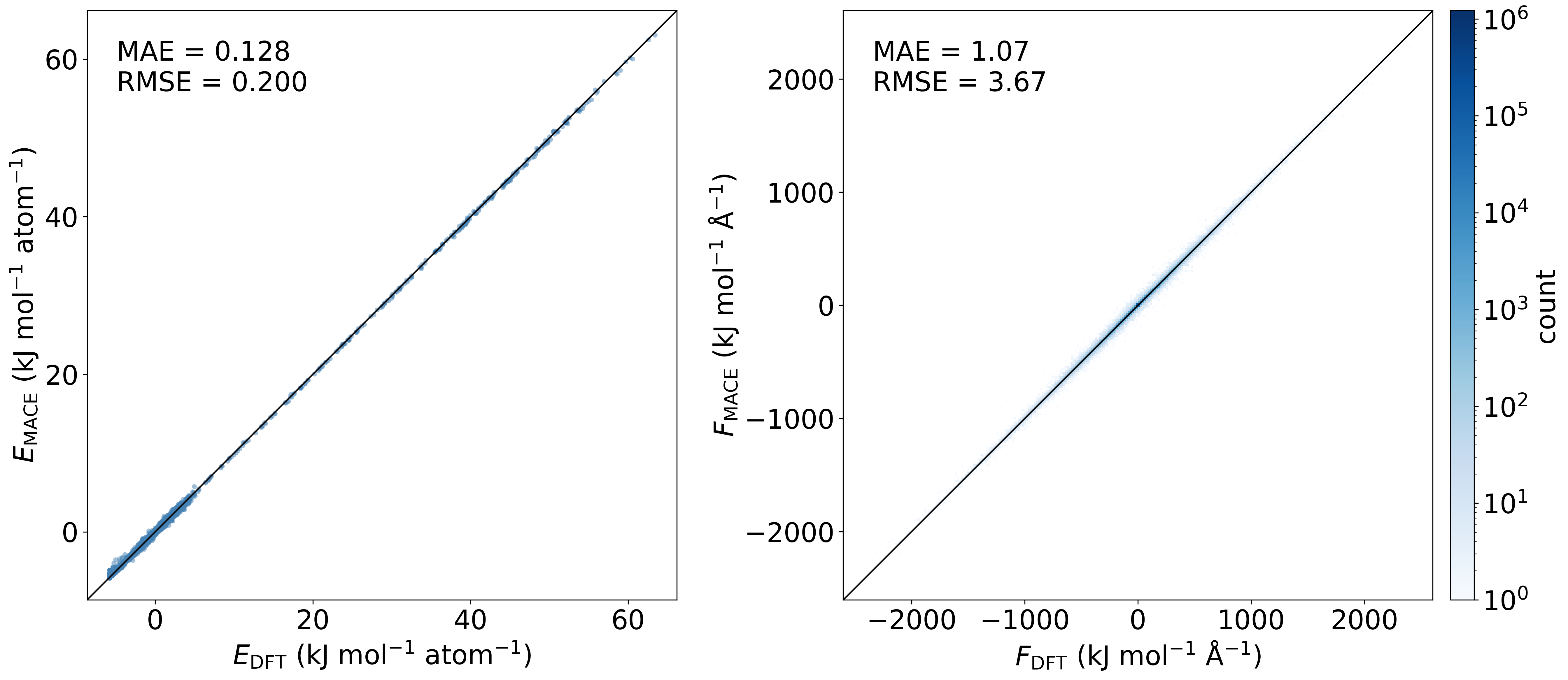}
  \caption{Parity between DFT and the acridine/anthracene MACE model on
  the held-out configurations: energies per atom, shifted by
  their mean (left), and force components (right, logarithmic colour
  scale).}
  \label{fig:si-parity-anth}
\end{figure}

\begin{figure}[H]
  \centering
  \includegraphics[width=0.8\linewidth]{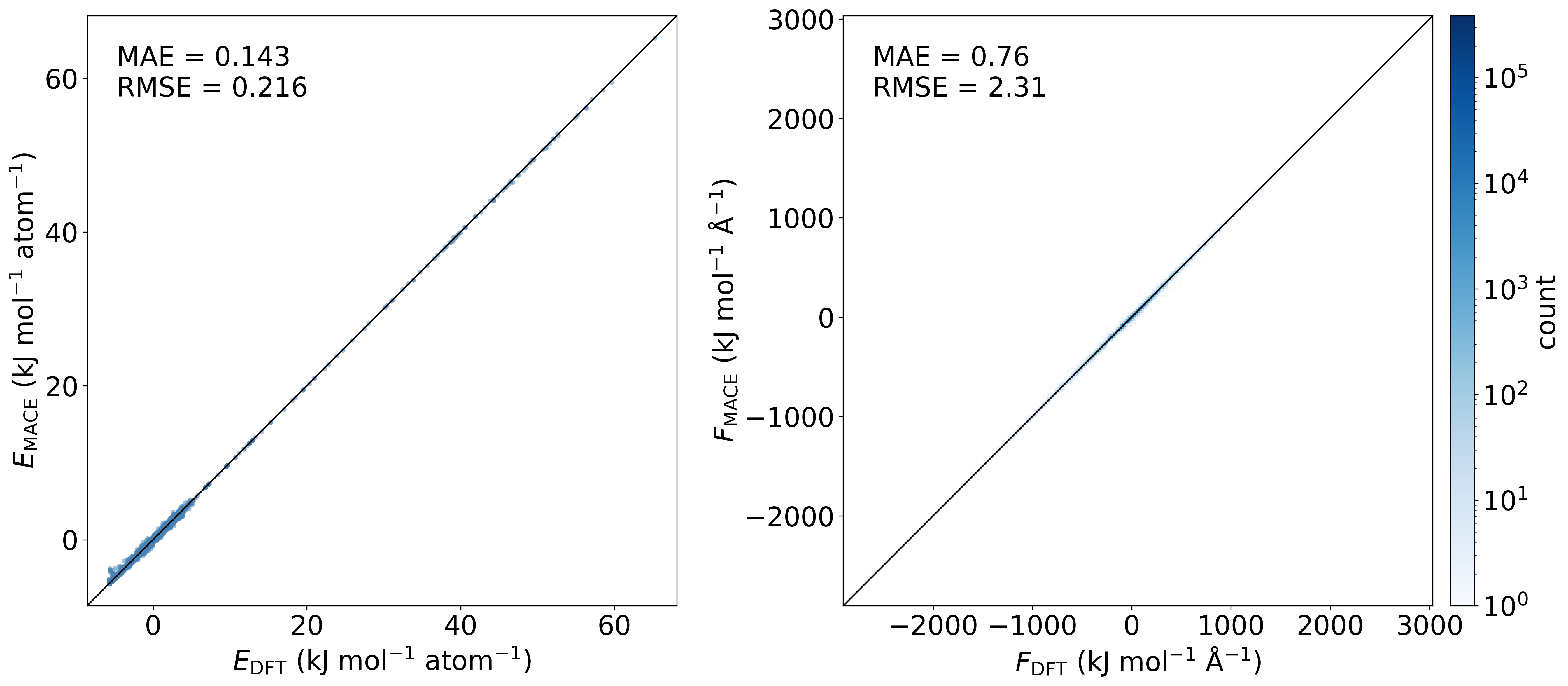}
  \caption{Parity between DFT and the acridine/2,7-DHN MACE model on
  the held-out configurations: energies per atom, shifted by their mean
  (left), and force components (right, logarithmic colour scale).}
  \label{fig:si-parity-dhn}
\end{figure}

\section{Regularised thermodynamic integration} 
\label{si:regti}

\subsection{Free energy decomposition}
The absolute Helmholtz free energy of each polymorph is computed along
a three-step Frenkel--Ladd path connecting an analytically solvable
Einstein (harmonic) reference to the target MACE crystal, with a
classical force field (GAFF) as an intermediate state. The total free
energy of polymorph $X$ is
\begin{equation}
F(X) = U_0(X) + F_{\mathrm{harm}}(X) + \Delta F_{\mathrm{A}}(X)
       + \Delta F_{\mathrm{B}}(X) + \Delta F_{\mathrm{C}}(X),
\label{eq:si-F}
\end{equation}
where $U_0$ is the static MACE energy at the cell-optimised geometry
and $F_{\mathrm{harm}}$ is the analytical free energy of the Einstein
crystal. Each atom is tethered to its equilibrium position
$\mathbf{r}_i^0$ by a spring of stiffness $\kappa$,
\begin{equation}
U_{\mathrm{harm}} = \kappa \sum_i \left| \mathbf{r}_i - \mathbf{r}_i^0 \right|^2 .
\label{eq:si-Uharm}
\end{equation}
The spring constant can be fitted per polymorph from the mean-square
displacement of a short MACE NVT run,
$\kappa = 3k_B T / \langle |\mathbf{r}_i - \mathbf{r}_i^0|^2 \rangle$,
or set to a common fixed value. Here a single
$\kappa = 0.70$~eV/\AA$^2$, representative of the average of the
per-polymorph fits, was used for all polymorphs so that the analytical
contribution largely cancels in the free-energy differences, with
\begin{equation}
F_{\mathrm{harm}} = -\frac{3N}{2} k_B T
  \ln\!\left( \frac{2\pi k_B T}{\kappa} \right).
\label{eq:si-Fharm}
\end{equation}

\subsection{Regularised coupling}
Steps A and B mix the two end states linearly in the coupling parameter $\lambda$,
\begin{equation}
U(\lambda) = (1-\lambda)\, U_{\mathrm{ref}} + \lambda\, U_{\mathrm{phys}},
\qquad
\frac{\partial U}{\partial \lambda} = U_{\mathrm{phys}} - U_{\mathrm{ref}} ,
\label{eq:si-switch}
\end{equation}
because the harmonic potential is present at both ends and keeps the integrand finite. In step C the harmonic potential is removed, and a switching function of order $m$ is used,
\begin{equation}
U(\lambda) = U_{\mathrm{MACE}} + (1-\lambda)^m U_{\mathrm{harm}},
\qquad
\frac{\partial U}{\partial \lambda} =
- m\,(1-\lambda)^{m-1} U_{\mathrm{harm}} ,
\label{eq:si-integrand}
\end{equation}
so that the integrand goes smoothly to zero at $\lambda = 1$. This avoids the divergence that appears when the harmonic potential is switched off linearly, and is the origin of the term \textit{regularised} TI. We used $m = 6$.

\paragraph{Step A: harmonic $+\,U_0 \rightarrow$ harmonic $+\,U_0 +$ GAFF.}
Starting from the harmonic reference, the intramolecular GAFF2 interactions are switched on:
\begin{equation}
\Delta F_{\mathrm{A}} = \int_0^1
  \left\langle U_{\mathrm{GAFF}} \right\rangle_\lambda d\lambda .
\end{equation}
Twelve $\lambda$ windows ($\lambda = 0.01, 0.02, 0.04, 0.06, 0.08, 0.10, 0.15, 0.20, 0.30, 0.50, 0.70, 0.90$) were used, with more points at small $\lambda$ where the integrand changes quickly.

\paragraph{Step B: harmonic $+\,U_0 +$ GAFF $\rightarrow$ harmonic $+$ MACE.}
GAFF2 is replaced by the MACE potential while the harmonic potential is kept:
\begin{equation}
\Delta F_{\mathrm{B}} = \int_0^1
  \left\langle U_{\mathrm{MACE}} - U_{\mathrm{GAFF}} - U_0
  \right\rangle_\lambda d\lambda ,
\end{equation}
where $U_0 = U_{\mathrm{MACE}}(\mathbf{q}_0)$ is the MACE energy of the reference structure. It removes the energy offset between the two models. The intermolecular interactions are introduced in this step, since GAFF2 is used for the intramolecular terms only. Six $\lambda$ windows ($\lambda = 0.05, 0.20, 0.40, 0.60, 0.80, 0.95$) were used.

\paragraph{Step C: harmonic $+$ MACE $\rightarrow$ MACE.}
The harmonic tether is removed to recover the pure MACE crystal:
\begin{equation}
\Delta F_{\mathrm{C}} = \int_0^1
  \left\langle -m\,(1-\lambda)^{m-1} U_{\mathrm{harm}} \right\rangle_\lambda d\lambda .
\end{equation}
Up to ten $\lambda$ windows ($\lambda = 0.02, 0.06, 0.10, 0.17, 0.25, 0.35, 0.45, 0.65, 0.85, 0.97$) were used. The points at $\lambda = 0.06, 0.17, 0.35$ were added to better describe the integrand at small $\lambda$ and reduce the integration error.

\subsection{Temperature propagation}
The 300~K anchor free energy is propagated across 280--340~K by
Gibbs--Helmholtz integration of the enthalpy difference,
\begin{equation}
\Delta G(T) = \frac{T}{T_0}\Delta G(T_0)
  - T \int_{T_0}^{T} \frac{\Delta H(T')}{T'^2}\,dT',
\label{eq:si-gh}
\end{equation}
and, in parallel, by MBAR reweighting of the NPT enthalpy samples to
obtain each polymorph's dimensionless free-energy shift $\delta f(T)$,
\begin{equation}
\Delta G_{I \rightarrow N}(T) = \frac{T}{T_0}\Delta G_{I \rightarrow N}(T_0)
  + RT\left[\delta f_N(T) - \delta f_I(T)\right].
\label{eq:si-mbar}
\end{equation}
The two methods agree to within $\sim$1~\kjmol\ across 280--340~K;
MBAR is reported as the primary estimate. The MBAR statistical
uncertainty, propagated in quadrature from the per-polymorph
\texttt{pymbar} covariance estimates, is below 0.005~\kjmol, so the
dominant uncertainty is systematic.

\subsection{NPT protocol}
Trajectories used a Nos\'e--Hoover chain thermostat
($\tau_T = 200$~fs) and a Parrinello--Rahman barostat
($\tau_P = 2000$~fs, assumed bulk modulus 20~GPa) at $P = 0$, with a
0.5~fs timestep, 20~ps equilibration and 100~ps production. A polymorph
was flagged as melted when its per-molecule volume exceeded
290~\AA$^3$ or its standard deviation exceeded 25~\AA$^3$; that
temperature and all higher ones were then excluded.

\begin{table}[htbp]
\centering
\caption{Reg-TI and NPT simulation parameters.}
\label{tab:si-regti-params}
\small
\begin{tabular}{lp{8.5cm}}
\toprule
\multicolumn{2}{l}{\textit{Reg-TI pathway}}\\
Einstein spring constant $\kappa$ & 0.70 eV\,\AA$^{-2}$ \\
Switching function order $m$ (step C) & 6 \\
Anchor temperature $T_0$ & 300 K \\
$\lambda$ windows, step A (12) & 0.01, 0.02, 0.04, 0.06, 0.08, 0.10, 0.15, 0.20, 0.30, 0.50, 0.70, 0.90 \\
$\lambda$ windows, step B (6) & 0.05, 0.20, 0.40, 0.60, 0.80, 0.95 \\
$\lambda$ windows, step C (10) & 0.02, 0.06, 0.10, 0.17, 0.25, 0.35, 0.45, 0.65, 0.85, 0.97 \\
Simulation length per window & 100 ps \\
Timestep & 0.5 fs \\
Thermostat & Nos\'e--Hoover chain (length 3, $\tau_T = 50$ fs) \\
Reference polymorph & Form II (ACRDIN04) \\
Intermediate force field & GAFF2 (intramolecular terms only) \\
\midrule
\multicolumn{2}{l}{\textit{NPT simulations}}\\
Thermostat & Nos\'e--Hoover chain \\
Thermostat relaxation time $\tau_T$ & 200 fs \\
Barostat & Parrinello--Rahman \\
Barostat relaxation time $\tau_P$ & 2000 fs \\
Assumed bulk modulus & 20 GPa \\
External pressure $P$ & 0 GPa \\
Timestep & 0.5 fs \\
Equilibration & 20 ps \\
Production & 100 ps \\
Save interval & 6 fs \\
Temperature range & 280--340 K \\
\midrule
\multicolumn{2}{l}{\textit{Melt detection}}\\
Volume threshold (per molecule) & 290 \AA$^3$ \\
Volume standard deviation threshold & 25 \AA$^3$ \\
\midrule
\multicolumn{2}{l}{\textit{Temperature propagation}}\\
Methods & Gibbs--Helmholtz and MBAR \\
Primary estimate & MBAR \\
Atoms per molecule & 23 (C$_{13}$H$_9$N) \\
MBAR $\delta f$ validity range & 3--21 \\
\bottomrule
\end{tabular}
\end{table}

\begin{figure}[H]
  \centering
  \includegraphics[width=0.6\linewidth]{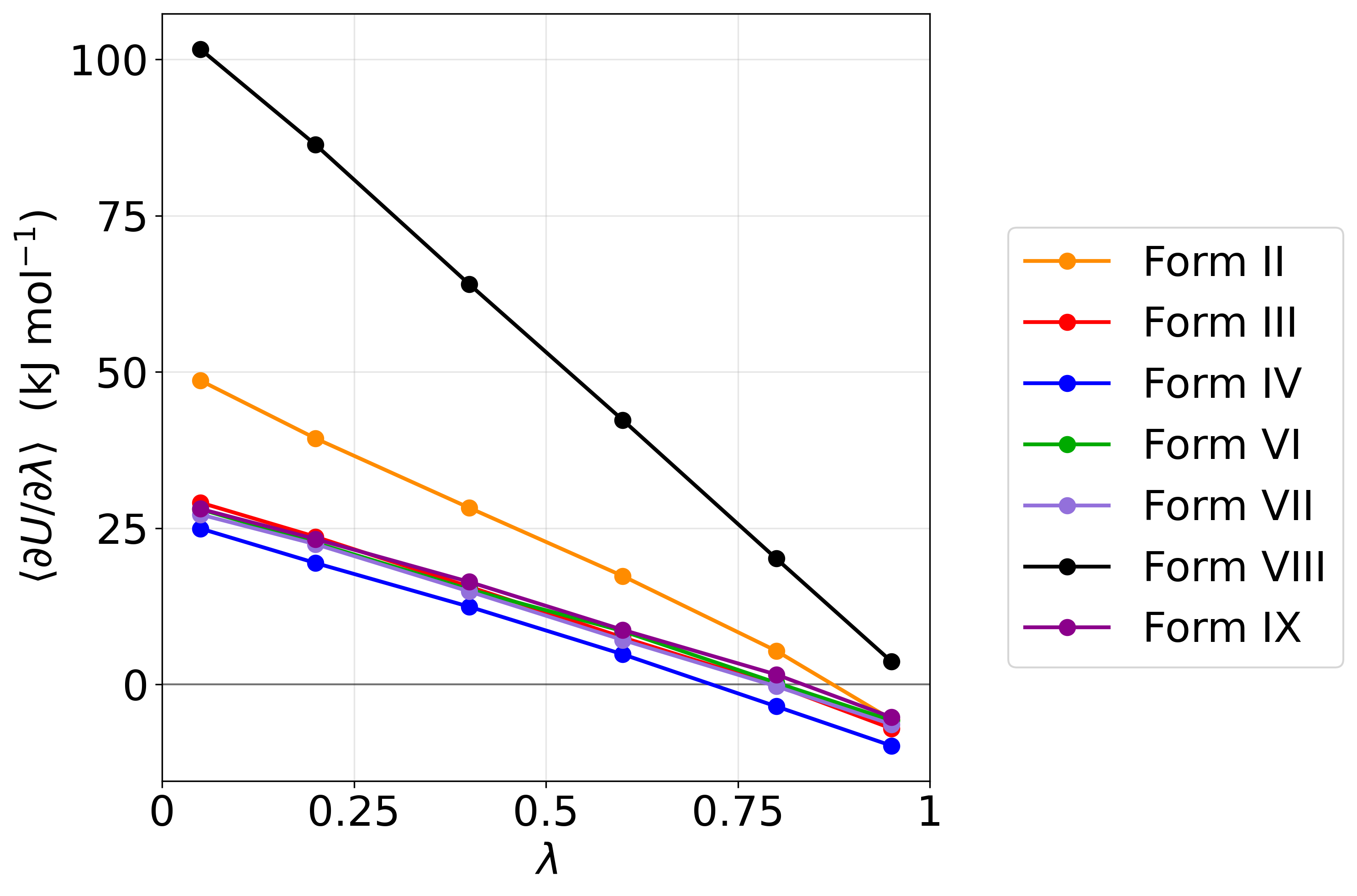}
  \caption{TI integrand $\langle \partial U/\partial\lambda \rangle$ for step B (GAFF $\rightarrow$ MACE) at 300~K for each acridine polymorph, in kJ\,mol$^{-1}$ per molecule. Markers show the six $\lambda$ windows used for the quadrature ($\lambda = 0.05, 0.20, 0.40, 0.60, 0.80, 0.95$).}
  \label{fig:si-integrand-B}
\end{figure}


\section{Harmonic and quasi-harmonic calculations}
\label{si:ha}

\subsection{Harmonic approximation}
Phonon calculations used the finite-displacement method as implemented in phonopy~\cite{phonopy}, starting from the DFT cell-optimised structures, with forces from (i) DFT (VASP) and (ii) the fine-tuned MACE potential. The DFT forces were computed with VASP at the PBE-D4 level, with a plane-wave cutoff of 850~eV, an electronic convergence of $10^{-7}$~eV, Gaussian smearing of 0.01~eV and $\Gamma$-centred $k$-point meshes (for example $5\times6\times5$ for Form II), using atomic displacements of 0.01~\AA. The unit cell was used for Forms II, III, VI, VII and VIII (92 atoms for Form II, 184 for the others) and a $1\times1\times2$ supercell (184 atoms) for Form IX. For the MACE calculations the lattice parameters were kept at their DFT values and only the atomic positions were relaxed, to a maximum force of $10^{-5}$~eV\,\AA$^{-1}$ in double precision, since the potential was trained without stress data; force constants were then built from displacements of 0.03~\AA{} in the unit cell. No acoustic sum rule or non-analytical term correction was applied. Vibrational free energies were obtained on $q$-point meshes of up to $20^3$, reduced for the larger cells, excluding modes below 0.5~THz, and are reported per molecule.

\subsection{Quasi-harmonic approximation}
The vibrational density of states (VDOS) was obtained from the velocity autocorrelation function (VACF) of MACE-driven NVT trajectories at 300~K. Each polymorph was simulated in a supercell (Table~\ref{tab:si-vacf-cells}) for 100~ps with a 0.5~fs timestep, using a Nos\'e--Hoover chain thermostat (chain length 3, time constant 50~fs). Velocities were stored at every step, which is needed to resolve the C--H stretching modes near 3000~cm$^{-1}$. After discarding the first 1~ps, the centre-of-mass velocity was removed and the mass-weighted VACF was computed up to a maximum lag of 5~ps, without a window function. The VDOS was obtained from a type-II discrete cosine transform of the zero-padded VACF, restricted to frequencies below 4000~cm$^{-1}$, with negative values set to zero, and normalised to $3N-3$ modes above 5~cm$^{-1}$. Vibrational free energies between 250 and 350~K were then computed from this single 300~K VDOS by weighting it with the quantum harmonic oscillator expression,

\begin{equation}
F_{\mathrm{vib}}(T) = k_B T \int g(\omega)\,
  \ln\!\left[ 2\sinh\!\left( \frac{\hbar\omega}{2 k_B T} \right)
  \right] d\omega ,
\end{equation}
which includes the zero-point energy. No further quantum correction was applied to the classical trajectories. Results are reported for Forms II, III, VI, VIII and IX; the VDOS for the other Forms did not converge.

\begin{table}[htbp]
  \centering
  \caption{Supercells used for the VACF trajectories.}
  \label{tab:si-vacf-cells}
  \begin{tabular}{lcc}
    \toprule
    Form & Supercell & Molecules \\
    \midrule
    II   & $3\times5\times2$ & 120 \\
    III  & $5\times2\times2$ & 160 \\
    IV   & $5\times2\times1$ & 120 \\
    VI   & $5\times1\times2$ & 80  \\
    VIII & $2\times2\times3$ & 96  \\
    IX   & $3\times2\times4$ & 96  \\
    \bottomrule
  \end{tabular}
\end{table}

\section{Mixed-crystal supercells and substitution energies}
\label{si:mixed}

Mixed-crystal supercells were built for Forms II, III, IV, VI, VII, VIII and IX from the lowest-energy DFT pure cell of each form, repeated to $N_\mathrm{mol} = 96$ molecules (Table~\ref{tab:si-mixed-cells}), so that all forms share the same concentrations: $n = 1$, 3, 5, 10, 14 and 24 additive molecules ($c = 1.04$ to 25.00~mol\,\%). For each form and concentration, ten patterns of $n$ randomly chosen acridine molecules were replaced by the additive, placed on the centre of the replaced molecule with the same orientation. For 2,7-DHN, each pattern was built with the hydroxyl groups on either side of the former nitrogen, giving 20 configurations. Energies are MACE single points with the model fine-tuned for each additive; configurations with a force component above 5~eV\,\AA$^{-1}$, which indicates a close contact, were removed. Gas-phase acridine and additive were relaxed with the same model to 0.01~eV\,\AA$^{-1}$ for $\Delta\mu$. Values are mean $\pm$ standard deviation over the configurations.
 
\begin{table}[H]
  \centering
  \caption{Supercells used for the mixed crystals ($N_\mathrm{mol} = 96$ for every form).}
  \label{tab:si-mixed-cells}
  \begin{tabular}{lcc}
    \toprule
    Form & Supercell & $N_{\mathrm{mol}}$ \\
    \midrule
    II   & $3\times4\times2$ & 96 \\
    III  & $6\times1\times2$ & 96 \\
    IV   & $4\times2\times1$ & 96 \\
    VI   & $6\times1\times2$ & 96 \\
    VII  & $6\times1\times2$ & 96 \\
    VIII & $2\times2\times3$ & 96 \\
    IX   & $3\times2\times4$ & 96 \\
    \bottomrule
  \end{tabular}
\end{table}
 
\begin{table}[H]
  \centering
  \caption{Substitution energies of anthracene in the acridine polymorphs (Eqs.~\ref{M-eq:esub} and \ref{M-eq:ddesub}) at additive concentrations $c$ from 1.04 to 25.00~mol\,\%, with $n$ anthracene molecules in 96-molecule supercells. $N_\mathrm{conf}$: number of configurations; $F_\mathrm{max}$: largest force component of each configuration, averaged over the configurations (pure cells: 0.01 to 0.22~eV\,\AA$^{-1}$); $\Delta\Delta E_\mathrm{sub}$ relative to Form~II at the same concentration.}
  \label{tab:si-subst-anth}
  \small
  \begin{tabular}{lrrrrcc}
    \toprule
    Form & $c$ (mol\,\%) & $n/N_\mathrm{mol}$ & $N_\mathrm{conf}$ & $F_\mathrm{max}$ (eV\,\AA$^{-1}$) & $\Delta E_\mathrm{sub}$ (kJ\,mol$^{-1}$) & $\Delta\Delta E_\mathrm{sub}$ (kJ\,mol$^{-1}$) \\
    \midrule
    II    &  1.04 &  1/96 & 10 & 0.62 & 0.72 $\pm$ 0.08 & \multicolumn{1}{c}{ref.} \\
          &  3.12 &  3/96 & 10 & 0.88 & 2.45 $\pm$ 0.20 & \multicolumn{1}{c}{ref.} \\
          &  5.21 &  5/96 & 10 & 0.78 & 4.22 $\pm$ 0.25 & \multicolumn{1}{c}{ref.} \\
          & 10.42 & 10/96 & 10 & 0.89 & 10.87 $\pm$ 0.29 & \multicolumn{1}{c}{ref.} \\
          & 14.58 & 14/96 & 10 & 0.88 & 16.62 $\pm$ 0.34 & \multicolumn{1}{c}{ref.} \\
          & 25.00 & 24/96 & 10 & 0.79 & 33.20 $\pm$ 0.30 & \multicolumn{1}{c}{ref.} \\
    \addlinespace
    III   &  1.04 &  1/96 & 10 & 0.33 & 1.34 $\pm$ 0.15 & 0.62 $\pm$ 0.17 \\
          &  3.12 &  3/96 & 10 & 0.49 & 4.46 $\pm$ 0.26 & 2.01 $\pm$ 0.32 \\
          &  5.21 &  5/96 & 10 & 0.49 & 7.26 $\pm$ 0.32 & 3.04 $\pm$ 0.41 \\
          & 10.42 & 10/96 & 10 & 0.49 & 15.88 $\pm$ 0.51 & 5.01 $\pm$ 0.59 \\
          & 14.58 & 14/96 & 10 & 0.48 & 23.29 $\pm$ 0.73 & 6.68 $\pm$ 0.81 \\
          & 25.00 & 24/96 & 10 & 0.48 & 42.33 $\pm$ 0.48 & 9.13 $\pm$ 0.56 \\
    \addlinespace
    IV    &  1.04 &  1/96 & 10 & 0.42 & 0.76 $\pm$ 0.12 & 0.05 $\pm$ 0.14 \\
          &  3.12 &  3/96 & 10 & 0.50 & 2.43 $\pm$ 0.24 & $-$0.02 $\pm$ 0.32 \\
          &  5.21 &  5/96 & 10 & 0.51 & 4.04 $\pm$ 0.36 & $-$0.18 $\pm$ 0.44 \\
          & 10.42 & 10/96 & 10 & 0.57 & 9.75 $\pm$ 0.45 & $-$1.12 $\pm$ 0.54 \\
          & 14.58 & 14/96 & 10 & 0.62 & 14.74 $\pm$ 0.69 & $-$1.88 $\pm$ 0.77 \\
          & 25.00 & 24/96 & 10 & 0.66 & 29.34 $\pm$ 0.52 & $-$3.85 $\pm$ 0.60 \\
    \addlinespace
    VI    &  1.04 &  1/96 & 10 & 0.26 & 1.33 $\pm$ 0.01 & 0.61 $\pm$ 0.08 \\
          &  3.12 &  3/96 & 10 & 0.25 & 4.18 $\pm$ 0.15 & 1.73 $\pm$ 0.25 \\
          &  5.21 &  5/96 & 10 & 0.27 & 6.97 $\pm$ 0.11 & 2.74 $\pm$ 0.28 \\
          & 10.42 & 10/96 & 10 & 0.31 & 14.87 $\pm$ 0.18 & 4.00 $\pm$ 0.35 \\
          & 14.58 & 14/96 & 10 & 0.34 & 21.64 $\pm$ 0.21 & 5.03 $\pm$ 0.40 \\
          & 25.00 & 24/96 & 10 & 0.34 & 39.27 $\pm$ 0.21 & 6.07 $\pm$ 0.36 \\
    \addlinespace
    VII   &  1.04 &  1/96 & 10 & 0.33 & 1.40 $\pm$ 0.11 & 0.68 $\pm$ 0.13 \\
          &  3.12 &  3/96 & 10 & 0.36 & 4.38 $\pm$ 0.29 & 1.93 $\pm$ 0.35 \\
          &  5.21 &  5/96 & 10 & 0.37 & 7.52 $\pm$ 0.20 & 3.29 $\pm$ 0.32 \\
          & 10.42 & 10/96 & 10 & 0.39 & 16.26 $\pm$ 0.31 & 5.38 $\pm$ 0.43 \\
          & 14.58 & 14/96 & 10 & 0.37 & 23.44 $\pm$ 0.47 & 6.82 $\pm$ 0.58 \\
          & 25.00 & 24/96 & 10 & 0.36 & 41.98 $\pm$ 0.42 & 8.78 $\pm$ 0.51 \\
    \addlinespace
    VIII  &  1.04 &  1/96 & 10 & 0.38 & 1.67 $\pm$ 0.05 & 0.95 $\pm$ 0.09 \\
          &  3.12 &  3/96 & 10 & 0.51 & 5.03 $\pm$ 0.10 & 2.58 $\pm$ 0.22 \\
          &  5.21 &  5/96 & 10 & 0.59 & 8.48 $\pm$ 0.09 & 4.26 $\pm$ 0.27 \\
          & 10.42 & 10/96 & 10 & 0.58 & 17.28 $\pm$ 0.16 & 6.41 $\pm$ 0.34 \\
          & 14.58 & 14/96 & 10 & 0.57 & 24.42 $\pm$ 0.27 & 7.81 $\pm$ 0.44 \\
          & 25.00 & 24/96 & 10 & 0.55 & 42.61 $\pm$ 0.22 & 9.42 $\pm$ 0.37 \\
    \addlinespace
    IX    &  1.04 &  1/96 & 10 & 0.59 & 1.32 $\pm$ 0.02 & 0.60 $\pm$ 0.08 \\
          &  3.12 &  3/96 & 10 & 0.78 & 4.25 $\pm$ 0.12 & 1.80 $\pm$ 0.24 \\
          &  5.21 &  5/96 & 10 & 0.87 & 7.20 $\pm$ 0.08 & 2.97 $\pm$ 0.27 \\
          & 10.42 & 10/96 & 10 & 0.87 & 15.94 $\pm$ 0.17 & 5.06 $\pm$ 0.34 \\
          & 14.58 & 14/96 & 10 & 0.80 & 23.16 $\pm$ 0.32 & 6.54 $\pm$ 0.47 \\
          & 25.00 & 24/96 & 10 & 0.77 & 42.51 $\pm$ 0.36 & 9.31 $\pm$ 0.46 \\
    \bottomrule
  \end{tabular}
\end{table}
 
\begin{table}[H]
  \centering
  \caption{Substitution energies of 2,7-dihydroxynaphthalene (2,7-DHN) in the acridine polymorphs at additive concentrations $c$ from 1.04 to 25.00~mol\,\%, as in Table~\ref{tab:si-subst-anth}. $N_\mathrm{conf}$: configurations kept out of 20 after removing those with a force component above 5~eV\,\AA$^{-1}$; $F_\mathrm{max}$ as in Table~\ref{tab:si-subst-anth} (pure cells: 0.02 to 0.33~eV\,\AA$^{-1}$).}
  \label{tab:si-subst-dhn}
  \small
  \begin{tabular}{lrrrrcc}
    \toprule
    Form & $c$ (mol\,\%) & $n/N_\mathrm{mol}$ & $N_\mathrm{conf}$ & $F_\mathrm{max}$ (eV\,\AA$^{-1}$) & $\Delta E_\mathrm{sub}$ (kJ\,mol$^{-1}$) & $\Delta\Delta E_\mathrm{sub}$ (kJ\,mol$^{-1}$) \\
    \midrule
    II    &  1.04 &  1/96 & 20 & 1.36 & 33.47 $\pm$ 0.17 & \multicolumn{1}{c}{ref.} \\
          &  3.12 &  3/96 & 20 & 1.85 & 81.86 $\pm$ 6.64 & \multicolumn{1}{c}{ref.} \\
          &  5.21 &  5/96 & 20 & 1.78 & 125.51 $\pm$ 5.69 & \multicolumn{1}{c}{ref.} \\
          & 10.42 & 10/96 & 20 & 2.07 & 158.71 $\pm$ 7.62 & \multicolumn{1}{c}{ref.} \\
          & 14.58 & 14/96 & 20 & 1.82 & 168.54 $\pm$ 5.70 & \multicolumn{1}{c}{ref.} \\
          & 25.00 & 24/96 & 20 & 1.34 & 160.42 $\pm$ 0.96 & \multicolumn{1}{c}{ref.} \\
    \addlinespace
    III   &  1.04 &  1/96 & 20 & 1.22 & 34.45 $\pm$ 0.46 & 0.99 $\pm$ 0.49 \\
          &  3.12 &  3/96 & 20 & 1.67 & 81.81 $\pm$ 12.64 & $-$0.05 $\pm$ 14.28 \\
          &  5.21 &  5/96 & 20 & 1.66 & 131.21 $\pm$ 14.54 & 5.70 $\pm$ 15.61 \\
          & 10.42 & 10/96 & 20 & 1.50 & 168.95 $\pm$ 9.34 & 10.24 $\pm$ 12.05 \\
          & 14.58 & 14/96 & 20 & 1.16 & 176.52 $\pm$ 8.00 & 7.97 $\pm$ 9.82 \\
          & 25.00 & 24/96 & 20 & 0.70 & 167.16 $\pm$ 1.55 & 6.73 $\pm$ 1.82 \\
    \addlinespace
    IV    &  1.04 &  1/96 & 20 & 2.37 & 24.62 $\pm$ 0.52 & $-$8.85 $\pm$ 0.55 \\
          &  3.12 &  3/96 & 20 & 2.81 & 59.58 $\pm$ 6.25 & $-$22.28 $\pm$ 9.12 \\
          &  5.21 &  5/96 & 20 & 2.50 & 86.33 $\pm$ 6.56 & $-$39.18 $\pm$ 8.69 \\
          & 10.42 & 10/96 & 20 & 2.20 & 112.77 $\pm$ 5.13 & $-$45.94 $\pm$ 9.19 \\
          & 14.58 & 14/96 & 20 & 1.72 & 114.59 $\pm$ 2.82 & $-$53.95 $\pm$ 6.36 \\
          & 25.00 & 24/96 & 20 & 0.79 & 99.98 $\pm$ 1.69 & $-$60.44 $\pm$ 1.94 \\
    \addlinespace
    VI    &  1.04 &  1/96 & 15 & 3.38 & 34.32 $\pm$ 1.15 & 0.86 $\pm$ 1.17 \\
          &  3.12 &  3/96 &  9 & 3.39 & 80.62 $\pm$ 11.63 & $-$1.24 $\pm$ 13.40 \\
          &  5.21 &  5/96 &  7 & 4.49 & 135.68 $\pm$ 6.17 & 10.17 $\pm$ 8.39 \\
          & 10.42 & 10/96 &  9 & 3.59 & 166.21 $\pm$ 3.21 & 7.51 $\pm$ 8.27 \\
          & 14.58 & 14/96 & 15 & 2.25 & 169.36 $\pm$ 9.39 & 0.82 $\pm$ 10.99 \\
          & 25.00 & 24/96 & 19 & 1.16 & 161.21 $\pm$ 1.36 & 0.78 $\pm$ 1.66 \\
    \addlinespace
    VII   &  1.04 &  1/96 & 20 & 1.77 & 36.36 $\pm$ 0.94 & 2.89 $\pm$ 0.95 \\
          &  3.12 &  3/96 & 20 & 1.89 & 85.21 $\pm$ 11.32 & 3.35 $\pm$ 13.13 \\
          &  5.21 &  5/96 & 20 & 1.90 & 136.08 $\pm$ 8.79 & 10.56 $\pm$ 10.47 \\
          & 10.42 & 10/96 & 20 & 1.47 & 171.98 $\pm$ 6.94 & 13.27 $\pm$ 10.31 \\
          & 14.58 & 14/96 & 20 & 1.17 & 177.77 $\pm$ 9.54 & 9.23 $\pm$ 11.11 \\
          & 25.00 & 24/96 & 20 & 0.59 & 168.55 $\pm$ 1.73 & 8.13 $\pm$ 1.97 \\
    \addlinespace
    VIII  &  1.04 &  1/96 & 20 & 1.80 & 29.29 $\pm$ 0.27 & $-$4.18 $\pm$ 0.32 \\
          &  3.12 &  3/96 & 20 & 2.03 & 74.89 $\pm$ 5.80 & $-$6.96 $\pm$ 8.82 \\
          &  5.21 &  5/96 & 20 & 1.98 & 106.67 $\pm$ 6.47 & $-$18.84 $\pm$ 8.62 \\
          & 10.42 & 10/96 & 20 & 1.84 & 150.07 $\pm$ 9.03 & $-$8.64 $\pm$ 11.82 \\
          & 14.58 & 14/96 & 19 & 1.55 & 161.87 $\pm$ 3.47 & $-$6.67 $\pm$ 6.67 \\
          & 25.00 & 24/96 & 20 & 0.88 & 158.60 $\pm$ 2.27 & $-$1.82 $\pm$ 2.46 \\
    \addlinespace
    IX    &  1.04 &  1/96 & 20 & 1.67 & 35.42 $\pm$ 0.12 & 1.95 $\pm$ 0.21 \\
          &  3.12 &  3/96 & 20 & 2.17 & 86.70 $\pm$ 8.11 & 4.84 $\pm$ 10.48 \\
          &  5.21 &  5/96 & 19 & 2.23 & 133.17 $\pm$ 4.18 & 7.66 $\pm$ 7.06 \\
          & 10.42 & 10/96 & 18 & 2.23 & 168.35 $\pm$ 5.71 & 9.64 $\pm$ 9.52 \\
          & 14.58 & 14/96 & 16 & 1.75 & 174.84 $\pm$ 5.50 & 6.30 $\pm$ 7.92 \\
          & 25.00 & 24/96 & 19 & 0.96 & 164.52 $\pm$ 1.80 & 4.10 $\pm$ 2.04 \\
    \bottomrule
  \end{tabular}
\end{table}

\section{CrystalMath topological CSP}
\label{si:crystalmath}

\subsection{Structure generation}
Structure generation in each space group begins by placing the
asymmetric unit at grid positions spanning the asymmetric unit of a
proposed unit cell, with a target grid spacing of 0.3~\AA\ adjusted to
adhere to the ZZPs~\cite{CrystalMath}, retaining structures that
satisfy geometric overlap criteria assessed by ellipsoidal
representation of the fragments and atomic clash checks. For $Z' = 1$,
the reference molecule is placed at each grid position within the
asymmetric unit and the full unit cell is constructed by applying the
symmetry operations of the target space group. For $Z' > 1$, the
asymmetric unit is assembled by combining pairs of independently
oriented single-molecule conformations, with the relative displacement
between the two molecules derived from a statistical analysis of
$Z' > 1$ structures in the Cambridge Structural Database;
inter-instance distances are validated before a structure enters the
pool. All structure generation steps are GPU-accelerated.

\begin{table}[htbp]
\centering
\caption{CrystalMath stage 1 and stage 2 statistics for acridine. Stage 1 structures are the
candidates built from the input cells and grid placements; stage 2 structures are those that
passed the stage 1 ellipsoid and atomic overlap checks and were optimized in stage 2.}
\label{tab:si-cm-stats}
\small
\begin{tabular}{lcc}
\toprule
Search & Stage 1 structures & Stage 2 structures \\
\midrule
$Z' = 1$ & 93{,}448  & 271 \\
$Z' = 2$ & 60{,}539  & 290 \\
$Z' = 3$ & 187{,}696 & 42  \\
\midrule
Total & 341{,}683 & 603 \\
\bottomrule
\end{tabular}

\vspace{4pt}
\footnotesize Total stage 1 wall-clock time $\approx$ 6.0~h on a
GPU-accelerated mobile workstation.
\end{table}

\begin{table}[htbp]
\centering
\caption{Space-group distribution of acridine structures through the CrystalMath pipeline for
$Z'=1$, 2 and 3. \emph{Gen.}: candidate structures built in stage~1 (input cells $\times$ grid
positions); \emph{S1}: structures passing the stage-1 ellipsoid and atomic overlap checks, all of
which were optimised and converged in stage~2; \emph{Acc.}: structures in the final accepted pool
after filtering, clustering and deduplication. A dash indicates that the search yielded no structures in that space group - a zero indicates that no structure survived a particular step. $P1$ and $P\bar{1}$ received no input cells at any $Z'$.}
\label{tab:sg_distribution}
\small
\setlength{\tabcolsep}{4pt}
\begin{tabular}{lr rrr rrr rrr}
\toprule
 & & \multicolumn{3}{c}{$Z'=1$} & \multicolumn{3}{c}{$Z'=2$} & \multicolumn{3}{c}{$Z'=3$} \\
\cmidrule(lr){3-5}\cmidrule(lr){6-8}\cmidrule(lr){9-11}
Space group & No. & Gen. & S1 & Acc. & Gen. & S1 & Acc. & Gen. & S1 & Acc. \\
\midrule
$P2_1$ & 4 & 3{,}712 & 0 & 0 & 4{,}432 & 0 & 0 & -- & -- & -- \\
$C2$ & 5 & 106 & 0 & 0 & 15 & 1 & \textbf{1} & -- & -- & -- \\
$Pc$ & 7 & 448 & 0 & 0 & 416 & 0 & 0 & -- & -- & -- \\
$Cc$ & 9 & 520 & 0 & 0 & 504 & 16 & \textbf{5} & -- & -- & -- \\
$P2_1/m$ & 11 & 508 & 0 & 0 & 15 & 0 & 0 & -- & -- & -- \\
$C2/m$ & 12 & 20 & 0 & 0 & 15 & 0 & 0 & -- & -- & -- \\
$P2/c$ & 13 & 1{,}182 & 0 & 0 & 15 & 0 & 0 & -- & -- & -- \\
$P2_1/c$ & 14 & 18{,}486 & 11 & 0 & 26{,}740 & 107 & \textbf{3} & -- & -- & -- \\
$P2_1/n$ & 14 & 18{,}486 & 188 & \textbf{3} & 26{,}740 & 166 & \textbf{2} & -- & -- & -- \\
$C2/c$ & 15 & 1{,}584 & 66 & 0 & 1{,}647 & 0 & 0 & -- & -- & -- \\
$P2_12_12$ & 18 & 6 & 0 & 0 & -- & -- & -- & 8 & 0 & 0 \\
$P2_12_12_1$ & 19 & 34{,}560 & 0 & 0 & -- & -- & -- & 139{,}776 & 42 & \textbf{1} \\
$Pca2_1$ & 29 & 960 & 0 & 0 & -- & -- & -- & 2{,}688 & 0 & 0 \\
$Pna2_1$ & 33 & 960 & 0 & 0 & -- & -- & -- & 2{,}688 & 0 & 0 \\
$Pbcn$ & 60 & 4{,}254 & 0 & 0 & -- & -- & -- & 5{,}032 & 0 & 0 \\
$Pbca$ & 61 & 7{,}650 & 6 & 0 & -- & -- & -- & 37{,}496 & 0 & 0 \\
$Pnma$ & 62 & 6 & 0 & 0 & -- & -- & -- & 8 & 0 & 0 \\
\midrule
Total & & 93{,}448 & 271 & \textbf{3} & 60{,}539 & 290 & \textbf{11} & 187{,}696 & 42 & \textbf{1} \\
\bottomrule
\end{tabular}
 
\vspace{2pt}
{\footnotesize Overall: 341{,}683 structures generated, 603 passed stage~1 and converged in stage~2, 15 accepted.Total stage 2 wall-clock time $\approx$ 73 min. on a GPU-accelerated mobile workstation.}
\end{table}

\subsection{Filtering and optimization}

Stage 1 structures were screened against geometric order parameter
constraints requiring heavy atoms to lie at or near preferred
crystallographic positions for the target space group. The surviving
structures were then relaxed with a flexible unit cell optimization 
that drives their packing coefficient and intermolecular close-contact
distances towards optimal values empirically derived from statistical 
analysis of CSD structures --- a physically motivated
substitute for energy minimization. A final filter based on the van der
Waals free volume removed candidates with anomalously large packing
voids, and the remainder were clustered by packing similarity and
deduplicated. The final accepted pool comprised 15 distinct structures:
eight in $P2_1/c$, five in $Cc$, one in $C2$ and one in $P2_12_12_1$.

Compared with the originally published protocol,~\cite{CrystalMath} the
version used here constructs candidate crystals from rigid molecular
fragments using CSD-derived geometric statistics for chemically
comparable fragments, rather than by random sampling, and relaxes and
ranks them with a revised cost function.



\section{Packing similarity comparisons}
\label{si:compack}

Packing similarity was assessed with Crystal Packing Similality feature in CSD Mercury version 2026.2, utilizing COMPACK~\cite{COMPACK} algorithm over
clusters of 20 molecules using 20\% distance tolerance and 20$^\circ$ angle tolerance.

\begin{sidewaystable}
\centering
\caption{Packing similarity between all 15 CrystalMath predicted
structures and the seven experimentally characterised acridine
polymorphs. Each cell gives the number of molecules overlaid out of 20
and, in parentheses, the \rmsdtw\ in \AA. Full 20/20 matches are shown
in bold. Structures are labelled by space group followed by
\textit{index}-c\textit{cell} as in Figure~2 of the main text (e.g.\
$P2_1/n$~71-c98 for structure 000071, cell 98).}
\label{tab:si-compack}
\scriptsize
\begin{tabular}{lccccccc}
\toprule
Predicted structure & Form II & Form III & Form IV & Form VI & Form VII & Form VIII & Form IX \\
\midrule
$P2_1/n$~71-c98       & \textbf{20/20 (0.364)} & 5/20 (0.334) & 6/20 (0.443) & 4/20 (0.957) & 8/20 (0.844) & 2/20 (0.542) & 2/20 (0.166) \\
$P2_1/c$~24-c10       & 5/20 (0.292) & \textbf{20/20 (0.273)} & 5/20 (0.429) & 7/20 (0.605) & 9/20 (0.493) & 2/20 (0.323) & 2/20 (0.215) \\
$P2_12_12_1$~0-c16    & 5/20 (0.291) & 6/20 (0.538) & \textbf{20/20 (0.267)} & 7/20 (0.387) & 8/20 (1.579) & 5/20 (6.959) & 3/20 (0.600) \\
$Cc$~10-c25           & 3/20 (0.465) & 6/20 (0.620) & 7/20 (0.905) & \textbf{20/20 (0.207)} & 15/20 (0.531) & 6/20 (5.133) & 2/20 (0.548) \\
$P2_1/n$~102-c35      & 6/20 (0.595) & 8/20 (0.493) & 8/20 (0.809) & 15/20 (0.431) & \textbf{20/20 (0.252)} & 5/20 (3.980) & 3/20 (0.597) \\
$Cc$~0-c13            & 2/20 (0.474) & 2/20 (0.287) & 3/20 (2.079) & 4/20 (3.424) & 3/20 (0.498) & \textbf{20/20 (0.234)} & 2/20 (0.389) \\
$P2_1/n$~0-c90        & 4/20 (2.679) & 2/20 (0.300) & 2/20 (0.620) & 3/20 (0.873) & 3/20 (0.842) & 2/20 (0.337) & \textbf{20/20 (0.381)} \\
$C2$~0-c19            & 4/20 (2.284) & 2/20 (0.186) & 2/20 (0.346) & 3/20 (1.402) & 2/20 (0.211) & 10/20 (0.201) & 4/20 (1.611) \\
$Cc$~2-c19            & 4/20 (2.290) & 2/20 (0.186) & 2/20 (0.348) & 3/20 (1.393) & 2/20 (0.211) & 10/20 (0.199) & 4/20 (1.612) \\
$P2_1/c$~0-c22        & 5/20 (1.226) & 2/20 (0.244) & 5/20 (1.000) & 3/20 (1.487) & 3/20 (0.819) & 7/20 (0.203) & 3/20 (0.859) \\
$Cc$~9-c31            & 3/20 (0.410) & 3/20 (0.200) & 5/20 (1.862) & 10/20 (0.266) & 5/20 (0.650) & 4/20 (3.336) & 2/20 (0.477) \\
$Cc$~13-c28           & 3/20 (0.447) & 3/20 (0.227) & 3/20 (0.146) & 7/20 (0.183) & 7/20 (0.327) & 4/20 (3.330) & 2/20 (0.623) \\
$P2_1/c$~17-c16       & 4/20 (2.061) & 3/20 (0.359) & 8/20 (0.816) & 3/20 (0.412) & 3/20 (0.370) & 5/20 (1.811) & 3/20 (1.032) \\
$P2_1/n$~53-c38       & 2/20 (0.519) & 1/20 (0.060) & 2/20 (0.520) & 2/20 (0.737) & 2/20 (0.279) & 1/20 (0.086) & 1/20 (0.071) \\
$P2_1/n$~106-c10      & 6/20 (0.376) & 13/20 (0.419) & 3/20 (0.094) & 5/20 (0.699) & 6/20 (0.452) & 2/20 (0.362) & 2/20 (0.224) \\
\bottomrule
\end{tabular}
\end{sidewaystable}

\section{Topological similarity measures}
\label{si:topo-sim}

\subsection{Definition of the comparison categories}
Each pair of structures (one acridine polymorph and one additive) is
compared across four independent categories, each probing a different
aspect of structural similarity. All metrics are quantitative: lower
Wasserstein distances or higher similarity scores indicate greater
resemblance. No arbitrary scoring weights are applied; the categories
are interpreted individually and then synthesised.
The centres of mass are expanded by the space-group operations over a
$5\times5\times5$ supercell and trimmed to a 15~\AA{} sphere. The RDF
covers 0--15~\AA{} in 0.1~\AA{} bins. Coordination shells are counted
within 6, 6--10 and 10--14~\AA{} and compared by their mean relative
difference. Neighbour-vector directions within 10~\AA{} are histogrammed
in 36 bins in $\varphi$ and in $\theta$. The RDF and angular Wasserstein
distances are evaluated over bin index, so one unit corresponds to
0.1~\AA, 10$^\circ$ ($\varphi$) and 5$^\circ$ ($\theta$). Steinhardt
$Q_l$ distributions ($l=2$--10, 12 nearest neighbours) are compared in
the same way.

\begin{description}\itemsep3pt
\item[Crystal packing, tier 1 (local environment).] How similar is the
arrangement of molecular centres of mass around each molecule? Measured
by the Wasserstein distance between radial distribution functions,
coordination shell counts and angular distributions.

\item[Crystal packing, tier 2 (global cell).] How similar are the
overall unit cell properties? Measured by packing coefficient
similarity, cell shape and angle distances, space group symmetry
element matching, and $Z$ / \Zp.
Packing coefficients are compared through the Gaussian similarity
$\exp[-(\Delta\mathrm{PC}/0.03)^2]$. Cell shape is compared by the
Wasserstein distance between the normalised eigenvalues of the metric
tensor. Cell angles are compared by the mean $|\cos\alpha_1-\cos\alpha_2|$.
Symmetry is compared by the Hamming similarity of a 10-element binary
fingerprint of the symmetry elements present (inversion, 2-fold and
$2_1$ axes, mirror and glide planes, 3- and 4-fold axes, C, I and A
centring).

\item[Fragment orientation.] How similarly are neighbouring molecules
oriented relative to one another? Computed from a supercell neighbour
search: for each reference molecule, the rotation required to align it
with each crystal neighbour is expressed as a rotation angle $\omega$
and a rotation axis direction $(\varphi, \theta)$, and the Wasserstein
distances between these distributions are the comparison metrics.
$\pi$--$\pi$ stacking geometry (interplanar angle, perpendicular
spacing, lateral offset, stacking motif) is analysed as a
sub-category.
Neighbours are all fragments within 12~\AA{} of the reference in a
$5\times5\times5$ supercell, which covers roughly the first two
coordination shells. Stacked pairs are those with a centroid--centroid
distance of at most 6~\AA. Each pair is classified as face-to-face,
slipped-parallel or non-stacking (interplanar angle $<15^\circ$, lateral
offset $<2$, 2--4 or $>4$~\AA), intermediate ($15$--$50^\circ$) or
edge-to-face ($\ge50^\circ$). Motif sets are compared by their Jaccard
index.

\item[Interaction networks.] How similar are the types and numbers of
intermolecular contacts? Measured by contact type Jaccard similarity
and contact density relative difference.
Contacts are typed by element pair, and hydrogen bonds are pooled into
one type. In the additive comparisons, the acridine C--N and H--N
contacts are relabelled C--O and H--O so that the ring nitrogen is
compared with the diol oxygen; these are the C--O(N) contacts of
Table~\ref{tab:si-cono}. For each shared contact type, the length distributions
(\AA) and the body-frame direction distributions are compared by
Wasserstein distance. The directionality of each contact type is
described by the anisotropy index $A=(\mu_1-\mu_3)/\mu_1$, where the
$\mu_i$ are the eigenvalues of the orientation tensor
$\langle\hat{\mathbf u}\hat{\mathbf u}^{\mathsf T}\rangle$ of the unit
contact vectors. Two structures are compared through $|A_1-A_2|$.
\end{description}

\subsection{Additives analysed}
Four representative additives were analysed in detail: anthracene,
naphthalene-2,7-diol (27DH), naphthalene-2,3-diol (23DH) and
\textit{trans}-azobenzene (AZO). The main text discusses anthracene and
27DH; the complete data for all four are given below.

\subsection{Anthracene}
\label{si:topo-anth}

\begin{table}[htbp]
\centering
\caption{Crystal packing: global cell properties. Anthracene reference:
$Z = 4$, $Z' = 1$, PC $= 0.722$, full symmetry set (inversion,
$2_1$ screw, glide).}
\label{tab:si-anth-t1}
\small
\begin{tabular}{lcccccl}
\toprule
Polymorph & $Z$ & \Zp & PC similarity & Cell shape W & Cell angle W & Sym.\ match / shared \\
\midrule
Form II   & 4  & 1 & 0.983 & 0.202 & 0.134 & 1.000 --- inversion, $2_1$, glide \\
Form III  & 8  & 2 & 0.254 & 0.218 & 0.159 & 1.000 --- inversion, $2_1$, glide \\
Form IV   & 12 & 3 & 0.042 & 0.092 & 0.189 & 0.800 --- $2_1$ screw only \\
Form VI   & 8  & 2 & 0.381 & 0.095 & 0.151 & 0.700 --- glide only \\
Form VII  & 8  & 2 & 0.775 & 0.108 & 0.155 & 1.000 --- inversion, $2_1$, glide \\
Form VIII & 8  & 2 & 0.512 & 0.278 & 0.042 & 0.700 --- glide only \\
Form IX   & 4  & 1 & 0.318 & 0.261 & 0.177 & 1.000 --- inversion, $2_1$, glide \\
\bottomrule
\end{tabular}
\end{table}

\begin{table}[htbp]
\centering
\caption{Crystal packing: local molecular environment (anthracene
reference).}
\label{tab:si-anth-t2}
\small
\begin{tabular}{lccccccc}
\toprule
Polymorph & RDF W & Shell $\Delta$ & Angular $\varphi$ W & Angular $\theta$ W & $Q_2$ W & $Q_4$ W & $Q_6$ W \\
\midrule
Form II   & 5.035  & 0.245 & 1.670 & 4.577 & 0.071 & 0.024 & 0.016 \\
Form III  & 5.204  & 0.174 & 1.112 & 3.455 & 0.036 & 0.140 & 0.016 \\
Form IV   & 6.152  & 0.203 & 0.877 & 3.675 & 0.017 & 0.145 & 0.023 \\
Form VI   & 6.334  & 0.239 & 0.792 & 3.579 & 0.012 & 0.169 & 0.022 \\
Form VII  & 6.220  & 0.203 & 0.816 & 3.241 & 0.042 & 0.112 & 0.018 \\
Form VIII & 8.725  & 0.344 & 2.003 & 3.363 & 0.016 & 0.148 & 0.024 \\
Form IX   & 10.297 & 0.360 & 3.020 & 4.066 & 0.021 & 0.246 & 0.071 \\
\bottomrule
\end{tabular}
\end{table}

\begin{table}[htbp]
\centering
\caption{Fragment orientation. Anthracene reference:
$\omega$ mean $= 119.2^\circ$.}
\label{tab:si-anth-t3}
\small
\begin{tabular}{lcccccccc}
\toprule
Polymorph & \Zp & $\omega$ W & $\omega$ mean & $\Delta$ mean & $\omega$ std & $\varphi$ W & $\theta$ W & Plan.\ W \\
\midrule
Form II   & 1 & 5.27  & $120.0^\circ$ & $0.8^\circ$  & 0.001 & 51.7 & 56.8 & 0.005 \\
Form IX   & 1 & 7.66  & $126.0^\circ$ & $6.8^\circ$  & 0.001 & 45.1 & 36.7 & 0.030 \\
Form IV   & 3 & 14.61 & $123.8^\circ$ & $4.6^\circ$  & 2.421 & 43.9 & 25.2 & 0.037 \\
Form III  & 2 & 18.79 & $126.2^\circ$ & $6.9^\circ$  & 0.832 & 44.4 & 20.5 & 0.036 \\
Form VII  & 2 & 21.87 & $120.9^\circ$ & $2.1^\circ$  & 1.001 & 33.3 & 17.0 & 0.025 \\
Form VI   & 2 & 40.29 & $98.9^\circ$  & $20.3^\circ$ & 0.001 & 52.2 & 19.4 & 0.014 \\
Form VIII & 2 & 69.79 & $49.5^\circ$  & $69.8^\circ$ & 0.001 & 75.2 & 23.8 & 0.032 \\
\bottomrule
\end{tabular}
\end{table}

\begin{table}[htbp]
\centering
\caption{$\pi$--$\pi$ stacking geometry. Anthracene reference: motifs
face-to-face and intermediate, angle $33.5^\circ$, perpendicular
spacing 3.22~\AA, offset 3.21~\AA.}
\label{tab:si-anth-t4}
\small
\begin{tabular}{lclccccc}
\toprule
Polymorph & $n$ & Motifs & Angle mean & Perp.\ mean & Offset mean & Angle W & Perp.\ W \\
\midrule
Form VIII & 5  & face-to-face             & $0.3^\circ$  & 3.389 \AA & 1.766 \AA & 33.26 & 1.163 \\
Form III  & 14 & edge-to-face, slipped-par. & $37.1^\circ$ & 1.858 \AA & 4.304 \AA & 17.46 & 1.426 \\
Form IV   & 9  & edge-to-face, intermediate & $59.9^\circ$ & 1.958 \AA & 4.552 \AA & 26.36 & 1.629 \\
Form IX   & 3  & slipped-parallel         & $0.0^\circ$  & 4.764 \AA & 3.318 \AA & 33.51 & 1.549 \\
Form II   & 12 & non-stacking, slipped-par. & $2.9^\circ$  & 0.364 \AA & 5.153 \AA & 30.58 & 2.875 \\
Form VII  & 14 & edge-to-face, slipped-par. & $60.8^\circ$ & 1.900 \AA & 4.124 \AA & 41.17 & 1.539 \\
Form VI   & 10 & edge-to-face only        & $72.7^\circ$ & 2.616 \AA & 3.398 \AA & 39.19 & 0.872 \\
\bottomrule
\end{tabular}

\vspace{3pt}
\footnotesize Motif Jaccard indices: Form VIII 0.500, Form IV 0.333,
all others 0.000.
\end{table}

\begin{table}[htbp]
\centering
\caption{Contact length Wasserstein distances (shared contact types
only). Anthracene reference lengths: C--C 3.794~\AA, C--H 3.152~\AA,
H--H 2.585~\AA.}
\label{tab:si-anth-t5a}
\small
\begin{tabular}{lccccccc}
\toprule
Polymorph & C--C W & C--C mean & C--H W & C--H mean & H--H W & H--H mean & Mean W \\
\midrule
Form VIII & 0.037 & 3.814 & 0.069 & 3.192 & 0.158 & 2.641 & 0.088 \\
Form VI   & 0.029 & 3.805 & 0.067 & 3.209 & 0.201 & 2.757 & 0.099 \\
Form III  & 0.031 & 3.805 & 0.101 & 3.231 & 0.177 & 2.748 & 0.103 \\
Form VII  & 0.028 & 3.801 & 0.072 & 3.215 & 0.212 & 2.794 & 0.104 \\
Form II   & 0.040 & 3.775 & 0.095 & 3.231 & 0.189 & 2.772 & 0.108 \\
Form IX   & 0.033 & 3.822 & 0.135 & 3.286 & 0.184 & 2.747 & 0.118 \\
Form IV   & 0.043 & 3.833 & 0.102 & 3.248 & 0.218 & 2.794 & 0.121 \\
\bottomrule
\end{tabular}
\end{table}

\begin{table}[htbp]
\centering
\caption{Contact anisotropy difference relative to anthracene --- the
most discriminating interaction-network metric. Lower values indicate
more similar contact geometry. Polymorphs are ranked by mean
$\Delta$.}
\label{tab:si-anth-t5b}
\small
\begin{tabular}{lcccc c}
\toprule
Polymorph & C--C $\Delta$ & C--H $\Delta$ & H--H $\Delta$ & Mean $\Delta$ & Rank (max $\Delta$)  \\
\midrule
Form II   & 0.081 & 0.084 & 0.022 & 0.062 & 1 \\
Form III  & 0.109 & 0.107 & 0.024 & 0.080 & 2 \\
Form IX   & 0.213 & 0.329 & 0.195 & 0.246 & 3 \\
Form VI   & 0.326 & 0.428 & 0.169 & 0.308 & 4 \\
Form VIII & 0.292 & 0.391 & 0.335 & 0.339 & 5 \\
Form VII  & 0.416 & 0.523 & 0.197 & 0.379 & 6 \\
Form IV   & 0.370 & 0.547 & 0.447 & 0.455 & 7 \\
\bottomrule
\end{tabular}
\end{table}


\clearpage
\subsection{2,7-dihydroxynaphthalene (27DH)}
\label{si:topo-27dh}
Tables~\ref{tab:si-27dh-t1}--\ref{tab:si-27dh-t5b} give the comparison of
2,7-dihydroxynaphthalene (NPHLDL01, fragment C$_{10}$H$_8$O$_2$) with the seven acridine
polymorphs, in the same format as the anthracene analysis
(Tables~\ref{tab:si-anth-t1}--\ref{tab:si-anth-t5b}). Unlike anthracene, 27DH shares five
contact types with acridine (C--C, C--H, H--H, C--O and H--O, the last two corresponding to
the relabelled acridine C--N and H--N contacts), and it forms hydrogen bonds, which acridine
does not (130 contacts per fragment, 6 of them hydrogen-bonded). The rotation angles $\omega$ of 27DH take only the
values 0$^\circ$, 90$^\circ$ and 180$^\circ$, so its reference $\omega$ distribution is
discrete. The C--O(N) anisotropies of the acridine polymorphs are listed per independent
molecule in Table~\ref{tab:si-cono}.

\begin{table}[htbp]
\centering
\caption{Crystal packing: global cell properties. 27DH reference: $Z=4$, $Z'=1$, PC $= 0.682$, space group $Pna2_1$ (symmetry elements: $2_1$, glide).}
\label{tab:si-27dh-t1}
\small
\begin{tabular}{lcccccl}
\toprule
Polymorph & $Z$ & \Zp & PC similarity & Cell shape $W$ & Cell angle $\overline{|\Delta\cos|}$ & Sym.\ match / shared \\
\midrule
Form II & 4 & 1 & 0.239 & 0.135 & 0.055 & 0.900 --- $2_1$, glide \\
Form III & 8 & 2 & 0.975 & 0.166 & 0.030 & 0.900 --- $2_1$, glide \\
Form IV & 12 & 3 & 0.817 & 0.041 & 0.000 & 0.900 --- $2_1$ \\
Form VI & 8 & 2 & 0.887 & 0.043 & 0.038 & 0.800 --- glide \\
Form VII & 8 & 2 & 0.507 & 0.056 & 0.034 & 0.900 --- $2_1$, glide \\
Form VIII & 8 & 2 & 0.771 & 0.211 & 0.147 & 0.800 --- glide \\
Form IX & 4 & 1 & 0.935 & 0.194 & 0.012 & 0.900 --- $2_1$, glide \\
\bottomrule
\end{tabular}
\end{table}

\begin{table}[htbp]
\centering
\caption{Crystal packing: local molecular environment (27DH reference).}
\label{tab:si-27dh-t2}
\small
\begin{tabular}{lccccccc}
\toprule
Polymorph & RDF $W$ & Shell $\Delta$ & Angular $\varphi$ $W$ & Angular $\theta$ $W$ & $Q_2$ $W$ & $Q_4$ $W$ & $Q_6$ $W$ \\
\midrule
Form II & 4.241 & 0.213 & 1.683 & 2.013 & 0.070 & 0.062 & 0.040 \\
Form III & 4.179 & 0.244 & 1.237 & 1.620 & 0.035 & 0.071 & 0.031 \\
Form IV & 5.041 & 0.295 & 2.171 & 1.195 & 0.018 & 0.076 & 0.058 \\
Form VI & 5.375 & 0.277 & 1.905 & 1.461 & 0.010 & 0.101 & 0.032 \\
Form VII & 4.678 & 0.266 & 2.051 & 1.778 & 0.044 & 0.043 & 0.039 \\
Form VIII & 7.333 & 0.292 & 1.328 & 1.809 & 0.010 & 0.080 & 0.018 \\
Form IX & 10.565 & 0.296 & 1.447 & 1.798 & 0.022 & 0.177 & 0.072 \\
\bottomrule
\end{tabular}
\end{table}

\begin{table}[htbp]
\centering
\caption{Fragment orientation, sorted by $\omega$ $W$. 27DH reference: $\omega$ mean $= 90.0^\circ$.}
\label{tab:si-27dh-t3}
\small
\begin{tabular}{lcccccccc}
\toprule
Polymorph & \Zp & $\omega$ $W$ & $\omega$ mean & $\Delta$ mean & $\omega$ std & $\varphi$ $W$ & $\theta$ $W$ & Plan.\ $W$ \\
\midrule
Form II & 1 & 30.00 & 120.0$^\circ$ & 30.0$^\circ$ & 0.000 & 9.8 & 8.7 & 0.006 \\
Form VII & 2 & 32.42 & 120.9$^\circ$ & 30.9$^\circ$ & 1.957 & 69.7 & 70.8 & 0.030 \\
Form VI & 2 & 34.15 & 98.9$^\circ$ & 8.9$^\circ$ & 0.000 & 56.7 & 56.7 & 0.017 \\
Form IV & 3 & 34.37 & 123.8$^\circ$ & 33.8$^\circ$ & 2.410 & 69.4 & 79.3 & 0.041 \\
Form IX & 1 & 36.00 & 126.0$^\circ$ & 36.0$^\circ$ & 0.000 & 25.0 & 29.9 & 0.034 \\
Form III & 2 & 36.18 & 126.2$^\circ$ & 36.2$^\circ$ & 4.561 & 60.0 & 62.4 & 0.039 \\
Form VIII & 2 & 40.61 & 49.5$^\circ$ & 40.5$^\circ$ & 0.003 & 43.5 & 43.5 & 0.036 \\
\bottomrule
\end{tabular}
\par\smallskip
{\footnotesize $\Delta$ mean is the mean over instance pairs of the absolute difference in mean $\omega$; $\omega$ std is the standard deviation of $\omega$ $W$ over instance pairs.}
\end{table}

\begin{table}[htbp]
\centering
\caption{$\pi$--$\pi$ stacking geometry. 27DH reference: 16 interactions, motifs edge-to-face, non-stacking, angle 45.5$^\circ$, perpendicular spacing 1.99~\AA, offset 4.70~\AA.}
\label{tab:si-27dh-t4}
\footnotesize
\begin{tabular}{lc>{\raggedright\arraybackslash}p{3.4cm}ccccc}
\toprule
Polymorph & $n$ & Motifs & Angle mean & Perp.\ mean & Offset mean & Angle $W$ & Perp.\ $W$ \\
\midrule
Form II & 12 & non-stacking, slipped-parallel & 2.9$^\circ$ & 0.364~\AA & 5.153~\AA & 42.58 & 1.631 \\
Form III & 14 & edge-to-face, slipped-parallel & 37.1$^\circ$ & 1.858~\AA & 4.304~\AA & 8.46 & 0.603 \\
Form IV & 9 & edge-to-face, intermediate & 59.9$^\circ$ & 1.958~\AA & 4.552~\AA & 28.67 & 0.947 \\
Form VI & 10 & edge-to-face & 72.7$^\circ$ & 2.616~\AA & 3.398~\AA & 27.19 & 0.888 \\
Form VII & 14 & edge-to-face, slipped-parallel & 60.8$^\circ$ & 1.900~\AA & 4.124~\AA & 19.59 & 0.460 \\
Form VIII & 5 & face-to-face & 0.3$^\circ$ & 3.389~\AA & 1.766~\AA & 45.36 & 1.394 \\
Form IX & 3 & slipped-parallel & 0.0$^\circ$ & 4.764~\AA & 3.318~\AA & 45.52 & 2.769 \\
\bottomrule
\end{tabular}
\par\smallskip
{\footnotesize Motif Jaccard indices: Form VI 0.500, Form II 0.333, Form III 0.333, Form IV 0.333, Form VII 0.333; Form VIII, Form IX 0.000.}
\end{table}

\begin{table}[htbp]
\centering
\caption{Contact length Wasserstein distances (shared contact types only). 27DH reference lengths: C--C 3.766~\AA, C--H 3.208~\AA, H--H 2.616~\AA, C--O 3.631~\AA, H--O 3.004~\AA. For acridine, C--O and H--O denote the relabelled C--N and H--N contacts.}
\label{tab:si-27dh-t5a}
\scriptsize
\begin{tabular}{lccccccccccc}
\toprule
 & \multicolumn{2}{c}{C--C} & \multicolumn{2}{c}{C--H} & \multicolumn{2}{c}{H--H} & \multicolumn{2}{c}{C--O} & \multicolumn{2}{c}{H--O} & \\\cmidrule(lr){2-3}\cmidrule(lr){4-5}\cmidrule(lr){6-7}\cmidrule(lr){8-9}\cmidrule(lr){10-11}
Polymorph & $W$ & mean (\AA) & $W$ & mean (\AA) & $W$ & mean (\AA) & $W$ & mean (\AA) & $W$ & mean (\AA) & Mean $W$ \\
\midrule
Form II & 0.030 & 3.775 & 0.048 & 3.231 & 0.176 & 2.772 & 0.027 & 3.619 & 0.095 & 3.026 & 0.075 \\
Form VI & 0.041 & 3.805 & 0.054 & 3.209 & 0.172 & 2.757 & 0.057 & 3.665 & 0.084 & 2.995 & 0.081 \\
Form IX & 0.058 & 3.822 & 0.080 & 3.286 & 0.141 & 2.746 & 0.048 & 3.627 & 0.097 & 2.920 & 0.085 \\
Form VII & 0.040 & 3.801 & 0.059 & 3.215 & 0.186 & 2.794 & 0.049 & 3.638 & 0.098 & 3.068 & 0.086 \\
Form IV & 0.070 & 3.833 & 0.067 & 3.248 & 0.202 & 2.794 & 0.075 & 3.664 & 0.076 & 3.030 & 0.098 \\
Form III & 0.042 & 3.805 & 0.082 & 3.231 & 0.158 & 2.747 & 0.073 & 3.702 & 0.141 & 3.083 & 0.099 \\
Form VIII & 0.053 & 3.814 & 0.074 & 3.192 & 0.175 & 2.641 & 0.057 & 3.682 & 0.194 & 2.811 & 0.110 \\
\bottomrule
\end{tabular}
\end{table}

\begin{table}[htbp]
\centering
\caption{Contact anisotropy difference relative to 27DH. Lower values indicate more similar contact geometry. 27DH anisotropies: C--C 0.483, C--H 0.431, H--H 0.987, C--O 0.948, H--O 0.920. Rank is by mean $\Delta$.}
\label{tab:si-27dh-t5b}
\small
\begin{tabular}{lccccccc}
\toprule
Polymorph & C--C $\Delta$ & C--H $\Delta$ & H--H $\Delta$ & C--O $\Delta$ & H--O $\Delta$ & Mean $\Delta$ & Rank \\
\midrule
Form VII & 0.138 & 0.230 & 0.210 & 0.078 & 0.060 & 0.143 & 1 \\
Form IX & 0.287 & 0.222 & 0.208 & 0.021 & 0.053 & 0.158 & 2 \\
Form VIII & 0.207 & 0.160 & 0.349 & 0.026 & 0.051 & 0.159 & 3 \\
Form VI & 0.237 & 0.303 & 0.182 & 0.069 & 0.061 & 0.170 & 4 \\
Form IV & 0.130 & 0.078 & 0.461 & 0.144 & 0.091 & 0.181 & 5 \\
Form III & 0.391 & 0.443 & 0.024 & 0.016 & 0.050 & 0.185 & 6 \\
Form II & 0.418 & 0.467 & 0.030 & 0.040 & 0.058 & 0.203 & 7 \\
\bottomrule
\end{tabular}
\end{table}

\begin{table}[htbp]
\centering
\caption{C--O(N) contact anisotropy per crystallographically independent molecule, which controls susceptibility to face-selective inhibition by the diol additives. Higher values indicate a more directional, more easily blocked nitrogen environment. Values are listed per independent molecule, in the order of the asymmetric unit.}
\label{tab:si-cono}
\small
\begin{tabular}{lccc}
\toprule
Polymorph & \Zp & C--O(N) anisotropy & Mean \\
\midrule
Form II & 1 & 0.908 & -- \\
Form III & 2 & 0.958, 0.969 & 0.964 \\
Form IV & 3 & 0.606, 0.971, 0.880 & 0.819 \\
Form VI & 2 & 0.826, 0.932 & 0.879 \\
Form VII & 2 & 0.811, 0.967 & 0.889 \\
Form VIII & 2 & 0.925, 0.919 & 0.922 \\
Form IX & 1 & 0.969 & -- \\
\bottomrule
\end{tabular}
\end{table}

\subsection{2,3-dihydroxynaphthalene (23DH)}
\label{si:topo-23dh}
Tables~\ref{tab:si-23dh-t1}--\ref{tab:si-23dh-t5b} give the comparison of
2,3-dihydroxynaphthalene (VOGSEP01, fragment C$_{10}$H$_8$O$_2$) with the seven acridine
polymorphs, in the same format as for anthracene and 27DH. 23DH crystallises in $P\bar{1}$
with $Z'=2$, so all comparisons are averaged over its two independent molecules. As for
27DH, it shares the C--C, C--H, H--H, C--O and H--O contact types with acridine and in
addition forms hydrogen bonds and O--O contacts, which acridine does not
(177 contacts per fragment, 9 of them hydrogen-bonded). Its rotation angles $\omega$ take only the values
0$^\circ$, 60$^\circ$ and 180$^\circ$. The C--O(N) anisotropies of the acridine
polymorphs, which do not depend on the additive, are listed in Table~\ref{tab:si-cono}.
 
\begin{table}[htbp]
\centering
\caption{Crystal packing: global cell properties. 23DH reference: $Z=4$, $Z'=2$, PC $= 0.724$, space group $P\bar{1}$ (symmetry elements: inversion).}
\label{tab:si-23dh-t1}
\small
\begin{tabular}{lcccccl}
\toprule
Polymorph & $Z$ & \Zp & PC similarity & Cell shape $W$ & Cell angle $\overline{|\Delta\cos|}$ & Sym.\ match / shared \\
\midrule
Form II & 4 & 1 & 0.957 & 0.120 & 0.182 & 0.800 --- inversion \\
Form III & 8 & 2 & 0.210 & 0.156 & 0.157 & 0.800 --- inversion \\
Form IV & 12 & 3 & 0.032 & 0.038 & 0.127 & 0.800 --- none \\
Form VI & 8 & 2 & 0.325 & 0.033 & 0.164 & 0.700 --- none \\
Form VII & 8 & 2 & 0.712 & 0.046 & 0.161 & 0.800 --- inversion \\
Form VIII & 8 & 2 & 0.447 & 0.196 & 0.273 & 0.700 --- none \\
Form IX & 4 & 1 & 0.267 & 0.179 & 0.139 & 0.800 --- inversion \\
\bottomrule
\end{tabular}
\end{table}
 
\begin{table}[htbp]
\centering
\caption{Crystal packing: local molecular environment (23DH reference).}
\label{tab:si-23dh-t2}
\small
\begin{tabular}{lccccccc}
\toprule
Polymorph & RDF $W$ & Shell $\Delta$ & Angular $\varphi$ $W$ & Angular $\theta$ $W$ & $Q_2$ $W$ & $Q_4$ $W$ & $Q_6$ $W$ \\
\midrule
Form II & 4.700 & 0.231 & 1.104 & 3.418 & 0.074 & 0.087 & 0.032 \\
Form III & 3.860 & 0.269 & 0.796 & 2.231 & 0.025 & 0.047 & 0.037 \\
Form IV & 4.266 & 0.307 & 0.865 & 2.699 & 0.025 & 0.051 & 0.022 \\
Form VI & 5.275 & 0.305 & 0.474 & 2.744 & 0.025 & 0.076 & 0.038 \\
Form VII & 4.340 & 0.291 & 0.684 & 2.315 & 0.025 & 0.018 & 0.036 \\
Form VIII & 7.589 & 0.314 & 1.397 & 2.356 & 0.027 & 0.055 & 0.049 \\
Form IX & 10.359 & 0.314 & 2.304 & 3.435 & 0.037 & 0.153 & 0.086 \\
\bottomrule
\end{tabular}
\end{table}
 
\begin{table}[htbp]
\centering
\caption{Fragment orientation, sorted by $\omega$ $W$. 23DH reference: $\omega$ mean $= 104.9^\circ$.}
\label{tab:si-23dh-t3}
\small
\begin{tabular}{lcccccccc}
\toprule
Polymorph & \Zp & $\omega$ $W$ & $\omega$ mean & $\Delta$ mean & $\omega$ std & $\varphi$ $W$ & $\theta$ $W$ & Plan.\ $W$ \\
\midrule
Form II & 1 & 15.13 & 120.0$^\circ$ & 15.1$^\circ$ & 0.905 & 57.0 & 69.5 & 0.046 \\
Form IX & 1 & 21.13 & 126.0$^\circ$ & 21.1$^\circ$ & 0.905 & 49.8 & 48.4 & 0.017 \\
Form IV & 3 & 27.76 & 123.8$^\circ$ & 19.0$^\circ$ & 2.631 & 61.8 & 26.9 & 0.011 \\
Form III & 2 & 32.56 & 126.2$^\circ$ & 21.3$^\circ$ & 0.886 & 69.2 & 27.4 & 0.012 \\
Form VII & 2 & 35.48 & 120.9$^\circ$ & 16.0$^\circ$ & 1.006 & 49.6 & 18.1 & 0.031 \\
Form VI & 2 & 53.70 & 98.9$^\circ$ & 5.9$^\circ$ & 0.105 & 70.1 & 29.7 & 0.034 \\
Form VIII & 2 & 58.70 & 49.5$^\circ$ & 55.4$^\circ$ & 0.540 & 87.8 & 34.6 & 0.016 \\
\bottomrule
\end{tabular}
\par\smallskip
{\footnotesize $\Delta$ mean is the mean over instance pairs of the absolute difference in mean $\omega$; $\omega$ std is the standard deviation of $\omega$ $W$ over instance pairs.}
\end{table}
 
\begin{table}[htbp]
\centering
\caption{$\pi$--$\pi$ stacking geometry. 23DH reference: 24 interactions, motifs intermediate, slipped-parallel, angle 25.0$^\circ$, perpendicular spacing 3.24~\AA, offset 3.58~\AA.}
\label{tab:si-23dh-t4}
\footnotesize
\begin{tabular}{lc>{\raggedright\arraybackslash}p{3.4cm}ccccc}
\toprule
Polymorph & $n$ & Motifs & Angle mean & Perp.\ mean & Offset mean & Angle $W$ & Perp.\ $W$ \\
\midrule
Form II & 12 & non-stacking, slipped-parallel & 2.9$^\circ$ & 0.364~\AA & 5.153~\AA & 22.10 & 2.880 \\
Form III & 14 & edge-to-face, slipped-parallel & 37.1$^\circ$ & 1.858~\AA & 4.304~\AA & 12.02 & 1.422 \\
Form IV & 9 & edge-to-face, intermediate & 59.9$^\circ$ & 1.958~\AA & 4.552~\AA & 34.97 & 1.931 \\
Form VI & 10 & edge-to-face & 72.7$^\circ$ & 2.616~\AA & 3.398~\AA & 47.67 & 0.880 \\
Form VII & 14 & edge-to-face, slipped-parallel & 60.8$^\circ$ & 1.900~\AA & 4.124~\AA & 35.73 & 1.557 \\
Form VIII & 5 & face-to-face & 0.3$^\circ$ & 3.389~\AA & 1.766~\AA & 24.93 & 1.722 \\
Form IX & 3 & slipped-parallel & 0.0$^\circ$ & 4.764~\AA & 3.318~\AA & 25.04 & 1.826 \\
\bottomrule
\end{tabular}
\par\smallskip
{\footnotesize Motif Jaccard indices: Form IX 0.500, Form II 0.333, Form III 0.333, Form IV 0.333, Form VII 0.333; Form VI, Form VIII 0.000.}
\end{table}
 
\begin{table}[htbp]
\centering
\caption{Contact length Wasserstein distances (shared contact types only). 23DH reference lengths: C--C 3.764~\AA, C--H 3.154~\AA, H--H 2.655~\AA, C--O 3.534~\AA, H--O 3.086~\AA. For acridine, C--O and H--O denote the relabelled C--N and H--N contacts.}
\label{tab:si-23dh-t5a}
\scriptsize
\begin{tabular}{lccccccccccc}
\toprule
 & \multicolumn{2}{c}{C--C} & \multicolumn{2}{c}{C--H} & \multicolumn{2}{c}{H--H} & \multicolumn{2}{c}{C--O} & \multicolumn{2}{c}{H--O} & \\\cmidrule(lr){2-3}\cmidrule(lr){4-5}\cmidrule(lr){6-7}\cmidrule(lr){8-9}\cmidrule(lr){10-11}
Polymorph & $W$ & mean (\AA) & $W$ & mean (\AA) & $W$ & mean (\AA) & $W$ & mean (\AA) & $W$ & mean (\AA) & Mean $W$ \\
\midrule
Form II & 0.035 & 3.775 & 0.085 & 3.231 & 0.126 & 2.772 & 0.095 & 3.619 & 0.100 & 3.026 & 0.088 \\
Form VII & 0.046 & 3.801 & 0.063 & 3.215 & 0.146 & 2.794 & 0.114 & 3.638 & 0.081 & 3.068 & 0.090 \\
Form VI & 0.048 & 3.805 & 0.064 & 3.209 & 0.135 & 2.757 & 0.132 & 3.665 & 0.108 & 2.995 & 0.097 \\
Form IV & 0.072 & 3.833 & 0.095 & 3.248 & 0.147 & 2.794 & 0.135 & 3.664 & 0.093 & 3.030 & 0.108 \\
Form III & 0.044 & 3.805 & 0.089 & 3.231 & 0.108 & 2.747 & 0.167 & 3.702 & 0.138 & 3.083 & 0.109 \\
Form IX & 0.058 & 3.822 & 0.132 & 3.286 & 0.116 & 2.746 & 0.109 & 3.627 & 0.170 & 2.920 & 0.117 \\
Form VIII & 0.052 & 3.814 & 0.049 & 3.192 & 0.158 & 2.641 & 0.147 & 3.682 & 0.275 & 2.811 & 0.136 \\
\bottomrule
\end{tabular}
\end{table}
 
\begin{table}[htbp]
\centering
\caption{Contact anisotropy difference relative to 23DH. Lower values indicate more similar contact geometry. 23DH anisotropies: C--C 0.853, C--H 0.897, H--H 0.958, C--O 0.959, H--O 0.996 (means over its independent molecules). Rank is by mean $\Delta$.}
\label{tab:si-23dh-t5b}
\small
\begin{tabular}{lccccccc}
\toprule
Polymorph & C--C $\Delta$ & C--H $\Delta$ & H--H $\Delta$ & C--O $\Delta$ & H--O $\Delta$ & Mean $\Delta$ & Rank \\
\midrule
Form II & 0.049 & 0.034 & 0.017 & 0.052 & 0.018 & 0.034 & 1 \\
Form III & 0.040 & 0.060 & 0.019 & 0.037 & 0.026 & 0.036 & 2 \\
Form IX & 0.083 & 0.244 & 0.180 & 0.037 & 0.023 & 0.113 & 3 \\
Form VIII & 0.162 & 0.306 & 0.320 & 0.039 & 0.029 & 0.171 & 4 \\
Form VI & 0.237 & 0.343 & 0.158 & 0.085 & 0.105 & 0.186 & 5 \\
Form VII & 0.287 & 0.438 & 0.182 & 0.093 & 0.063 & 0.212 & 6 \\
Form IV & 0.240 & 0.462 & 0.432 & 0.156 & 0.145 & 0.287 & 7 \\
\bottomrule
\end{tabular}
\end{table}

\clearpage
\subsection{\textit{trans}-Azobenzene (AZO)}
\label{si:topo-azo}
Tables~\ref{tab:si-azo-t1}--\ref{tab:si-azo-t5b} give the comparison of
\textit{trans}-azobenzene (AZOBEN03) with the seven acridine polymorphs, in the same format as
for the other additives.
 
\paragraph{Fragment decomposition.} Azobenzene is not a single rigid unit, so it is
represented by its two phenyl rings, each treated as a rigid C$_6$H$_5$ fragment; the azo
nitrogens are not part of either fragment. AZOBEN03 crystallises in $P2_1/c$ with two
independent half-molecules, each on an inversion centre, giving four phenyl instances in the
asymmetric-unit description. All comparisons are averaged over these instances. This choice
has two consequences:
\begin{itemize}
  \item The relative-orientation angle $\omega$ takes only seven values (0$^\circ$,
  90$^\circ$, 97.1$^\circ$, 102.1$^\circ$, 142.2$^\circ$, 151.2$^\circ$ and 180$^\circ$). A
  discrete $\omega$ set is expected for any rigid fragment, one value per relative
  orientation of symmetry-related fragments (three values for 27DH and for 23DH); AZO has
  more because its four phenyl rings adopt four distinct orientations.
  \item The two rings of the same molecule are counted as neighbours of each other. Their
  centroids are 6.33 and 6.35~\AA{} apart in the two independent molecules, so the
  centre-of-mass RDF contains an intramolecular contribution at about 6.3~\AA{} that has no
  counterpart in acridine, and the coordination counts in the first two shells are inflated
  accordingly.
\end{itemize}
 
\paragraph{Contact types.} Unlike the diol comparisons, in which the acridine C--N and H--N
contacts are relabelled C--O and H--O to match the hydroxyl oxygen of the additive, the AZO
comparison uses the element labels directly. Contacts to the acridine ring nitrogen are
therefore compared with contacts from the phenyl rings of AZO to the azo nitrogens of
neighbouring molecules, and all five contact types (C--C, C--H, H--H, C--N and H--N) are
shared, giving a contact-type Jaccard index of 1 for every polymorph. AZO forms 49.5 contacts
per phenyl fragment. The nitrogen contacts of AZO are, however, sparse: 4 C--N and
8 H--N contacts in total over the four phenyl fragments, compared with 9--20 nitrogen
contacts per acridine molecule. Their contact-length and anisotropy comparisons
(Tables~\ref{tab:si-azo-t5a} and \ref{tab:si-azo-t5b}) should therefore be read with caution;
in particular, the anisotropy of AZO for these two types is 1 by construction.
AZO forms no hydrogen bonds.
 
\begin{table}[htbp]
\centering
\caption{Crystal packing: global cell properties. AZO reference: $Z=4$, $Z'=1$, PC $= 0.681$, space group $P2_1/c$ (symmetry elements: inversion, $2_1$, glide).}
\label{tab:si-azo-t1}
\small
\begin{tabular}{lcccccl}
\toprule
Polymorph & $Z$ & \Zp & PC similarity & Cell shape $W$ & Cell angle $\overline{|\Delta\cos|}$ & Sym.\ match / shared \\
\midrule
Form II & 4 & 1 & 0.221 & 0.065 & 0.072 & 1.000 --- inversion, $2_1$, glide \\
Form III & 8 & 2 & 0.965 & 0.096 & 0.097 & 1.000 --- inversion, $2_1$, glide \\
Form IV & 12 & 3 & 0.840 & 0.062 & 0.127 & 0.800 --- $2_1$ \\
Form VI & 8 & 2 & 0.867 & 0.048 & 0.089 & 0.700 --- glide \\
Form VII & 8 & 2 & 0.481 & 0.036 & 0.093 & 1.000 --- inversion, $2_1$, glide \\
Form VIII & 8 & 2 & 0.746 & 0.141 & 0.020 & 0.700 --- glide \\
Form IX & 4 & 1 & 0.919 & 0.124 & 0.115 & 1.000 --- inversion, $2_1$, glide \\
\bottomrule
\end{tabular}
\end{table}
 
\begin{table}[htbp]
\centering
\caption{Crystal packing: local molecular environment (AZO reference).}
\label{tab:si-azo-t2}
\small
\begin{tabular}{lccccccc}
\toprule
Polymorph & RDF $W$ & Shell $\Delta$ & Angular $\varphi$ $W$ & Angular $\theta$ $W$ & $Q_2$ $W$ & $Q_4$ $W$ & $Q_6$ $W$ \\
\midrule
Form II & 4.458 & 0.390 & 0.456 & 1.733 & 0.086 & 0.212 & 0.030 \\
Form III & 3.008 & 0.492 & 1.510 & 0.864 & 0.045 & 0.079 & 0.045 \\
Form IV & 3.383 & 0.548 & 1.427 & 1.120 & 0.034 & 0.074 & 0.021 \\
Form VI & 4.453 & 0.522 & 1.321 & 1.121 & 0.021 & 0.050 & 0.034 \\
Form VII & 3.553 & 0.499 & 1.361 & 0.924 & 0.056 & 0.107 & 0.030 \\
Form VIII & 5.202 & 0.533 & 0.689 & 1.135 & 0.022 & 0.071 & 0.049 \\
Form IX & 8.714 & 0.556 & 1.381 & 2.002 & 0.027 & 0.028 & 0.054 \\
\bottomrule
\end{tabular}
\end{table}
 
\begin{table}[htbp]
\centering
\caption{Fragment orientation, sorted by $\omega$ $W$. AZO reference: $\omega$ mean $= 118.9^\circ$.}
\label{tab:si-azo-t3}
\small
\begin{tabular}{lcccccccc}
\toprule
Polymorph & \Zp & $\omega$ $W$ & $\omega$ mean & $\Delta$ mean & $\omega$ std & $\varphi$ $W$ & $\theta$ $W$ & Plan.\ $W$ \\
\midrule
Form VII & 2 & 12.24 & 120.9$^\circ$ & 2.3$^\circ$ & 1.027 & 37.2 & 19.9 & 0.042 \\
Form III & 2 & 14.50 & 126.2$^\circ$ & 7.3$^\circ$ & 1.976 & 31.8 & 16.3 & 0.023 \\
Form IV & 3 & 14.59 & 123.8$^\circ$ & 5.0$^\circ$ & 3.093 & 41.0 & 23.9 & 0.021 \\
Form IX & 1 & 16.75 & 126.0$^\circ$ & 7.1$^\circ$ & 0.559 & 54.6 & 41.1 & 0.028 \\
Form II & 1 & 17.56 & 120.0$^\circ$ & 1.1$^\circ$ & 0.558 & 61.6 & 61.8 & 0.056 \\
Form VI & 2 & 25.55 & 98.9$^\circ$ & 19.9$^\circ$ & 0.559 & 50.2 & 19.9 & 0.045 \\
Form VIII & 2 & 69.42 & 49.5$^\circ$ & 69.4$^\circ$ & 0.594 & 67.1 & 30.1 & 0.026 \\
\bottomrule
\end{tabular}
\par\smallskip
{\footnotesize $\Delta$ mean is the mean over instance pairs of the absolute difference in mean $\omega$; $\omega$ std is the standard deviation of $\omega$ $W$ over instance pairs.}
\end{table}
 
\begin{table}[htbp]
\centering
\caption{$\pi$--$\pi$ stacking geometry. AZO reference: 327 interactions, motifs edge-to-face, face-to-face, intermediate, non-stacking, angle 35.8$^\circ$, perpendicular spacing 2.81~\AA, offset 3.42~\AA.}
\label{tab:si-azo-t4}
\footnotesize
\begin{tabular}{lc>{\raggedright\arraybackslash}p{3.4cm}ccccc}
\toprule
Polymorph & $n$ & Motifs & Angle mean & Perp.\ mean & Offset mean & Angle $W$ & Perp.\ $W$ \\
\midrule
Form II & 12 & non-stacking, slipped-parallel & 2.9$^\circ$ & 0.364~\AA & 5.153~\AA & 32.83 & 2.463 \\
Form III & 14 & edge-to-face, slipped-parallel & 37.1$^\circ$ & 1.858~\AA & 4.304~\AA & 8.86 & 1.022 \\
Form IV & 9 & edge-to-face, intermediate & 59.9$^\circ$ & 1.958~\AA & 4.552~\AA & 27.82 & 1.639 \\
Form VI & 10 & edge-to-face & 72.7$^\circ$ & 2.616~\AA & 3.398~\AA & 36.94 & 0.983 \\
Form VII & 14 & edge-to-face, slipped-parallel & 60.8$^\circ$ & 1.900~\AA & 4.124~\AA & 25.00 & 1.250 \\
Form VIII & 5 & face-to-face & 0.3$^\circ$ & 3.389~\AA & 1.766~\AA & 35.64 & 1.506 \\
Form IX & 3 & slipped-parallel & 0.0$^\circ$ & 4.764~\AA & 3.318~\AA & 35.76 & 2.183 \\
\bottomrule
\end{tabular}
\par\smallskip
{\footnotesize Motif Jaccard indices: Form IV 0.500, Form VI 0.250, Form VIII 0.250, Form II 0.200, Form III 0.200, Form VII 0.200; Form IX 0.000.}
\end{table}
 
\begin{table}[htbp]
\centering
\caption{Contact length Wasserstein distances (shared contact types only). AZO reference lengths: C--C 3.828~\AA, C--H 3.224~\AA, H--H 2.785~\AA, C--N 3.730~\AA, H--N 3.126~\AA.  The C--N and H--N contacts are those to the ring nitrogen in acridine and to the azo nitrogens in AZO; AZO has only 4 C--N and 8 H--N contacts in total over its four phenyl fragments (0--2 per fragment), so its values for these two types rest on very few contacts.}
\label{tab:si-azo-t5a}
\scriptsize
\begin{tabular}{lccccccccccc}
\toprule
 & \multicolumn{2}{c}{C--C} & \multicolumn{2}{c}{C--H} & \multicolumn{2}{c}{H--H} & \multicolumn{2}{c}{C--N} & \multicolumn{2}{c}{H--N} & \\\cmidrule(lr){2-3}\cmidrule(lr){4-5}\cmidrule(lr){6-7}\cmidrule(lr){8-9}\cmidrule(lr){10-11}
Polymorph & $W$ & mean (\AA) & $W$ & mean (\AA) & $W$ & mean (\AA) & $W$ & mean (\AA) & $W$ & mean (\AA) & Mean $W$ \\
\midrule
Form VII & 0.050 & 3.801 & 0.060 & 3.215 & 0.090 & 2.794 & 0.103 & 3.638 & 0.150 & 3.068 & 0.091 \\
Form III & 0.045 & 3.805 & 0.057 & 3.231 & 0.091 & 2.747 & 0.078 & 3.702 & 0.212 & 3.083 & 0.097 \\
Form IV & 0.037 & 3.833 & 0.057 & 3.248 & 0.084 & 2.794 & 0.122 & 3.664 & 0.194 & 3.030 & 0.099 \\
Form VI & 0.050 & 3.805 & 0.074 & 3.209 & 0.094 & 2.757 & 0.128 & 3.665 & 0.166 & 2.995 & 0.103 \\
Form II & 0.076 & 3.775 & 0.066 & 3.231 & 0.082 & 2.772 & 0.113 & 3.619 & 0.209 & 3.026 & 0.109 \\
Form IX & 0.040 & 3.822 & 0.086 & 3.286 & 0.126 & 2.746 & 0.103 & 3.627 & 0.244 & 2.920 & 0.120 \\
Form VIII & 0.039 & 3.814 & 0.061 & 3.192 & 0.175 & 2.641 & 0.116 & 3.682 & 0.315 & 2.811 & 0.141 \\
\bottomrule
\end{tabular}
\end{table}
 
\begin{table}[htbp]
\centering
\caption{Contact anisotropy difference relative to AZO. Lower values indicate more similar contact geometry. AZO anisotropies: C--C 0.804, C--H 0.784, H--H 0.928, C--N 1.000, H--N 1.000. Rank is by mean $\Delta$. Because AZO has at most two C--N or H--N contacts per fragment, its anisotropy for these types is 1 by construction, so the C--N and H--N $\Delta$ values reduce to one minus the acridine value; ranked on the C--C, C--H and H--H contacts alone, the order is Form II, III, IX, VIII, VI, VII, IV.}
\label{tab:si-azo-t5b}
\small
\begin{tabular}{lccccccc}
\toprule
Polymorph & C--C $\Delta$ & C--H $\Delta$ & H--H $\Delta$ & C--N $\Delta$ & H--N $\Delta$ & Mean $\Delta$ & Rank \\
\midrule
Form III & 0.151 & 0.167 & 0.066 & 0.036 & 0.030 & 0.090 & 1 \\
Form II & 0.150 & 0.158 & 0.066 & 0.092 & 0.022 & 0.098 & 2 \\
Form IX & 0.150 & 0.188 & 0.149 & 0.031 & 0.027 & 0.109 & 3 \\
Form VIII & 0.179 & 0.223 & 0.289 & 0.078 & 0.029 & 0.160 & 4 \\
Form VII & 0.268 & 0.363 & 0.151 & 0.111 & 0.067 & 0.192 & 5 \\
Form VI & 0.267 & 0.346 & 0.167 & 0.121 & 0.109 & 0.202 & 6 \\
Form IV & 0.206 & 0.352 & 0.401 & 0.181 & 0.149 & 0.258 & 7 \\
\bottomrule
\end{tabular}
\end{table}

\clearpage
\section{Supplementary figures}
\label{si:figures}

\begin{figure}[htbp]
\centering
\includegraphics[width=0.9\linewidth]{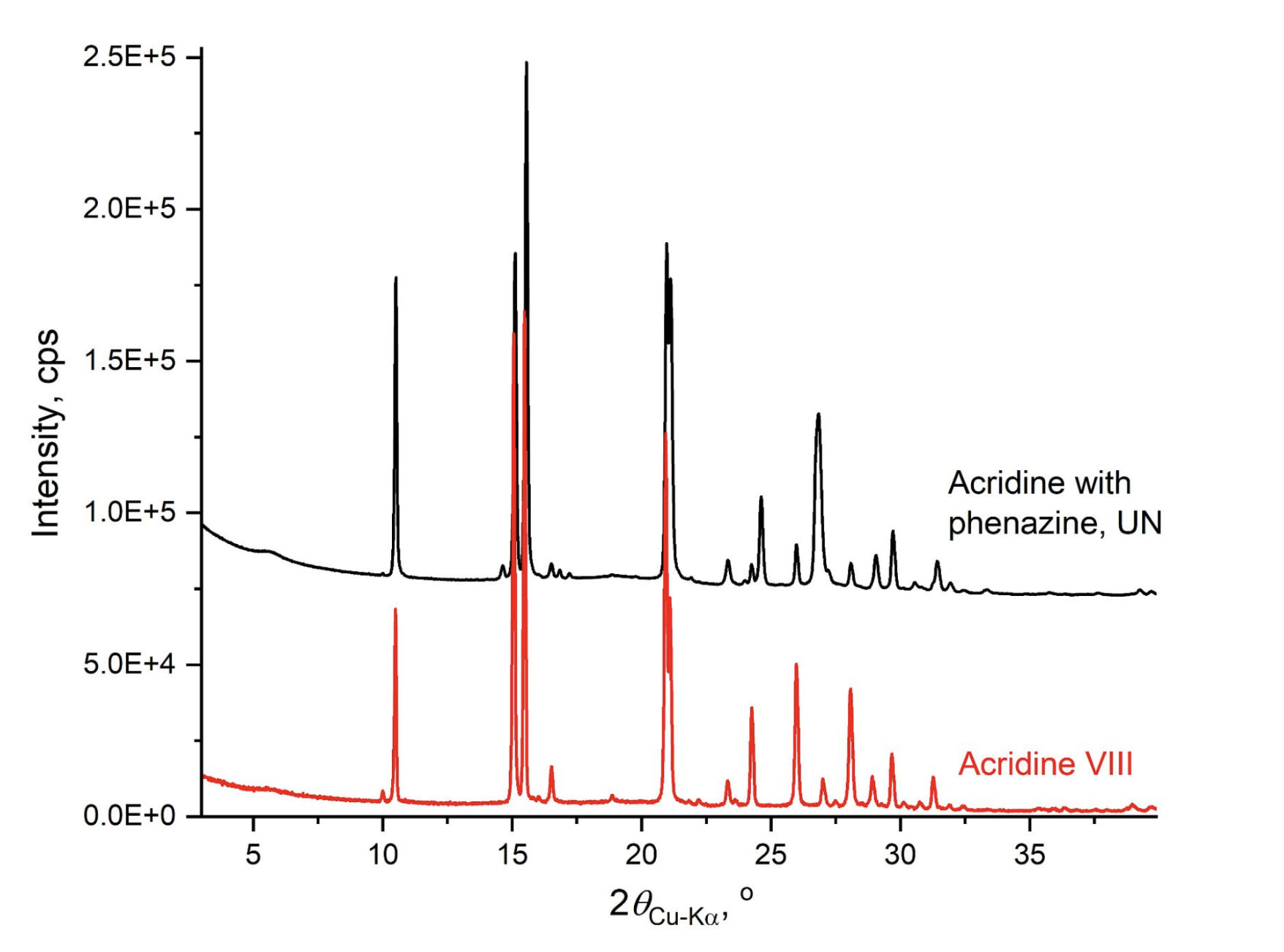}
\caption{Powder diffraction patterns of acridine Form VIII and the new
form UN crystallized from the melt of acridine containing 5~wt.\%
phenazine. } 
\label{fig:si-pxrd}
\end{figure}

\begin{figure}[htbp]
\centering
\includegraphics[width=\linewidth]{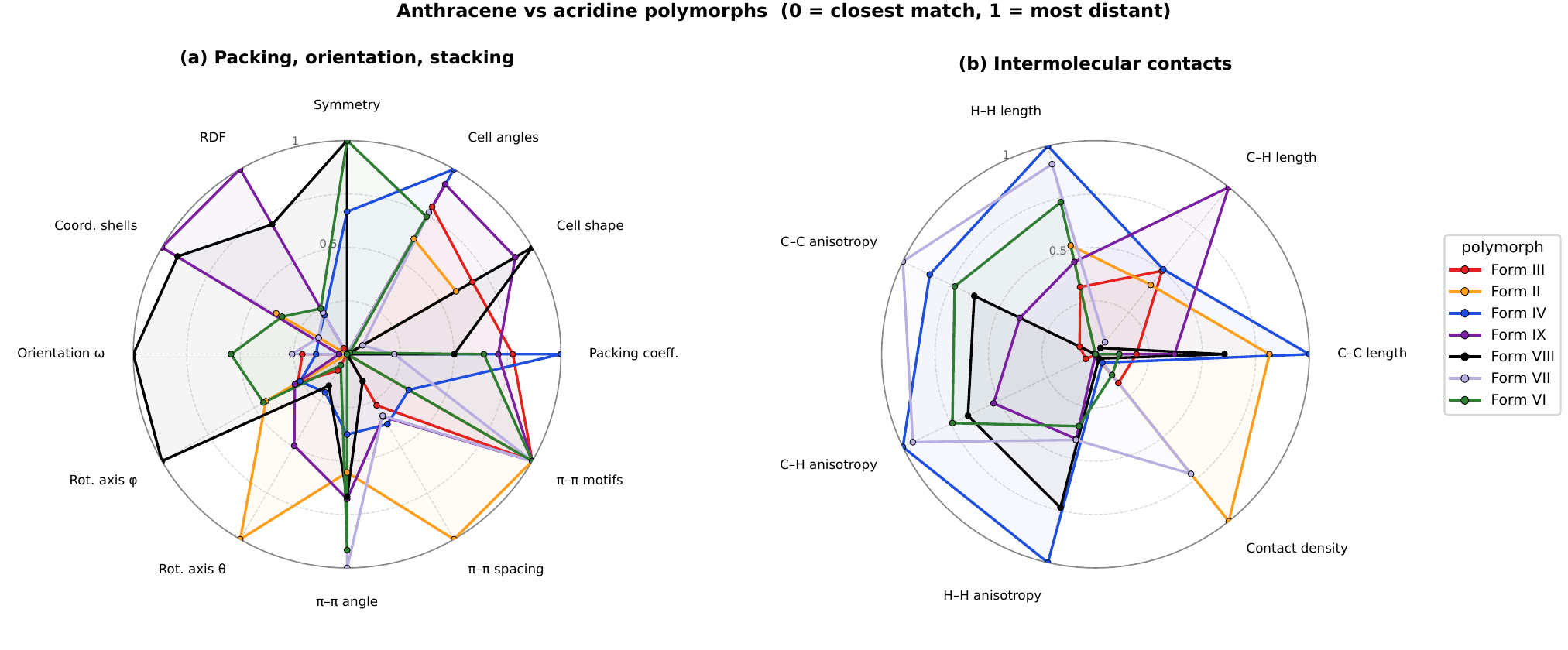}
\caption{Radar plots of the topological similarity between anthracene (ANTCEN17) and the seven
acridine polymorphs. (a) Crystal packing (packing coefficient, cell shape, cell angles, symmetry,
COM radial distribution function, coordination shells), fragment orientation ($\omega$ and the
rotation-axis angles $\phi$, $\theta$) and $\pi$--$\pi$ stacking (interplanar angle, spacing,
motif set). (b) Intermolecular contacts: length distributions and anisotropy of the C--C, C--H
and H--H contacts, and contact density. Each metric is min--max scaled independently across the
seven polymorphs, so that the centre (0) marks the polymorph closest to anthracene and the rim (1)
the most distant; similarity measures (packing-coefficient similarity, symmetry Hamming similarity,
motif Jaccard index) are converted to $1-s$ before scaling. The scaled values rank the polymorphs
within each metric only and are not combined into a score; raw values are given in
Tables~\ref{tab:si-anth-t1}--\ref{tab:si-anth-t5b}}
\label{fig:si-radar-anth}
\end{figure}


\begin{figure}[htbp]
\centering
\includegraphics[width=\linewidth]{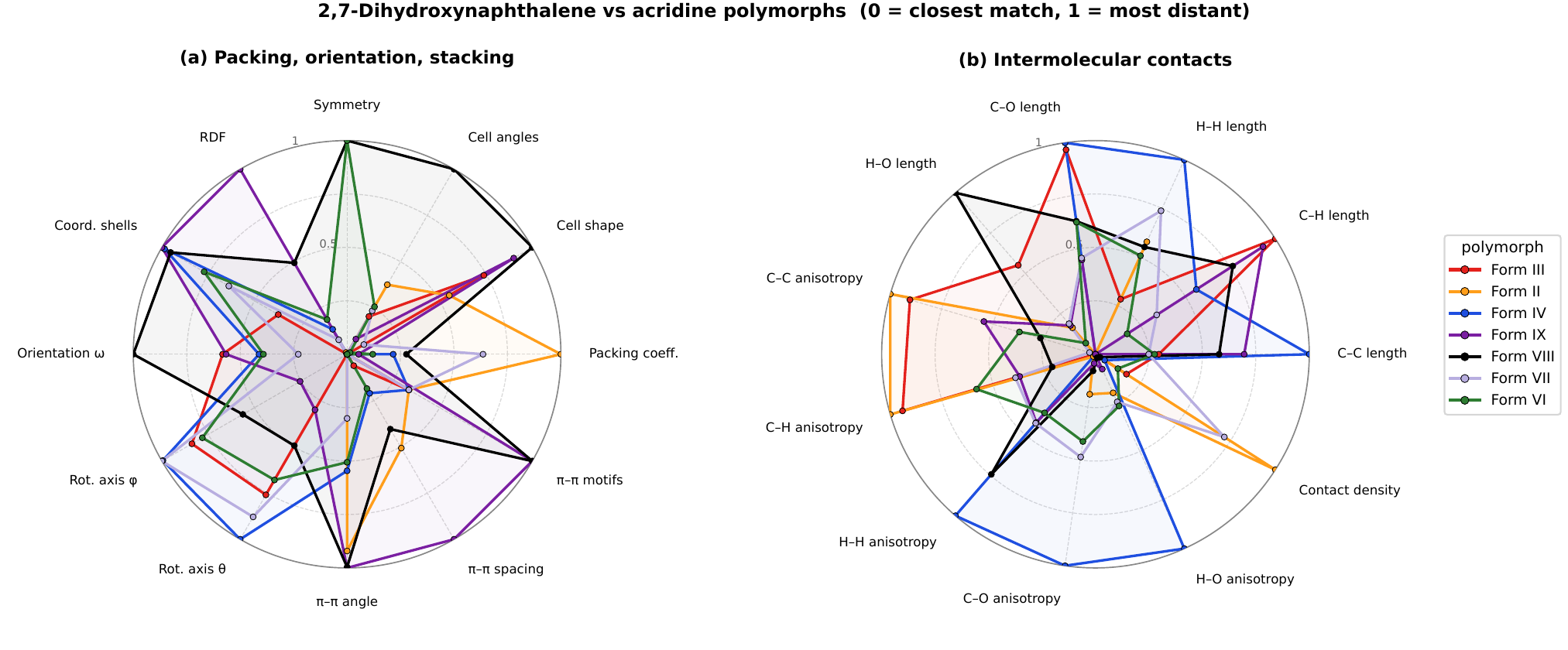}
\caption{Radar plots of the topological similarity between naphthalene-2,7-diol (27DH, NPHLDL01)
and the seven acridine polymorphs. (a) Crystal packing, fragment orientation and $\pi$--$\pi$
stacking metrics, as in Figure~\ref{fig:si-radar-anth}. (b) Intermolecular contacts: length
distributions and anisotropy of the C--C, C--H, H--H, C--O and H--O contacts, and contact density.
For the acridine polymorphs, C--O and H--O denote the C--N and H--N contacts, relabelled so that
the ring nitrogen of acridine is compared with the hydroxyl oxygen of 27DH. Each metric is min--max
scaled independently across the seven polymorphs (centre, 0 = closest to 27DH; rim, 1 = most
distant), with similarity measures converted to $1-s$ before scaling. The scaled values are not
combined into a score; raw values are given in
Tables~\ref{tab:si-27dh-t1}--\ref{tab:si-27dh-t5b}.}
\label{fig:si-radar-27dh}
\end{figure}


\begin{figure}[htbp]
\centering
\includegraphics[width=\linewidth]{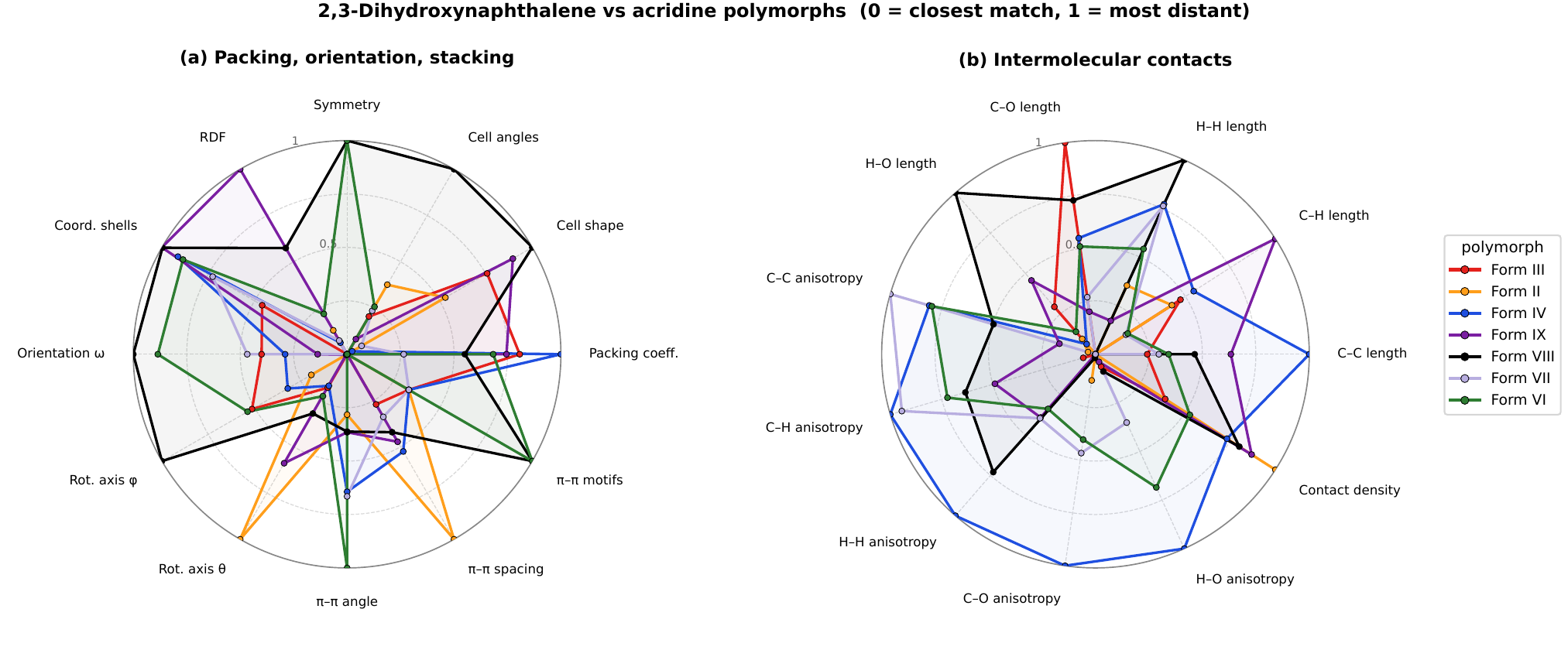}
\caption{Radar plots of the topological similarity between naphthalene-2,3-diol (23DH, VOGSEP01)
and the seven acridine polymorphs. (a) Crystal packing, fragment orientation and $\pi$--$\pi$
stacking metrics, as in Figure~\ref{fig:si-radar-anth}. (b) Intermolecular contacts: length
distributions and anisotropy of the C--C, C--H, H--H, C--O and H--O contacts, and contact density.
For the acridine polymorphs, C--O and H--O denote the C--N and H--N contacts, relabelled so that
the ring nitrogen of acridine is compared with the hydroxyl oxygen of 23DH. Each metric is min--max
scaled independently across the seven polymorphs (centre, 0 = closest to 23DH; rim, 1 = most
distant), with similarity measures converted to $1-s$ before scaling. The scaled values are not
combined into a score; raw values are given in
Tables~\ref{tab:si-23dh-t1}--\ref{tab:si-23dh-t5b}.}
\label{fig:si-radar-23dh}
\end{figure}


\begin{figure}[htbp]
\centering
\includegraphics[width=\linewidth]{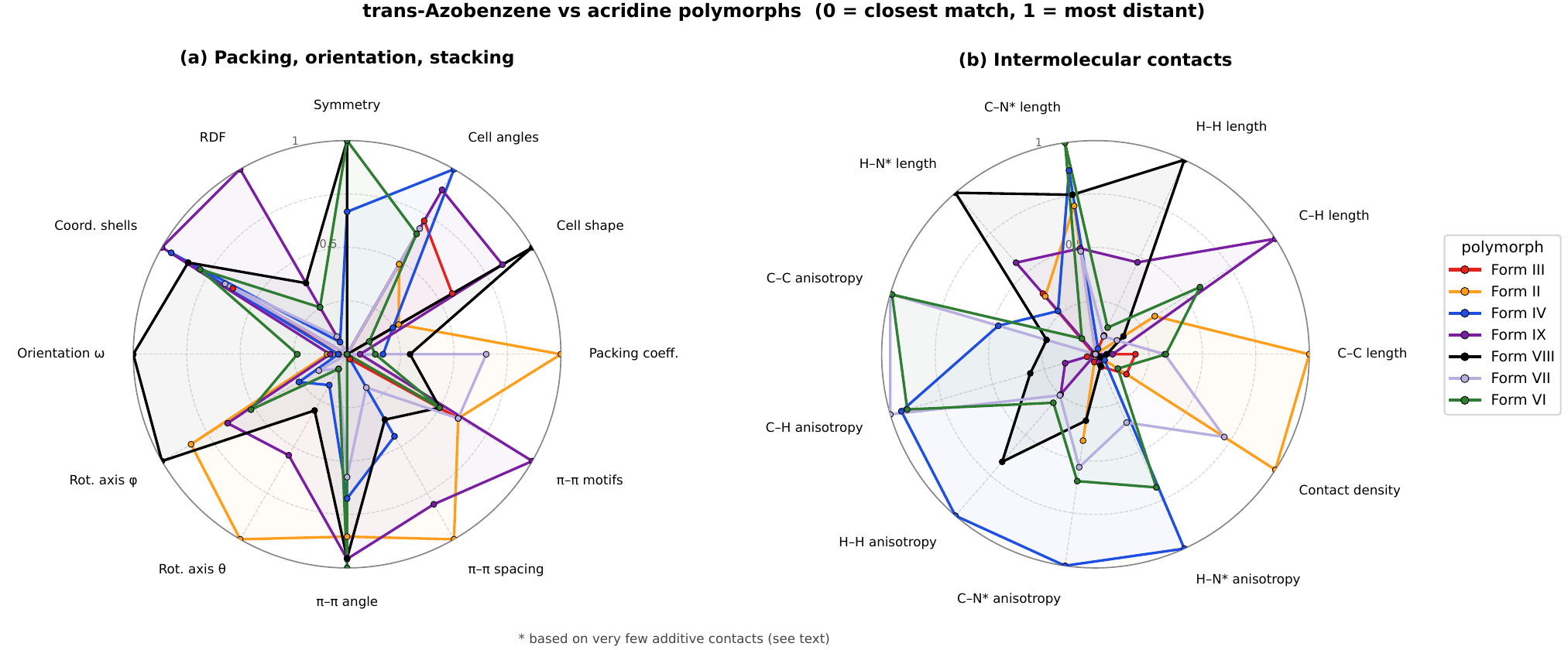}
\caption{Radar plots of the topological similarity between \textit{trans}-azobenzene (AZO,
AZOBEN03) and the seven acridine polymorphs, with each AZO phenyl ring (C$_6$H$_5$) treated as a
separate fragment. (a) Crystal packing, fragment orientation and $\pi$--$\pi$ stacking metrics, as
in Figure~\ref{fig:si-radar-anth}. (b) Intermolecular contacts: length distributions and
anisotropy of the C--C, C--H, H--H, C--N and H--N contacts, and contact density. No contact-type
relabelling is applied: C--N and H--N are contacts to the ring nitrogen in acridine and to the azo
nitrogens in AZO. Axes marked * rest on very few AZO contacts (4 C--N and 8 H--N in total, at most
two per fragment), so the AZO anisotropy for these types is 1 by construction and these axes
should not be over-interpreted. Each metric is min--max scaled independently across the seven
polymorphs (centre, 0 = closest to AZO; rim, 1 = most distant), with similarity measures converted
to $1-s$ before scaling. The scaled values are not combined into a score; raw values are given in
Tables~\ref{tab:si-azo-t1}--\ref{tab:si-azo-t5b}.}
\label{fig:si-radar-azo}
\end{figure}



\clearpage 
\section{Predicted structures}
\label{si:cifs}
The 15 CrystalMath predicted structures are provided as CIF files.
\begin{sidewaystable}
\centering
\caption{The 15 CrystalMath predicted structures provided as CIF files. Structures are labelled
\textit{index}-c\textit{cell} as in Figures~2 and~3 of the main text and Table~\ref{tab:si-compack};
file paths are relative to the \texttt{accepted\_structures} folder. Densities are computed from the
cell volume and the symmetry-expanded cell contents; cost per molecule is the total CrystalMath
topological cost divided by $Z'$. The last column gives the experimental form matched at the 20/20
level by COMPACK (Table~\ref{tab:si-compack}); the lower block lists unassigned structures.}
\label{tab:si-cif-list}
\scriptsize
\setlength{\tabcolsep}{2.8pt}
\begin{tabular}{l l c c c r r r r r c r c}
\toprule
Structure & File & Space group & $Z'$ & $Z$ & $a$ (\AA) & $b$ (\AA) & $c$ (\AA) & $\beta$ ($^\circ$) & $V$ (\AA$^3$) & $\rho$ (g\,cm$^{-3}$) & Cost/mol. & Form \\
\midrule
71-c98 & \texttt{zprime\_1/P21n/struct\_000071\_cell98\_gp0.1211\_0.0159\_0.1274.cif} & $P2_1/n$ & 1 & 4 & 10.840 & 5.719 & 13.251 & 99.55 & 810.1 & 1.469 & 511.6 & II \\
24-c10 & \texttt{zprime\_2/P21c/struct\_000024\_cell10\_gp0.2474\_0.2492\_0.0506.cif} & $P2_1/c$ & 2 & 8 & 5.922 & 18.383 & 15.714 & 95.16 & 1703.8 & 1.397 & 279.9 & III \\
0-c16 & \texttt{zprime\_3/P212121/struct\_000000\_cell16\_gp-0.0143\_0.2250\_0.2689.cif} & $P2_12_12_1$ & 3 & 12 & 6.028 & 15.278 & 28.599 & 90.00 & 2633.9 & 1.356 & 207.3 & IV \\
10-c25 & \texttt{zprime\_2/Cc/struct\_000010\_cell25\_gp0.0000\_0.0685\_-0.0000.cif} & $Cc$ & 2 & 8 & 6.107 & 23.456 & 12.462 & 96.50 & 1773.7 & 1.342 & 350.1 & VI \\
102-c35 & \texttt{zprime\_2/P21n/struct\_000102\_cell35\_gp0.4389\_0.1743\_0.0500.cif} & $P2_1/n$ & 2 & 8 & 5.935 & 22.347 & 12.916 & 95.93 & 1703.8 & 1.397 & 310.3 & VII \\
0-c13 & \texttt{zprime\_2/Cc/struct\_000000\_cell13\_gp-0.0000\_0.2477\_-0.0000.cif} & $Cc$ & 2 & 8 & 12.321 & 13.457 & 11.557 & 116.08 & 1721.0 & 1.383 & 48.2 & VIII \\
0-c90 & \texttt{zprime\_1/P21n/struct\_000000\_cell90\_gp0.2700\_0.1135\_0.0200.cif} & $P2_1/n$ & 1 & 4 & 10.744 & 11.663 & 6.667 & 92.04 & 834.9 & 1.426 & 38.7 & IX \\
\midrule
53-c38 & \texttt{zprime\_1/P21n/struct\_000053\_cell38\_gp0.3926\_0.4295\_0.0996.cif} & $P2_1/n$ & 1 & 4 & 10.106 & 10.845 & 9.764 & 116.09 & 961.1 & 1.239 & 217.7 & -- \\
106-c10 & \texttt{zprime\_2/P21n/struct\_000106\_cell10\_gp0.2395\_0.2621\_0.0530.cif} & $P2_1/n$ & 2 & 8 & 6.102 & 18.805 & 16.484 & 95.13 & 1883.9 & 1.264 & 321.2 & -- \\
0-c22 & \texttt{zprime\_2/P21c/struct\_000000\_cell22\_gp0.2468\_0.0014\_0.2138.cif} & $P2_1/c$ & 2 & 8 & 13.093 & 13.641 & 11.517 & 116.15 & 1846.4 & 1.289 & 63.6 & -- \\
17-c16 & \texttt{zprime\_2/P21c/struct\_000017\_cell16\_gp0.2528\_0.4862\_0.2394.cif} & $P2_1/c$ & 2 & 8 & 12.785 & 14.437 & 11.138 & 116.08 & 1846.4 & 1.289 & 102.1 & -- \\
13-c28 & \texttt{zprime\_2/Cc/struct\_000013\_cell28\_gp0.0000\_0.0757\_0.0000.cif} & $Cc$ & 2 & 8 & 6.084 & 23.156 & 12.671 & 96.47 & 1773.7 & 1.342 & 357.9 & -- \\
9-c31 & \texttt{zprime\_2/Cc/struct\_000009\_cell31\_gp0.0000\_0.0743\_0.0000.cif} & $Cc$ & 2 & 8 & 6.034 & 22.622 & 12.946 & 96.45 & 1755.9 & 1.356 & 344.7 & -- \\
0-c19 & \texttt{zprime\_2/C2/struct\_000000\_cell19\_gp0.2493\_0.0000\_0.2501.cif} & $C2$ & 2 & 8 & 12.444 & 13.322 & 11.677 & 116.11 & 1738.4 & 1.370 & 53.4 & -- \\
2-c19 & \texttt{zprime\_2/Cc/struct\_000002\_cell19\_gp0.0000\_0.2503\_0.0000.cif} & $Cc$ & 2 & 8 & 12.458 & 13.287 & 11.695 & 116.11 & 1738.4 & 1.370 & 51.8 & -- \\
\bottomrule
\end{tabular}
\end{sidewaystable}

\clearpage
\bibliographystyle{unsrtnat}
\bibliography{references}